\documentclass{aa}  

\usepackage{multirow}
\usepackage{afterpage}
\usepackage{graphicx}
\usepackage{subcaption}
\usepackage{txfonts}
\usepackage{natbib,twoopt}
\bibpunct{(}{)}{;}{a}{}{,} 
\makeatletter
\newcommandtwoopt{\citeads}[3][][]{\href{http://adsabs.harvard.edu/abs/#3}%
        {\def\hyper@linkstart##1##2{}%
                \textbf\hyper@linkend\@empty\citealp[#1][#2]{#3}}}
\newcommandtwoopt{\citepads}[3][][]{\href{http://adsabs.harvard.edu/abs/#3}%
        {\def\hyper@linkstart##1##2{}%
                \textbf\hyper@linkend\@empty\citep[#1][#2]{#3}}}
\newcommandtwoopt{\citetads}[3][][]{\href{http://adsabs.harvard.edu/abs/#3}%
        {\def\hyper@linkstart##1##2{}%
                \textbf\hyper@linkend\@empty\citet[#1][#2]{#3}}}
\newcommandtwoopt{\citeyearads}[3][][]%
{\href{http://adsabs.harvard.edu/abs/#3}
        {\def\hyper@linkstart##1##2{}%
                \textbf\hyper@linkend\@empty\citeyear[#1][#2]{#3}}}
\makeatother

\begin{document}

   \title{[Ultra] Luminous Infrared Galaxies selected at 90 $\mu$m in the AKARI deep field: a study of AGN types contributing to their infrared emission }
   \titlerunning{[U]LIRGs - on the trail of AGN's types}

   \author{K. Ma\l{}ek \inst{1} \and M. Bankowicz \inst{2}  \and A. Pollo\inst{1,2}   \and V.~Buat \inst{3}     \and  T.~T.~Takeuchi \inst{4}        \and D.~Burgarella \inst{3}     \and T. Goto \inst{5}  
   \and {M.~Malkan} \inst{6} \and {H.~Matsuhara} \inst{7} 
          }

  \institute{{National Centre for Nuclear Research, ul. Ho\.za 69, 00-681 Warszawa, Poland, } 
  \email{Katarzyna.Malek@ncbj.gov.pl}
  \and{The Astronomical Observatory of the Jagiellonian University, ul.\ Orla 171, 30-244 Krak\'{o}w, Poland } 
  \and{Laboratoire d'Astrophysique de Marseille, OAMP, Universit$\acute{e}$ Aix-Marseille, CNRS, 38 rue Fr$\acute{e}$d$\acute{e}$ric Joliot-Curie, 13388 Marseille, cedex 13, France} 
  \and{Department of Particle and Astrophysical Science, Nagoya University, Furo-cho, Chikusa-ku, 464-8602 Nagoya, Japan} 
  \and{Dark Cosmology Centre, Niels Bohr Institute, University of Copenhagen, Juliane Maries Vej 30, 2100 Copenhagen, Denmark} 
  \and{Department of Physics and Astronomy, University of California, Los Angeles, CA 90024} 
  \and{Institute of Space and Astronautical Science, Japan Aerospace Exploration Agency, Sagamihara, 229-8510 Kanagawa, Japan} 
}

   \date{Received September 15, 1996; accepted March 16, 1997\\}

 
  \abstract
 {}
  {The aim of this work is to characterize physical properties of Ultra Luminous Infrared Galaxies (ULIRGs) and Luminous Infrared Galaxies (LIRGs) detected in the far-infrared (FIR) 90~$\mu$m band in the AKARI Deep Field-South (ADF-S) survey. 
  In particular, we want to estimate the  active galactic nucleus (AGN) contribution to the LIRGs and ULIRGs' infrared emission and which types of AGNs are related to their activity. }
 {We  examined 69 galaxies at redshift $\geq$ 0.05 detected at 90~$\mu$m by the AKARI satellite in the AKARI Deep-Field South (ADF-S), with optical counterparts and spectral coverage from the ultraviolet to the FIR. 
We used two independent spectral energy distribution  fitting codes:  one fitting the SED from FIR to FUV (CIGALE) (we use the results from CIGALE as a reference) and gray-body + power spectrum fit for the infrared part of the spectra (\texttt{CMCIRSED}) in order to identify  a subsample of ULIRGs and LIRGs, and to estimate their properties.}
 {Based on the CIGALE SED fitting, we have found that LIRGs and ULIRGs selected at the 90~$\mu$m AKARI band compose $\sim$ 56\% of our sample (we found 17 ULIRGs and {22} LIRGs, spanning over the redshift range 0.06$<$z$<$1.23). 
 Their physical parameters, such as stellar mass, star formation rate (SFR), and specific SFR are consistent with the ones found for other  samples selected at infrared wavelengths. 
 We have detected a significant  AGN contribution to the mid-infrared luminosity for 63\% of  LIRGs and ULIRGs. 
 Our LIRGs contain Type~1, Type~2, and intermediate types of AGN, whereas for  ULIRGs, a majority (more than 50\%) of AGN emission originates from Type~2 AGNs. 
 The temperature--luminosity  and temperature--mass  relations for the dust component of ADF--S  LIRGs and ULIRGs   
 indicate that these  relations are shaped by the dust mass and not by the increased dust heating.  
 }
 {We conclude that LIRGs contain Type~1, Type~2, and intermediate types of AGNs, with an AGN contribution to the mid infrared emission at the median level of 13 $\pm$ 3\%, whereas the majority of  ULIRGs contain  Type~2 AGNs, with a median AGN fraction equal to 19 $\pm$ 8\%.}

  \maketitle
%

\section{Introduction}

The first infrared all-sky survey performed by the satellite IRAS (\textit{The Infra-Red Astronomical Satellite}, \citealp{soifer87, neugebauer84}) in the early 1980s established the existence of galaxies emitting very brightly in infrared (IR) wavelengths with total IR luminosities greater than $\rm{10^{12}\mbox{ }[L_{\odot}]}$. These were thus named Ultra Luminous Infrared Galaxies (ULIRGs; \citealp[eg., ][]{Houck1985,soifer86}). 
\cite{Sanders88} reported a discovery of  ten infrared objects with luminosities $\rm{L_{(8-1000\mu m)}\geqslant 10^{12}\mbox{ }[L_{\odot}]}$ in the IRAS Bright Galaxy Sample. 
Analysis of this first sample of ULIRGs has shown that these sources are often interacting galaxies, and in the near-infrared  colors they appear to be a  mixture of starburst and active galactic nuclei (AGN).  

The properties of this sample of galaxies have been the subject of numerous analyses. 
Many physical properties, such as the  dust temperature, the star formation rate, and the mass-luminosity relationship were derived based on the local examples of ULIRG samples, but their global evolution remains unclear \citep{Lonsdale2006}. 
Previous studies indicate that in the local Universe ULIRGs are rather rare objects \citep{kim98}, but at higher redshift they become more common \citep{takeuchi05a, symeonidis11, symeonidis13}. 
A  group of sources, ten times fainter in the IR, known as Luminous
Infrared Galaxies (LIRGs) and characterized by a total  IR luminosity  between $\rm{10^{11}\mbox{ }and\mbox{ }10^{12}\mbox{ }[L_{\odot}]}$, \citealp{soifer86}), is very often analyzed together with ULIRGs.

There is no straightforward explanation of the nature of   LIRGs and ULIRGs. 
It  appears that the majority of local  ULIRGs are merger systems of gas-reach disk galaxies \citep{murphy1996,Clements1996,Veilleux02,Lonsdale2006}, whereas local LIRGs have more varied morphologies \citep{Hung14}; 
they emit most of their energy (over 90\%) in the IR, which implies that they are  heavily obscured by dust \citep{sanders96}. 
Despite a very luminous dust component,  LIRGs and ULIRGs   have moderate luminosities in optical bands. 
For this reason, to explain their very  high IR luminosities, these objects must have ongoing, extreme star-formation processes, forming new stars at rates of an order of 100 $\rm{M_{\odot} yr^{-1}}$ (based on the \citealp{Kennicutt1998} equation, and results published by \citealp[e.g.,][]{Noeske2007,daCunha10,daCunha15,howell10,podigachoski15}).
This very violent star formation is almost undetectable at wavelengths different from IR.

As demonstrated by previous studies, on average, more than 25\% \citep{veilleux97,Ichikawa2014}, 30\% \citep{Clements1996, u12}, 40\% \citep{Farrah07}, 50\% \citep{alonsoherrero12,Carpineti15}, and 70\% \citep{nardini10}  of local  LIRGs and ULIRGs  host an AGN.  
\citealp{Veilleux09} found that all local ULIRGs have some AGN contribution to the bolometric luminosity. 
Based on the results presented above it is possible to conclude that  the percentage of  LIRGs and ULIRGs  hosting an AGN depends on the sample selection and the diagnostic method used to identify the AGN. 
Previous studies have also shown that the relative  contribution of AGNs  to the bolometric luminosities is only a few percent (from 7 to 10\%) of the total luminosity of local LIRGs \citep{PereiraSantaella11, Petric11}, and $\sim$20\% of local ULIRGs \citep{Farrah07,Nardini09}. 
Consequently, the most likely dominant contributor to the total IR emission in most ULIRGs is star formation. 
\cite{Veilleux09}, based on the IRAS 1~Jy sample,  found that the AGN contribution to local ULIRGs may range from 7\% to even 95\% with an average contribution of 35-40\%, and that this value strongly depends on the dust luminosity. 
\cite{Lee11} show that ULIRGs more often host Type~2 AGNs than Type~1 (hereafter: Type 2~ULIRGs, and Type  1~ULIRGs), and the percentage of  type 2~ULIRGs increases with infrared luminosity. 
Their sample of 115 ULIRGs includes only eight broad-line AGNs, and 49 narrow-line AGNs (activity of 58 was found to be non-AGN--related).   
\cite{Kim95} came to a similar conclusion. 
Analogous results were presented by \citealp{Veilleux09} (12 Type~2 ULIRGs and 9 Type~1 ULIRGs), but in this case the difference is insignificant. 
The fact that these findings were based on almost the same selection of objects (IRAS 1~Jy and Spitzer~1~Jy surveys) poses the question as to whether or not   ULIRGs selected in a different way have similar properties. 

Previous studies \citep{CharyElbaz01,lefloch05,PerezGonzalez05,magnelli09,goto10} show that galaxies with total IR luminosity $\rm{>10^{11}\mbox{ }[L_{\odot}]}$  are the major contributors to the star-formation  density at redshifts of $\sim$ 1-2.  
The co-moving number density of LIRGs from the \cite{magnelli09} sample has increased by a factor of approximately $ $100 between $z\sim0$ and $z\sim $ 1.
A similar result was presented by \cite{goto10} based on the AKARI NEP-deep data.  
Their advantage over previous works was a continuous filter coverage in
the mid-IR wavelengths (2.4, 3.2, 4.1, 7, 9, 11, 15, 18, and 24~$\mu$m).  
They found that the ULIRGs contribution increases by a factor of 10  from z = 0.35 to z = 1.4, suggesting that IR-bright galaxies are more dominant sources of total infrared density at higher redshift. 

A very important role of  LIRGs and ULIRGs in galaxy evolution was also shown by \cite{caputi07}, for example. 
Based on the Spitzer GOODS data, they analyzed the evolution of the co-moving bolometric IR luminosity density with redshift for the {LIRG and ULIRG} populations.  
They found that the relative contributions of  LIRGs and ULIRGs   to the total IR luminosity density increase from $\sim 28\%$ for zero  redshift to almost 80\% for a redshift~of one. 
A very similar result,  based on the Spitzer MIPS data, was shown by \cite{lefloch05}, who found that the  IR luminosity density of {LIRG and ULIRG} populations for a redshift of one is equal to 75\%. 
This implies that  LIRGs and ULIRGs play a significant  role in galaxy evolution, and that the detailed  studies of  LIRGs and ULIRGs' physical properties are crucial to trace back the evolution of massive galaxies.

\cite{symeonidis11}, based on the data collected by the submm and IR surveys, showed that different selection processes can result in a significantly different final sample of  LIRGs and ULIRGs. 
The long-wavelength surveys are more sensitive to cold objects, while the selection in the shorter wavelengths is more complete with respect to  all types of LIRGs  and ULIRGs in a large temperature interval. 
Cold luminous infrared galaxies   have the IR peak placed {at} wavelengths longer than  90~$\mu m$, while the warm  LIRGs and ULIRGs   have the IR peak located at wavelengths shorter than this threshold.  
This conclusion illustrates how complicated the global analysis of LIRGs and ULIRGs is, and how many different parameters need to be taken into account to draw final conclusions.

The infrared properties of LIRGs and ULIRGs, as well as those of  AGNs themselves, can smooth the path for the understanding of the star-formation history in the Universe, by  combining information about the galaxy formation and evolution with information about the central black hole masses  \citep{schweitzer06}. 
Unfortunately, the lack of photometric data impedes statistical studies of  LIRGs and ULIRGs \citep[see, e.g., ][]{u12}. 
For this reason, all new analyses, even partial, based on the newest infrared data are very important to broaden our knowledge about most actively star-forming galaxies in the Universe. 

In this paper we present a sample of 39 LIRGs and ULIRGs selected from  one of the deep  surveys created by  the infrared Japanese satellite named AKARI \citep{Murakami07}.  
We have used the far-infrared survey centered  on  the south ecliptic pole, AKARI Deep Field South (ADF--S). 
All sources used for the presented analysis have optical and near infrared counterparts, and based on the multi-wavelength information across ultraviolet--to--far-infrared spectra, we are able to constrain the fraction of AGNs which contributes to their infrared luminosity. 
We have also drawn the luminosity-temperature relationship based on the double-checked selection of ADF--S  LIRGs and ULIRGs, that is, objects selected as  LIRGs and ULIRGs  using two different  spectral energy distribution (SED)  fitting methods:  
one based on the SED fitted from far ultraviolet  to far infrared by  CIGALE code \citep{noll09}, 
and gray body + power law  fitting of the infrared part of the spectrum only, based on the method presented by \cite{casey12}. 

The aim of our work is to estimate the fraction of AGNs in  LIRGs' and ULIRGs'   mid-infrared light, and to determine which types of AGNs contribute to their mid-infrared emission. 
At the same time, ADF-S   allows us to analyze global physical properties of  LIRGs and ULIRGs, such as, star formation rate, stellar mass and dust luminosity,  and also permits comparison of our results with those already presented in the literature for differently selected sources. 
For example, dust temperature and dust luminosity obtained from the \cite{casey12} model can be easily compared to the \cite{symeonidis13} relation for  LIRGs and ULIRGs, where the authors have shown that the dust temperature increases with the dust luminosity but that the relation is not very steep. 
\cite{symeonidis13} claim that the increase in dust temperature is related to the dust mass, and the $\rm{L_{dust}-T_{dust}}$ relation is close to the limiting scenarios of  the Stefan-Boltzman law, where  $\rm{L\propto M_{dust}}$ with $\rm{T_{dust}}$ being a constant value (we refer  the reader to  \citealp{symeonidis13} for more details).  
We  verify whether or not the \cite{symeonidis13} relation  is fulfilled for the 90~$\mu$m  selected sample.

The paper is laid out as follows: the description of the data can be found in Section~\ref{sec:data}; 
in Section~\ref{sec:sampleselection} we present the sample selection. 
Section~\ref{sec:methodology} presents the spectral energy distributions (SED) fitting method implemented in the  CIGALE, and \texttt{CMCIRSED} codes and applied models. 
Section~\ref{sec:SEDresults} contains the discussion of  the main physical properties of the ADF--S  LIRGs and ULIRGs, including dust temperature; dust mass/luminosity relation.
In Section~\ref{sec:AGNs} we present the results concerning fractional AGN contribution to the  LIRGs' and ULIRGs' infrared emission and the types of AGNs related to  LIRGs' and ULIRGs' activity.  
A summary of results obtained from our analysis is presented in Section~\ref{sec:Results}.

\section{Data}
\label{sec:data}
We used the ADF-S for our analysis. 
ADF--S  was covered by four far--infrared (FIR) AKARI bands, centered at 65, 90, 140, and 160~$\mu$m, with a continuous filter coverage, which allows for very precise estimation of the dust luminosity and  temperature, eliminating uncertainties caused by gaps in the filter coverage. 
Observations were made by the Far-Infrared Surveyor (FIS: \citealp{kawada07}), and the field was centered at $\mathrm{RA}=4^{\rm{h}}44^{\rm{m}}00^{\rm{s}}$, $\mathrm{DEC=-53^{\rm{\circ}}20^{'}00^{''}.0}$  J2000 \citep{shirahata09b}.
More than 2~000 sources were detected in the area of $\sim$ 12.3~deg$^2$. 
Based on a sample of sources detected at 90~$\mu$m WIDE-S band and brighter than 0.0301~Jy (which corresponds  to $\sim$~10$\sigma$ detection level), we created a multi-wavelength catalog \citep{malek10}, which contains 545 sources with optical counterparts found in public databases ({NED}, SIMBAD, IRSA). 

Originally, for the identifications of  ADF-S sources, the search for counterparts was performed within a radius of 40'' around each WIDE-S source. 
The estimated accuracy of the position in the slow-scanned images for the Short Wavelength AKARI Detector (SW, bands N60 and WIDE-S) is equal to $\sim$7$''$, and for Long Wavelength AKARI Detector (LW, bands WIDE-L and N160) equals $\sim$10$''$ (I. Yamamura, 2016, private communication).
Nevertheless, the point spread functions of the slow-scanned images at each AKARI band are well  represented  by a double-Gaussian profile, which includes  $\sim$80\% of the flux power \citep{shirahata09}.
The standard deviation of the narrower component of the  90~$\mu$m WIDE-S band is equal to  $30''\pm 1''$. 
However, most of the identified ADF--S sources have counterparts closer than 20'' and the search angle has been reduced to this value, as the identifications at angular distances $>$20'' are relatively chance coincidences. 
The median value of angular distance of the source from a counterpart is 9.7 arcsec.  

This sample was previously used for a more general analysis of properties of active, star forming FIR sources \citep[]{malek10,malek13,malek14}. 
In the present paper, new measurements, mainly from GALEX, Spitzer/MIPS sample \citep[][taken from public database and for a part of ADF--S sources, measured directly from images by ourselves]{Rieke04},  WISE \citep{wright10}, and ATCA-ADFS survey \citep{white12}, as well as new spectroscopic data  not included in our previous studies \citep{sedgwick11,Jones09}, were added to create the extended version  of our multi-wavelength catalog. 

We used 545 galaxies in total. 
In our sample, 98 galaxies have measurements in UV (GALEX), 100 sources have 15, 24 and/or 70~$\mu$m Spitzer measurements, and 280 sources have WISE data (W1, W2, and/or W3; we did not use W4 data as the errors of the photometric measurements were too great to make them usable for the SED fitting).
Detailed information with the name of the survey, the band name, and effective wavelength, as well as number of sources correlated with the ADF-S database, can be found in \citealp[ Table~1]{malek14}. 

To improve our analysis, we have also used \textit{Herschel/SPIRE} photometric measurements \citep{griffin2010} taken from the HerMES  survey \citep{Olivier2012} second data release located in the HeDaM Herschel Database in Marseille\footnote{http://hedam.lam.fr/}. 
The search for \textit{Herschel/SPIRE} counterparts was performed within a radius of 10'', but  most of the identified sources have counterparts closer than 5'' (mean angular distance equal to 2.48~$\pm$~1.91''; our result is in agreement with the typical astrometric accuracy of SPIRE maps as given by {Swinyard, 2010, and Bendo, 2013, for example}). 
\textit{Herschel/SPIRE} measurements allow us to compute  dust temperature and dust luminosity for our sample of galaxies with a very good precision. 
Thanks to the far-infrared data we are able to model the infrared part of the {LIRG and ULIRG} spectra including the influence of the warm dust located in the galaxy.   


The spectroscopic redshift information is available for 183 from 545 galaxies from the sample. 
As the redshift is an essential parameter to compute the SED,  we have also used an additional 106 of the 113 photometric redshifts estimated with the aid of the CIGALE code \citep{noll09}, and published by \cite{malek14}. 
Nine estimated photometric redshifts from this sample  were replaced by spectroscopic redshifts from the 6dF Galaxy Survey \citep{jones04}. 
For nine objects from the 6dF Galaxy Survey  we found that the  redshift accuracy calculated between spectroscopic redshifts and redshifts estimated by \cite{malek14}  equal to 0.074, with no catastrophic errors.   

Photometric redshifts were estimated for galaxies with at least six photometric measurements in the whole spectral range. 
In \cite{malek14} we performed a series of tests of the accuracy of the photometric redshift estimated by CIGALE. 
As representative parameters describing the accuracy of our method, we use: 
\begin{itemize}
\item the percentage of catastrophic errors (CE), which meet the condition:
\begin{equation}
\rm{CE:= \frac{|z_{spect.}-z_{photo}|}{1+z_{spect.}}>CE_{AKARI} }
.\end{equation}
\item the estimated redshift accuracy, calculated  as: $\rm{|z_{spect}-z_{photo}|/(1+z_{spect.}).}$
\end{itemize}
CE was defined by \cite{ilbert06} as the difference between photometric and spectroscopic redshifts largely greater than the expected uncertainty for the sample.  
The limiting redshift accuracy ($\rm{CE=0.15}$) was calculated by \cite{ilbert06} for the statistically significant sample of 3~241  VVDS galaxies and based on the scatter between spectroscopic and photometric redshifts. 
For this sample the number of catastrophic errors increases with decreasing magnitude (in the case of the VVDS the redshift accuracy decreases by a factor of two between observed magnitude ranges  17.5$<$i'$<$21.5 and  23.5$<$i'$<$24). 
Regarding the ADF–S survey, as the number of spectroscopic measurements is very low, we are not able to calculate any statistically significant specific threshold for our sample.
We decided to adapt an equation given by \cite{Rafelski2015}, where the limiting value for the CE is calculated as:
\begin{equation}
\label{CE_AKARI_PART1}
\rm{CE_{AKARI}=5*\sigma_{NMAD\mbox{ }AKARI},}
\end{equation}
where
\begin{equation}
\label{CE_AKARI_PART2}
\rm{\sigma_{NMAD\mbox{ }AKARI}=1.48\times\mid\frac{(z_{spect}-z_{photo})-median(\Delta z)}{1+z_{spec}}\mid.}
\end{equation}
$\rm{CE_{AKARI}}$ calculated from  Eqs.~\ref{CE_AKARI_PART1} and~\ref{CE_AKARI_PART2} is equal to  0.222, and:  
(1) is much bigger than the CE estimated by \cite{ilbert06} for the VVDS data, widely used in literature for different surveys \cite[i.e.,][]{Speagle2016,Moutard2016,Castellano2016,Simm2015,Dahlen2013}, and 
(2) we have only two galaxies with redshift accuracy $>$ 0.222 (which means almost no CEs in our sample). 
As our sample is not numerous, and the statistical significance of the obtained $\rm{CE_{AKARI}}$ might be weak,  we decided to  use  the threshold  given by \cite{ilbert06} equal to 0.15 instead of the calculated
value of  0.222, even though it is more conservative than ours. 
The number of CEs calculated with $\rm{CE_{AKARI}}$=0.15  increases to eight (10\%) in total. 

Our mean redshift accuracy is equal to 0.056 (calculated as the normalized median absolute deviation\footnote{$\sigma_{MED}=1.48\times|\Delta z/(1+z_{spectr})|$}=0.048), 
much lower than  the one obtained by \cite{ilbert06} with Photometric Analysis for Redshift Estimate (Le PHARE, \citealp{arnouts99,ilbert06}), but this might be attributed to the  much poorer statistics of our sample (95 galaxies) when compared to \citealp{ilbert06,ilbert13} (3241 and 9389 galaxies, respectively).  
For a more detailed description of the ADF--S photometric redshift estimation process, we refer the reader to \cite{malek14}. 

\begin{figure}[t!]
        \begin{center}
                \includegraphics[width=0.5\textwidth, clip]{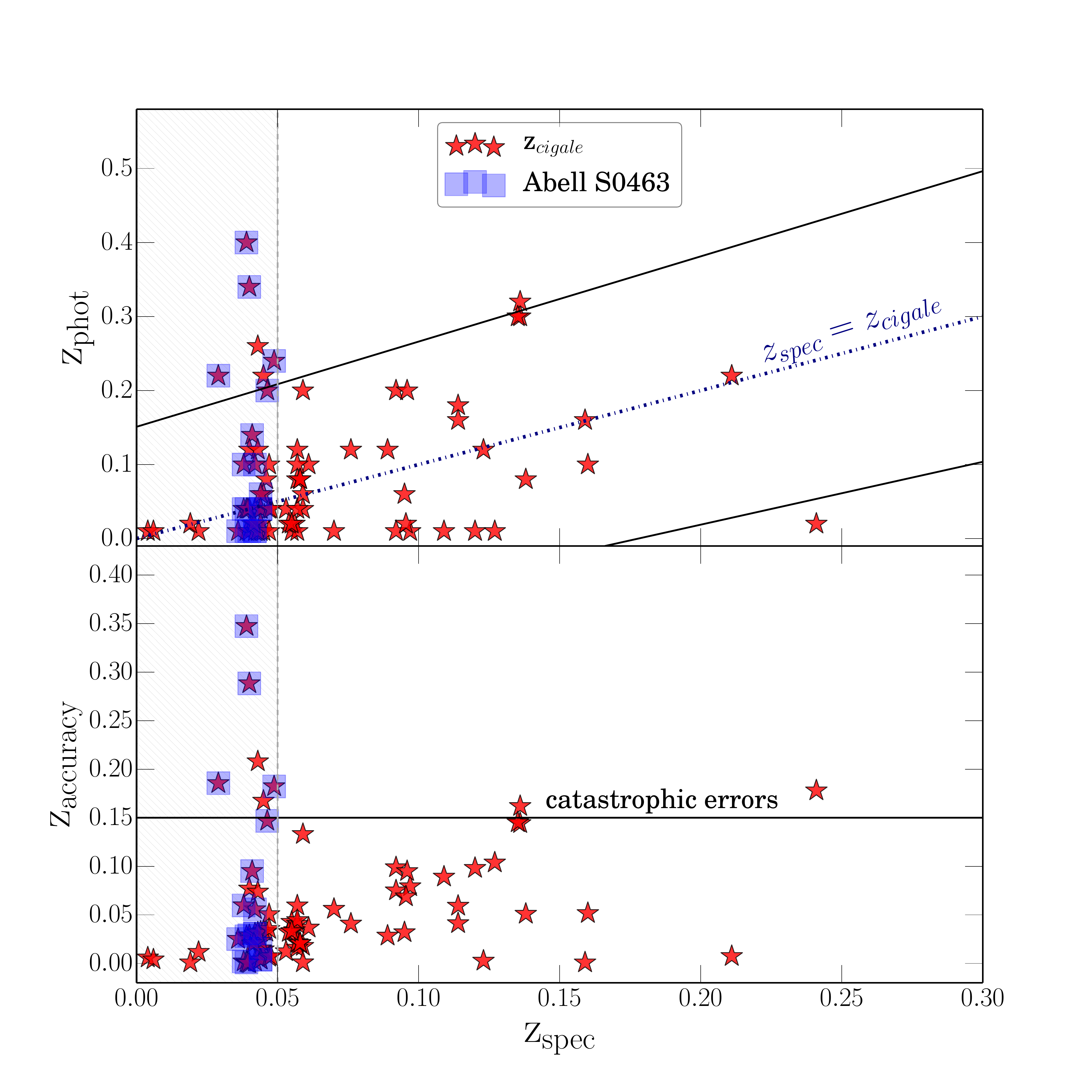}
                \caption{Photometric versus spectroscopic redshifts  and redshift accuracy versus spectroscopic redshift for a sample of 84 galaxies (red stars) for which the photometric redshift was calculated by \cite{malek14} using the CIGALE tool. 
                Blue squares correspond to the Abell S0463 cluster.  
                The region of CEs, defined as $|z_{spec} -  z_{photo}|/(1+z_{spec}) > 0.15$ \citep{ilbert06}, is marked by a solid black line. 
                The navy dashed-dotted line corresponds to the $z_{phot} = z_{spec}$. 
                Shadowed areas represent the redshift range rejected from our analysis.}
                \label{fig:0}
        \end{center}
\end{figure}

Fig.~\ref{fig:0} shows the comparison of the full ADF--S spectroscopic sample (with nine additional sources  not presented in \citealp{malek14}) with their photometric estimation. 
This figure presents a large group of galaxies with spectroscopic redshift and a failed estimation of the photometric value  at $z\sim$0.04. 
These galaxies are members of a nearby cluster, Abell~S0463, one of several parts of the ADF--S field extensively observed in the past \citep{dressler80a,dressler80b,abell89}.
Incidentally, it is a redshift where almost all the existing tools for estimation of photometric redshifts  experience a serious degeneracy  (i.e., \citealp[Fig.~10 of][]{Bolzonella2000} for Hyperz tool,  \citealp[Fig.~3 of ][]{ilbert06},  \citealp[Fig.~1 of][]{ilbert13} and \citealp[Fig.~1 of][]{Heinis2016} for lePhare tool). 
One of the reasons for the high fraction of CEs for galaxies at low redshifts could be the mismatch between the Balmer break and the intergalactic Lyman-alpha forest depression \citep[for a more detailed discussion we refer the reader to ][]{ilbert06}.

The scatter for a low $\rm{z_{spec}}$ sample, visible in Fig.~\ref{fig:0}, can induce a large error in the luminosity distance ($\rm{D_L}$), and finally, cause  possible inaccuracy of our analysis. 
To avoid this problem, we rejected all objects with spectroscopic or photometric  redshift  $<$0.05. 
With such a cut, the median difference between the luminosity distance, calculated for the spectroscopic and photometric redshifts for the presented sample as $\rm{D_{L}(z_{spec})/D_{L}(z_{phot})}$, is as low as  1.33 [Mpc] (the Median Absolute Deviation = 0.78).  
The redshift accuracy for the sample of galaxies with $\rm{z_{spec}\geq} $0.05 calculated as an arithmetic (as the distribution of the redshift accuracy is not Gaussian, and the number of objects is not high, we cannot calculate the mean and $\sigma$ values) mean of $\rm |z_{spect_i}-z_{photo_i}|/(1+z_{spect_i})$ (where $i$ stands for each galaxy) equals 0.054$\pm$0.024, and 78\% of the sources have a redshift accuracy below this value (the redshift accuracy calculated as the normalized median absolute deviation is equal to 0.060), and in this case  the number of CEs dropped down to two. 
We stress that, for the final sample, we used  the same threshold  both for spectroscopic and estimated photometric redshifts, and the obtained  results are even better (redshift accuracy is equal to  0.050 with only one catastrophic error). 
To check how the redshift accuracy influences our final results, we ran the same analysis for all galaxies changing the value of the redshift $\pm$ 0.05 (meaning that we do not show the worst effects of photometric redshift errors, but the mean influence of the uncertainty of the estimated $\rm{z_{phot}}$). 
This test is presented in Appendix~\ref{app:A}.

\section{Sample selection and the properties of the final catalogue}
\label{sec:sampleselection}

To construct the final ADF--S sample  we restrict our analysis to sources with the best quality photometry available to fit SED models with the highest confidence. 
The main criterion was availability of redshift information (spectroscopic or photometric) in addition to at least six photometric measurements spanning the whole spectra. 
Additionally, we cut all objects with $z<$0.05.  
In total, we selected {78} FIR-bright galaxies ({22} galaxies with known spectroscopic redshift and {56} galaxies with redshifts estimated by CIGALE), with an average number of photometric measurements equal to 11.

Our sample is based on 90~$\mu$m detection and, therefore, each galaxy has at least one measurement in the FIR. 
What is very important for our method is that, in  total, 27 {ADF--S} sources from the final sample have \textit{Herschel/SPIRE} counterparts, allowing us to obtain very precise models of the  far--infrared  spectra, and to measure the real total dust temperature.  
All galaxies have measurements from WISE. 
45 sources have Spitzer data (24 and 70~$\mu$m), and 56 have counterparts in the  2MASS catalog \citep{skrutskie06}.  
Unlike \cite{malek10}, we did not use the IRAS data, as the uncertainties are too great and do not improve the SED fitting. 
We have also used the third release of the DENIS data \citep{Paturel03}, and  have found 55 counterparts for our selected sources. 
Almost 20\% of our sample was  detected in the UV GALEX bands.
Thus, we are able to perform a multi-wavelength analysis of the ADF--S sources in order to build detailed SEDs from FUV to FIR, to get a range of physical 
properties of our sample. 

Eight of our sources have 20~cm counterparts from the ACTA--AKARI survey. 
Unfortunately, two of them are unresolved (according to \citealp{white12}, the ratio of the integrated flux density and peak flux density for unresolved sources is  lower than 1.3). 
As for our SED fitting method, every additional model requires additional parameters. We decided to reject radio data from our analysis, since only six sources (3\% of our sample) have reliable measurements in the 20~cm band.

\begin{figure}[h]
        \begin{center}
                \includegraphics[width=0.5\textwidth, clip]{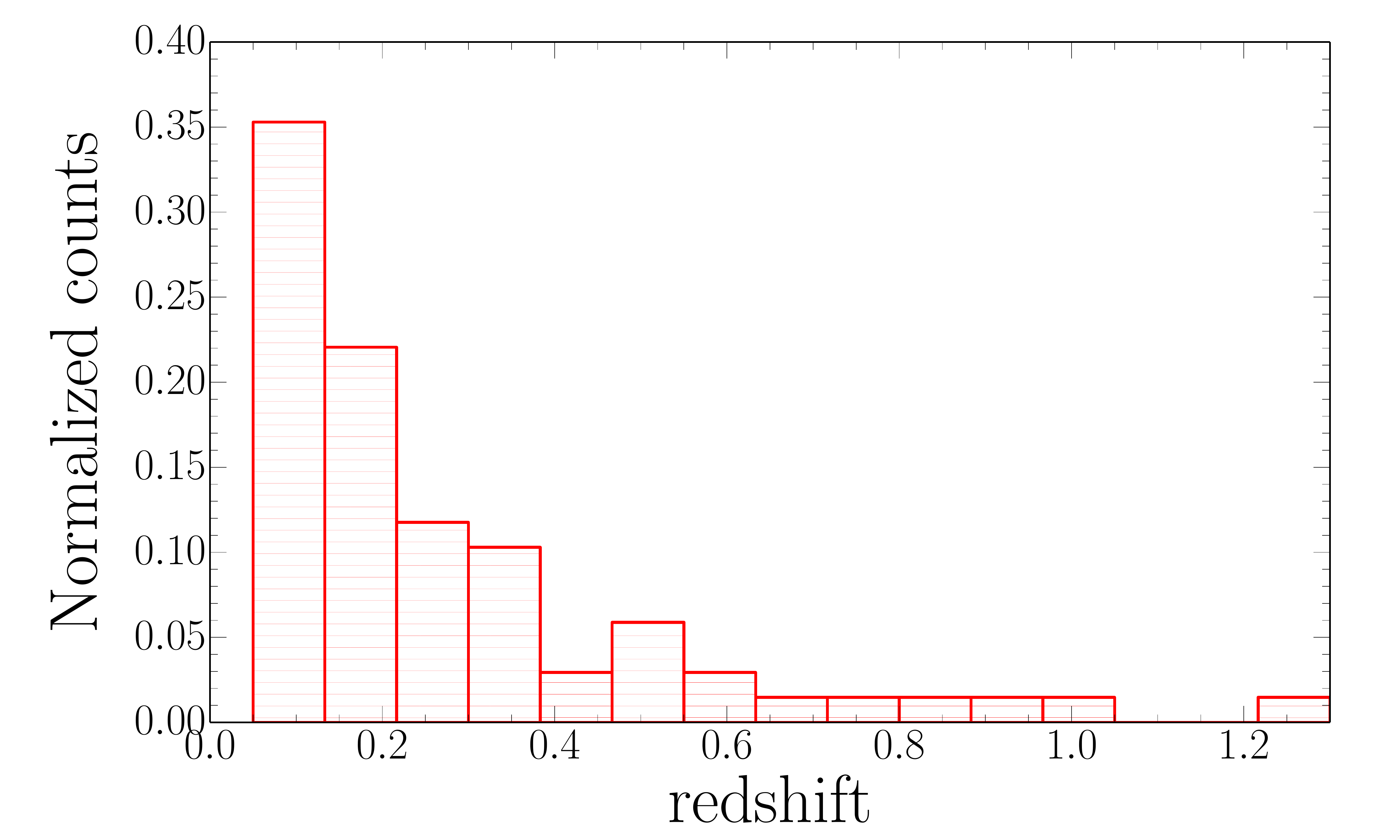}
                \caption{Redshift distribution of the final sample of 78 ADF--S  galaxies used for our analysis.}
                \label{fig:1}
        \end{center}
\end{figure}

The redshift distribution of our sample is shown in Fig.~\ref{fig:1}. 
From previous studies \citep{malek10,malek14} we know that the {ADF-S} sample contains mostly nearby galaxies, whose properties are similar to the properties of the local population of optically bright, star-forming galaxies, except for having an unusually high ratio of peculiar ($\sim$10\%) or interacting objects \citep{malek10}.

\section{Methodology}
\label{sec:methodology}
Our main aim is to select and study  LIRGs and ULIRGs to determine the main characteristics of their physical properties, and to use normal galaxies; the remaining sources from our catalog of {78} galaxies, not classified as {LIRGs or ULIRGs},  were used as a control sample.  
To achieve this goal it was very important to take into account the properties of the ADF--S galaxies along the whole wavelength range. 
We determined the main physical properties of FIR-bright galaxies form the AKARI survey by fitting their spectral energy distributions (SEDs), taking into account photometric information from ultraviolet (UV) to far infrared. 
We also decided to use additional, much simpler modeling based on the IR data only as a double check for ADF--S  LIRGs and ULIRGs, and to derive the dust mass and dust temperature (not given from the \citealp{dale14} model, used for the UV-FIR SED fitting). Also, we used CIGALE results as a reference.

\subsection{UV to FIR SED-fitting}

The SED fitting was performed using version~v0.5 of the CIGALE code (Code Investigating GALaxy Emission)\footnote{http://cigale.lam.fr/} developed
with PYTHON, which provides physical information about galaxies by fitting SEDs that combine  UV-optical stellar SED with a dust component emitting in the IR.
CIGALE conserves the energy balance between the dust-absorbed stellar emission and its  re-emission in the IR. 
The list of input parameters of  CIGALE  is shown in Table~\ref{tab:input}. 
We refer the reader to  the CIGALE web page for a detailed description of the code.

\subsubsection{Models}

\textit{Star formation history}

The star formation histories  (SFHs) used (optionally) by CIGALE  are:  
(1) the double decreasing exponential star-formation history module (\texttt{sfh2exp}), which implements a SFH composed of two decreasing exponentials, 
(2) the delayed tau model, which  implements a SFH described as a delayed rise of the SFR up to a maximum, followed by an exponential decrease, 
and the (3) custom module, which can be read from the file. 
  
Recently,  \citet{buat14} published a comparison between  SED fits obtained with two stellar populations.
The comparison includes an exponentially decreasing/increasing SFR and a more complex SFR with two components: 
(1) the decreasing SFR for an old stellar population with and without a fixed value for the age of the oldest population,  and (2) the younger burst of constant star formation. 
The conclusion was that  when two stellar populations are introduced, fixing or not fixing the age of the oldest population, has only a modest impact, and two populations still give the best fit to the data. 

Thus, we assume a similar SFH to \cite{buat14}. 
In our analysis we adopted an exponentially decreasing star-formation rate for the old and young stellar populations.   
A similar approach was used by \citealp[ ][]{erb06,lee09,buat11,giovannoli11}, and \citealp{ciesla15}.

\textit{Single stellar population}

CIGALE uses the stellar population synthesis models either by \cite{maraston05} or \cite{bruzal03}. 
For this work we adopted the stellar population synthesis models of \cite{maraston05}, as they consider the thermally pulsating asymptotic giant branch (TP-AGB) stars.
This model includes young stellar population tracks  from the Geneva database, and the Frascati database for older populations  (for a more detailed discussion we refer the reader to \citealp{maraston05}, and \citealp{maraston09}).
In general, \cite{maraston05} models  provide reliable information on the NIR-bright stellar population of galaxies, which is very important for our analysis, since our sources have been detected in the NIR bands.
To calculate the initial mass function (IMFs), CIGALE has two built-in algorithms based on  \cite{salpeter55} and \cite{kroupa01} models.

\textit{Attenuation curve}

The galaxy attenuation curve adopted by CIGALE is based on the Calzetti law \citep{calzetti00} with some modifications \citep[for a more detailed discussion we refer the reader to ][]{noll09}.
CIGALE allows the user to alter the steepness of the attenuation curve and/or to add a UV bump centered at 2175~\AA{} \citep{noll09}. 
We did not modify the Calzetti law, but we use three different values for E(B-V), which represent the attenuation of the youngest population, and the reduction factors are applied to the color excess of stellar populations older than $\rm{10^7}$ years.

\textit{Dust emission}

The new version of CIGALE includes  four different models to calculate dust properties: \cite{casey12},  \cite{dale07}, \cite{dale14}, and an updated \cite{draine07} model.
Since we decided to use the \cite{casey12} model for  a double-check of  CIGALE, the main dust analysis was based on the \cite{dale14} model, which  is the most  suitable for very luminous infrared sources. 
The results given by \cite{dale14} are 
\begin{itemize}
\item the fraction/contribution of an AGN to the mid infrared spectrum, 
\item the slope of the infrared part of the spectrum -- parameter $\alpha$ -- empirically determined by \cite{dale02} based on the ratio of IRAS fluxes detected at 60 and 100~$\mu$m  \citep[the model was improved by ][]{dale14}. It describes the progression of the far-infrared peak toward shorter wavelengths for increasing global heating intensities. 
Parameter $\alpha$ ranges from 0.0625 to 4, as the dust heating changes from strong to quiescent \citep{dale05}. 
The lower the value of $\alpha$ , the more actively star forming  a galaxy is.  
Normal galaxies display 1 $< \alpha <$ 2.6, while $\alpha$ $\sim$2.5 characterizes  an already  quiescent galaxy \citep{dale01, dale02}. 
Values of $\alpha$ $\sim$4 are typically fitted for galaxies where the FIR emission peak appears at even longer wavelengths than for the most quiescent, well studied, galaxies and  corresponds to the coolest and most quiescent galaxies, 
\item the total IR luminosity between 8 and 1000 microns ($\rm{L_{dust}}$),  defined as the sum of stellar and AGN luminosities re-processed by dust. 
\end{itemize}

\cite{dale14} presented an updated version of the model originally proposed by \cite{dale02}.
This model describes the progression of the far-infrared peak toward shorter wavelengths for increasing global heating intensities \citep{dale01}.  
CIGALE adopts 64 templates of \cite{dale14}  models.
All models are parametrized by the $\alpha$ parameter.

The main improvements between versions presented by \cite{dale02} and \cite{dale14}  are:  
(1) the mid-infrared parts of the models,  originally based on the ISOPHOT data from the \textit{Infrared Space Observatory}  were rebuilt  using the data from the \textit{Spitzer Space Telescope}, and 
(2) the AGN component has been included to compose the total infrared luminosity. 
For the updated mid-infrared (MIR) spectrum the 5-34 $\mu$m "pure" star-forming curve from \cite{Spoon07} was adopted; it makes use of  the Spitzer data to model a sequence of MIR  spectral shapes for AGNs,  LIRGs and ULIRGs, and star-forming galaxies. 

The \cite{dale14} model can also derive a fraction of AGN that contributes to the galaxy  dust luminosity in MIR. 
Unfortunately, this model is more suitable for broad-line AGNs, and additionally,  does not include any absorption features (e.g., 9.7~$\mu$m, characteristic for ULIRGs), and thus, the AGN components estimated by this model can be misleading. 
A lack of the 9.7~$\mu$m absorption line does not affect an $\alpha$ parameter as the  AKARI ADF--S sample has very good coverage between MIR and FIR wavelengths, sufficient to calculate a proper slope. 
Since our aim is to look for the presence of both Type~1, Type~2, and intermediate types of AGNs in our sources, we decided to use \cite{dale14} templates to estimate global heating intensities and dust luminosity values only. 
We have adopted the \cite{fritz06} model to check the fractional contribution of AGNs to the mid-infrared emission. 

\textit{AGN emission}

\begin{figure}[t]
        \centering
         \includegraphics[width=0.48\textwidth, clip]{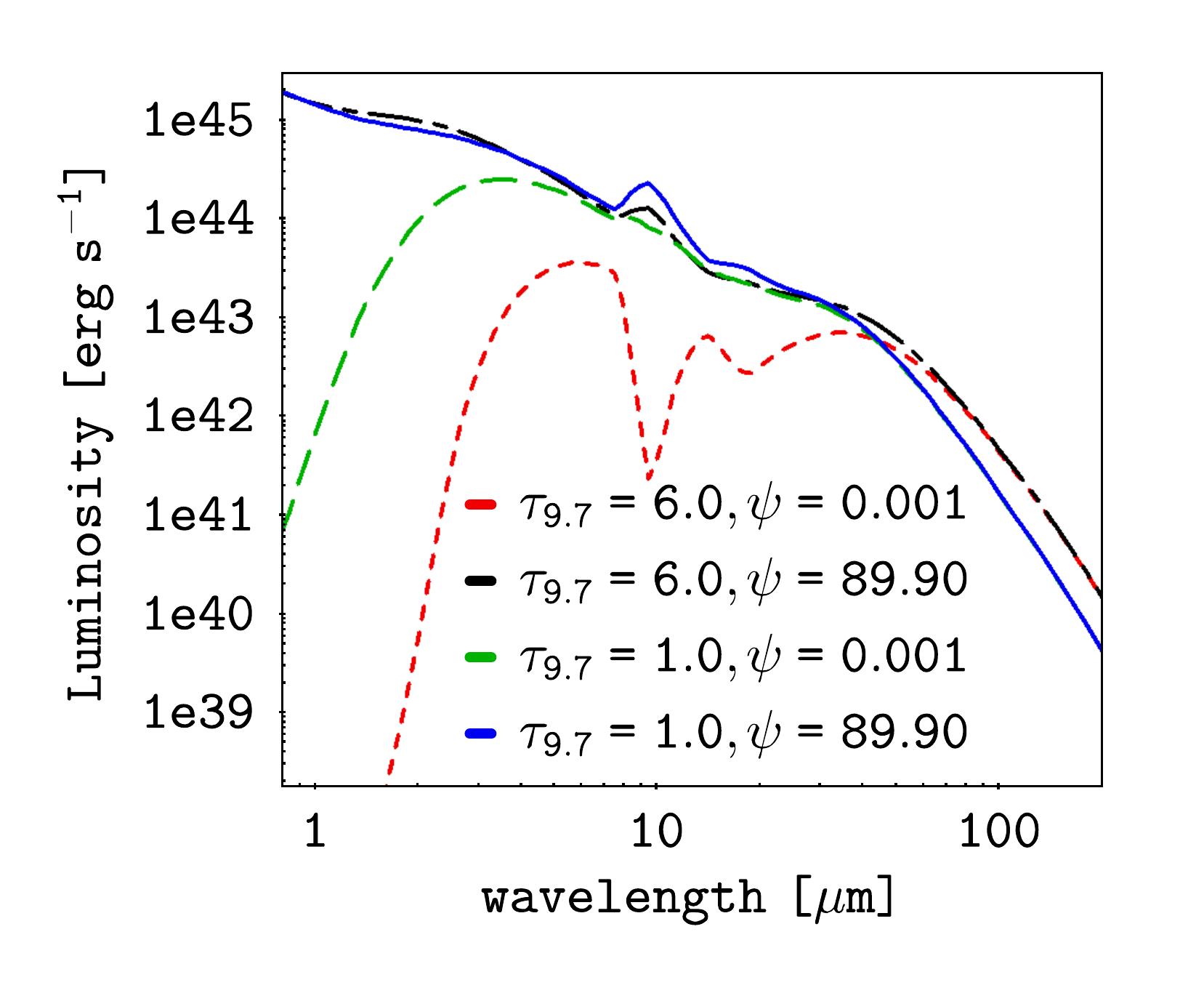}
         \caption{\cite{fritz06}  AGNs' SEDs used  in our analysis. 
         These templates are built based on two optical depth $\tau_{9.7}$ values (1.0 and 6.0), and the extreme values of the angle between the equatorial axis and the line of sight ($\psi$). 
         The red dotted line represents $\tau_{9.7}$=1.0, and $\psi$=89.900~[deg]; the green dashed line corresponds to the $\tau_{9.7}$=1.0, and $\psi$=0.001~[deg]; the black dashed-dotted line to $\tau_{9.7}$=6.0, and $\psi$=89.900~[deg];  and the blue solid line to $\tau_{9.7}$=6.0, and $\psi$=0.001~[deg].}
         \label{fig:AGN}
\end{figure}

The latest release of CIGALE includes \cite{fritz06} models of AGN emission. 
\cite{fritz06} created an improved model for the emission from dusty torus  {heated by a central AGN}.  
In this model a flared torus is defined by its inner  and outer radii and the total opening angle ($\rm{\theta}$). 
The adopted density of dust grain distribution depends on the torus radial coordinate and polar angle, and the optical depth is computed taking into account the different sublimation temperatures for silicate and graphite grains. 
The templates are computed at different lines of sight with respect to the torus equatorial plane ($\rm{\psi}$) in order to account for both Type~1 and Type~2 emission, from $\rm{0^{\circ}}$ to $\rm{90^{\circ}}$, respectively,  in steps of $\rm{10^{\circ}}$. 
Very important for our analysis is that \cite{fritz06} models preserve the energy balance which  perfectly matches the method used by CIGALE. 

This model gives a good estimation  for Type~1 and Type~2, as well as for the intermediate types  of mid-infrared SED of AGNs, even for  non-numerous photometric data \cite[][and references therein]{Feltre12}. 
According to the convention used by \cite{fritz06}, the angle $\rm{\psi}$ between the AGN and the line of sight for the extreme cases of Type~1 and Type~2 AGNs is equal to  $\rm{90^{\circ}}$ and  $\rm{0^{\circ}}$, respectively. 
For a detailed description of the \cite{fritz06} model we refer the reader to the original paper. 

The first usage of \cite{fritz06} templates in CIGALE was reported by \cite{Buat15}, and \cite{ciesla15}. 
\cite{Buat15} used \cite{fritz06} templates as an additional model to built SEDs of real galaxies from the AKARI North Ecliptic Pole Deep Field, while \cite{ciesla15} analyzed simulated realistic SEDs of Type~1, Type~2, and intermediate types of AGNs to estimate  the ability of CIGALE  to retrieve the physical properties of the galaxy, and to calculate possible under- and over-estimations of AGN fraction, stellar masses, and star formation parameters obtained from the SED fitting. 

\cite{ciesla15}  showed the difference between flux density ratio of SEDs with and without AGN emission for Type~1, Type~2 and intermediate types of AGNs \citep[][Fig.~4]{ciesla15}, which can be reckoned as the main description of different types of AGN influence on the full spectra fitting.
As it was shown by \cite{ciesla15}, based on the model introduced by \cite{fritz06} for the UV-FIR SED fitting method, Type~1 AGNs are chosen  based on the higher emission in the UV and MIR rest-frame spectra, compared to the same SED without the AGN component. 
The main characteristic assumption for selection of Type~2 AGNs 
is the amplified MIR and FIR emission. 
The ADF-S sample was selected based on the FIR bands, and all galaxies used for the analysis have measurements spanning the whole spectral range. 
As a result, parts of the spectrum needed for AGN identification are sufficiently well covered, thus, Type~1,  Type~2, and intermediate types of  AGNs can be easily recognized based on the SED fitting.  

For our analysis, we have used four \cite{fritz06} templates built based on average parameters from previous studies of the model \cite[i.e., ][]{Hatziminaoglou08,Hatziminaoglou09,Buat15,ciesla15}. 
These parameters are: (1) the ratio of the outer to inner radius of the torus fixed to 60.0 , (2) the radial and angular dust  distribution in the torus equal to -0.5, and 0.0, respectively, and (3) the angular opening angle of the torus fixed to 100~[deg]. 
To follow the low and the high optical depth at 9.7$\mu$m, two values of  ($\rm{\tau_{9.7}}$) were chosen: 1.0 and 6.0. 
These values of $\rm{\tau_{9.7}}$  are consistent with results presented by  \cite{ciesla15} and  \cite{Buat15}, as well as our internal tests. 
To consider the most extreme cases of AGNs (Type~1 and Type~2) we used the  values of an angle between equatorial axis and the  line of sight equal to 0.001 and  89.990 degrees.  
Four selected templates used in the further analysis  are plotted in Fig.~\ref{fig:AGN}.

\begin{figure*}[ht]
        \begin{center}
                \includegraphics[scale=0.4,clip]{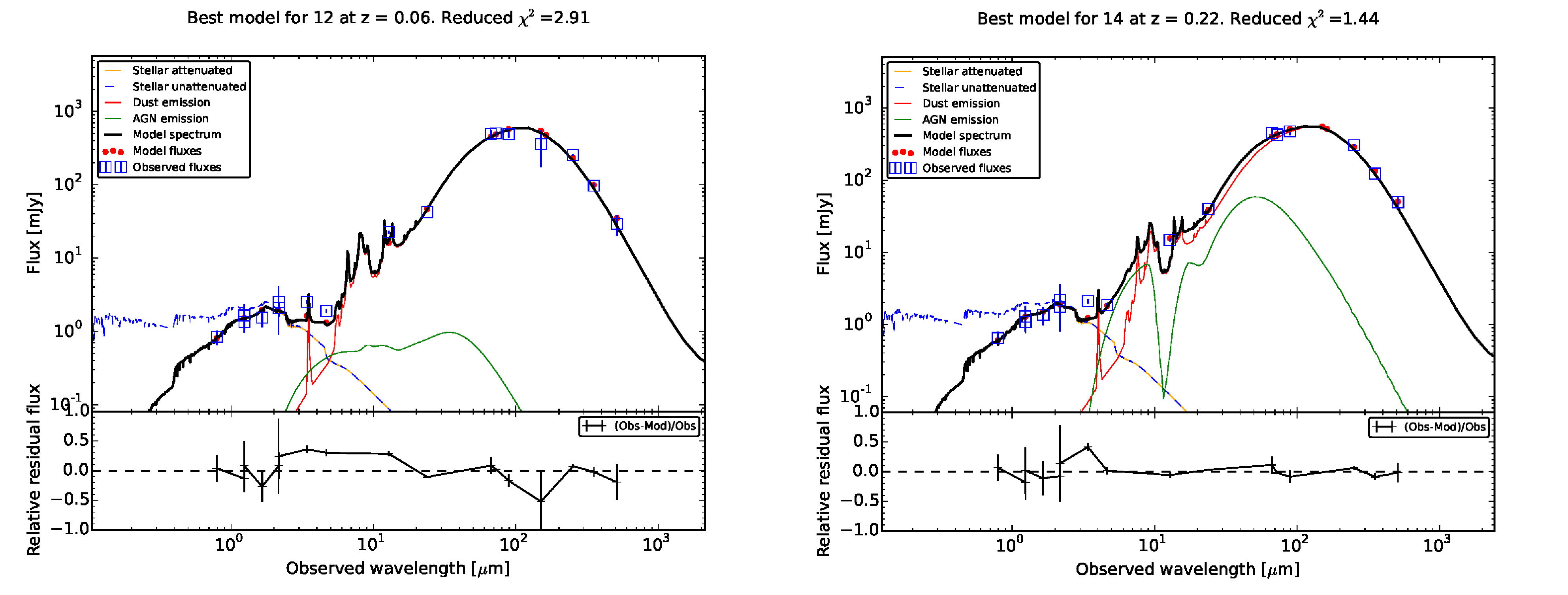}
        \end{center}
        \caption{Examples of the best fit models (CIGALE SED fitting code). 
        The left-hand side spectrum is a star-forming galaxy, while the spectrum shown on the right-hand side is ULIRG.  
        Observed fluxes are plotted with open blue squares. 
        Filled red circles correspond to the model fluxes. 
        The final model is plotted as a solid black line. 
        The remaining three lines correspond to the stellar, dust, and AGN components. 
        The relative residual fluxes are plotted at the bottom of each spectrum.
        } \label{fig:2}
\end{figure*}
 
 \begin{figure*}[t]
        \begin{center}
                \includegraphics[width=0.85\textwidth,clip]{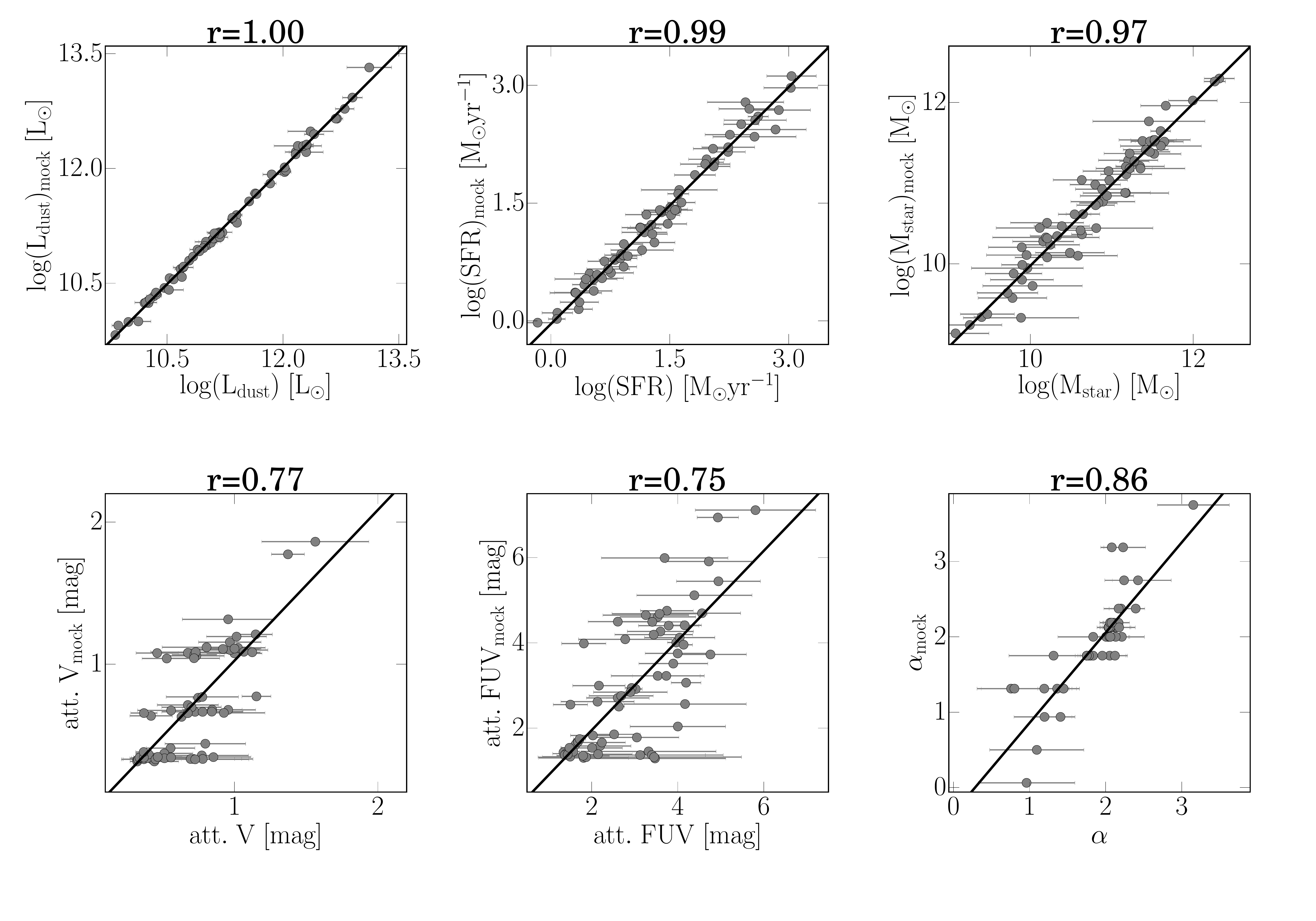}
        \end{center}
        \caption{Pearson product-moment correlation coefficient (\texttt{r}) calculated between ADF--S galaxies and the corresponding mock catalog (created from the input values) for dust luminosity,  star-formation rate, stellar mass,  dust attenuation in V and FUV bands, and  $\alpha$ parameter of the model of \cite{dale14}. 
        The calculated values of \texttt{r} are written above each plot.}
        \label{fig:3}
 \end{figure*}

\subsubsection{Modeled SEDs}

The quality of the fitted SEDs is calculated making use of the $\chi^2$ value of the best model, marginalized over
all parameters except the one assigned for the further physical analysis. 
In the next step, the values of the probability distribution function (PDF) of the derived parameters of interest (in the case of this paper: $\rm{M_{star}}$, $\rm{L_{dust}}$, AGN fraction, the AGN torus angle with respect to the line of sight, $\alpha$ slope of the model \cite{dale14}, and SFR are calculated. 
The final output values of analyzed parameters are calculated as the mean  and standard deviation determined from the PDFs   \citep[][Burgarella et al., in prep; Boquien et al., in prep., see also \cite{Walcher08} for more detailed explanation of the $\chi^2$ -- PDF method]{ciesla15,buat14}.

Based on the distribution of $\chi^2$ values for the ADF--S sample  we decided to restrict our analysis to the modeled SEDs with the  reduced $\chi^2$ value lower than five.
The threshold value was chosen as a mean $\rm{\chi^2}$ + 3$\rm{\sigma}$. 
We found 71 galaxies  that fulfill this condition for the $\rm{\chi^2}$.

After visual verification of all single fits, we decided to remove  two objects with a satisfying $\chi^2$ from the further analysis because of  a poor photometric coverage of the spectra.
Namely, (1) object $\rm{ID_{ADFS}=148}$ has the minimal number of  photometric measurements required for the analysis (six) but only one located in the infrared part of the spectrum (from the 90$\mu$m AKARI band).
This object has no Herschel counterpart  and only one photometric point from the WISE;
(2) object $\rm{ID_{ADFS}=281}$ has a similar distribution of photometric data but with a total number of nine measurements. 
In both cases, the stellar part  is very well fitted, but the infrared part is rather an estimate of the real shape of the spectrum. 

Finally, we restricted the final sample to {69} galaxies (among them, 20 objects with known spectroscopic redshift and {49} galaxies with photometric redshift) that fulfill the $\chi^2$ and visual inspection conditions.

\begin{table*}[]
        \begin{center} 
                \caption[]{{List of the input parameters of the code CIGALE.}}
                \label{tabela_par}
                \begin{tabular}{p{0.57\linewidth}| p{0.36\linewidth}} \hline \hline  
                        \label{tab:input}
                        \multirow{2}{*}{paramater} & \multirow{2}{*}{values} \\ 
                        &  \\ \hline 
                        \multicolumn{2}{c}{star formation history} \\   \hline 
                        $\tau$ e-folding time of the main stellar population model [Myr] & 6000, 4000, 3000 \\
                        $\tau$ e-folding time of the late starburst population model [Myr] & 35, 30, 25, 15, 10, 8\\
                        mass fraction of the late burst population & 0.3, 0.2, 0.1, 0.001\\
                        \hline
                        \multicolumn{2}{c}{single stellar population } \\       \hline 
                    initial mass function & \cite{salpeter55} \\
                    metallicities (solar metallicity) & 0.2 \\
                    age of the separation between the young and the old star population [Myr] & 10 \\ \hline 
                        \multicolumn{2}{c}{attenuation curve } \\       \hline 
                        slope correction of the Calzetti law & 0.0 \\
                        color excess of the stellar continuum light for the yought stellar population & 0.15, 0.3, 0.5 \\ \hline
                        \multicolumn{2}{c}{dust emission} \\    \hline 
                        \multirow{2}{*}{IR power-law slope} & 0.0625, 0.5, 0.9375, 1.3125, 1.75, 2.0, 2.1250, 2.1875, 2.3750, 2.75, 3.1875, 3.7500 \\ \hline
                        \multicolumn{2}{c}{AGN emission } \\    \hline
                        ratio of the maximum to minimum radius of the dust torus & 60 \\
                        optical depth at 9.7 microns & 1.0, 6.0\\
                        radial dust distribution in the torus & -0.5 \\
                        angular dust distribution in the torus & 0.0 \\
                        angular opening angle of the torus [deg]& 100.0 \\
                        angle between equatorial axis and the line of sight [deg] & 0.001, 89.900 \\
                        fractional contribution of AGN & 0.01, 0.15, 0.25, 0.3, 0.35,  0.4, 0.45,0.55\\ \hline
                \end{tabular} 
        \end{center}
\end{table*}

Examples of the best fit models for the ADF--S galaxy sample are shown in Fig.~\ref{fig:2}.

\subsubsection{Reliability check}
\label{sec:Reliability_check}

The mock catalog was generated to check the reliability of the computed parameters. 
To perform this test we used an option included in CIGALE, which allows for creation of mock objects for each galaxy. 
The final step of verification of estimated parameters is to run CIGALE on the mock sample using the same set of input parameters as for the real catalog, and to compare the output physical parameters of the artificial catalog with the real ones.
A similar reliability check was performed for example by \cite{giovannoli11}, \cite{yuan11}, \cite{boquien12}, \cite{malek14}, and \cite{ciesla15}. 

Fig.~\ref{fig:3} presents the comparison of  the output parameters of the mock catalogue created from the input parameters versus values estimated by the code for our real galaxy sample. 
The determinant value, which can characterize  the reliability of the obtained values, is the Pearson product-moment correlation coefficient (\texttt{r}). 
The comparison between the results from the mock and real catalogs shows that CIGALE gives a very good estimation of main  physical parameters which we use  for the final analysis ($\rm{L_{dust}}$, SFR, $\rm{M_{star}}$, fraction of AGN, and $\alpha$ parameter from the \cite{dale14} model).
For dust attenuation in V and FUV  bands, the correlation is not obvious. 
Consequently, we only used the results to present a  general trend of how the dust component affects the stellar part of the spectra.

\begin{figure}[t]
        \centering
        \includegraphics[ height=0.25\textheight, clip]{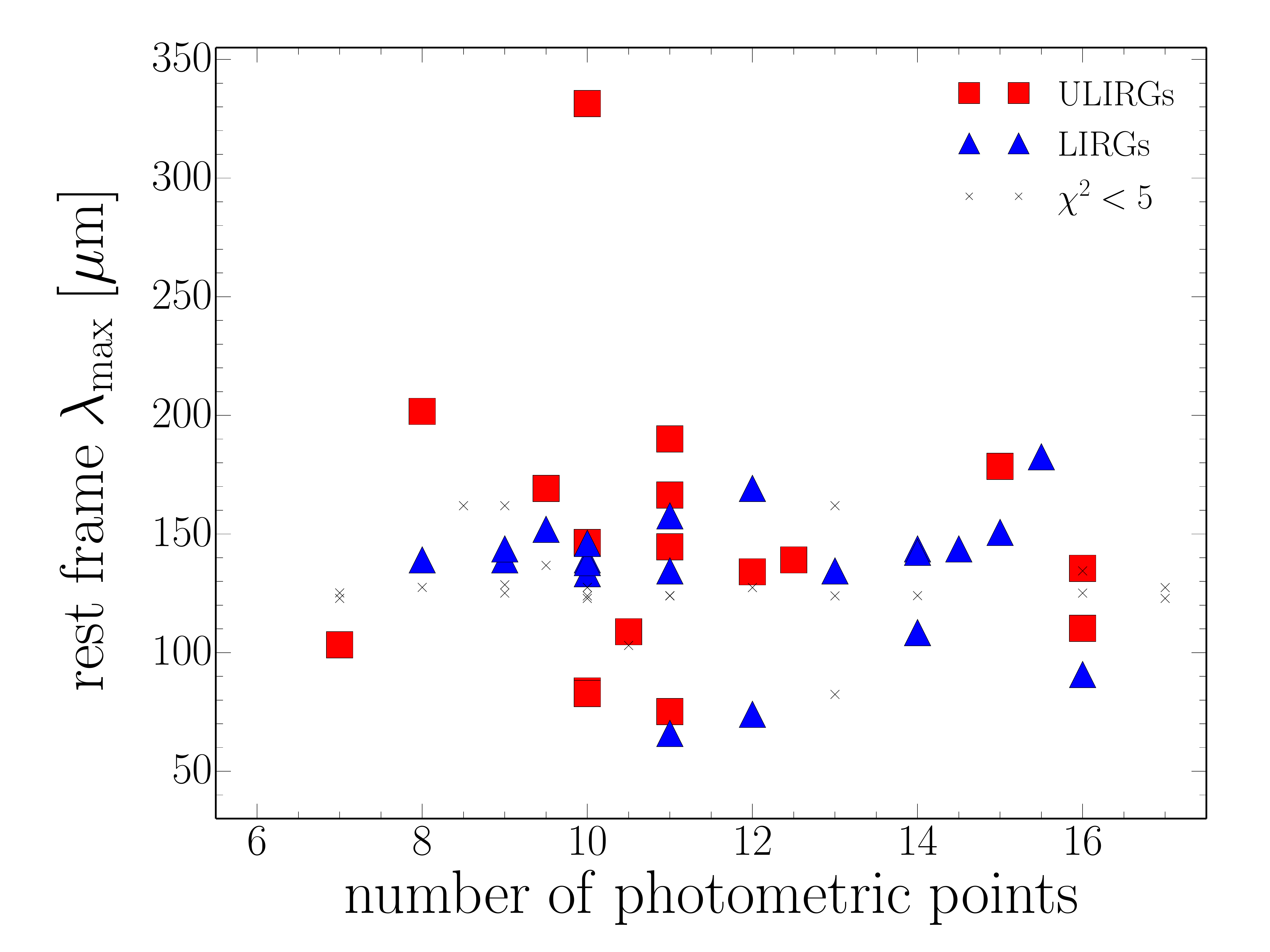}
        \caption{Number of photometric points used for CIGALE fitting  vs the peak of the dust part of the spectra ($\lambda_{max}$ for 156 ADF--S objects fitted with $\chi^2 <$ 10). 
        Filled red squares represent ULIRGs, filled blue traingles represent LIRGs, and the gray "x"-s correspond to normal star-forming galaxies. 
        As the $\lambda_{max}$ corresponds to the dust temperature, we conclude that we do not find any degeneracy with respect to the number of photometric points in the estimation of the peak of the dust  part of the spectrum.  }
        \label{fig:z}
\end{figure}

We have also examined how the peak of the dust part of the spectra ($\lambda_{max}$, related to the dust temperature) depends on the number of photometric points used for the SED fitting. 
Fig.~\ref{fig:z} shows the relation between the number of photometric measurements  and $\lambda_{max}$ relation for our sample of 69 galaxies with a marked dependence on the total dust luminosity. 
We did not find any degeneracy in this relation. 
Based on this figure we conclude that the significant percentage of LIRGs and ULIRGs selected based on the 90$\mu$m AKARI band compose cold sources, as only five of them ($\sim$12\%) have $\lambda_{max}<$90$\mu$m. Our results are in agreement with \cite{symeonidis11}.

We have also examined how the number of photometric points affects other physical properties estimated by CIGALE, and we did not find any significant correlation.  
We performed a similar test to examine whether or not the photometric redshifts estimated by \cite{malek14}  depend on the number of photometric measurements, but we found (similarly to the $\lambda_{max}$ parameter) a flat relation without any significant dependence on the  number of photometric points that fulfill our initial conditions. 
We conclude that our final selection (galaxies with photometric points covering the whole spectrum from  FUV to FIR, with more than two  measurements between 8 and 1000 $\mu$m, and at least three measurements in the UV-to-optical part of the spectrum for the final sample of 69 ADF--S objects) gives us very good coverage of the spectral energy distribution, sufficient for a reliable fit. 

In the  ADF--S sample, we found 17 galaxies  (24\% of the sample) that can be classified as ULIRGs. 
They are mostly galaxies with photometric redshifts; only one galaxy (HE 0435-5304) has a known spectroscopic redshift; it is the most distant galaxy from our sample, detected at z=1.23 \citep{Wisotzki00} and classified as quasar by \citealp{veron06}. 
The interesting point about HE 0435-5304 is that  \cite{Keeney2013} found a different spectroscopic redshift using the data from  the Cosmic Origins Spectrograph installed aboard the Hubble Space Telescope; 
based on the Ly$\alpha$ and NV emission lines, the spectroscopic redshift for HE 0435-5304 was found to be equal to 0.425. 
Now, with additional data we were able to estimate $\rm{z_{phot}}$, and we obtain $\rm{z_{phot}=0.30}$, and the redshift accuracy calculated for HE~0435-5304 between \cite{Keeney2013} and our photometric value equals 0.125.  
For $\rm{z_{spec}}=1.23$ we obtain an extremely  bright ULIRG (or the lower limit HLIRG, taking into account the uncertainties, with $\rm{log(L_{dust})=13.12\pm0.29}\mbox{ }[L_{\odot}]$). 
The physical properties of this object, based on the `official' high redshift, always make it an outlier with respect to the whole ULIRG sample as seen   from Figs.~\ref{fig:7}, \ref{fig:13}, and \ref{fig:aa}, for example. 
Properties of the same object, calculated with a new redshift z=0.43, give us a typical LIRG  ($\rm{log(L_{dust})=11.12\pm0.09}\mbox{ }[L_{\odot}]$) with  physical parameters characteristic for the LIRG sample.
As the previous redshift (1.23) is given by the NED and other public databases, we decided to present the results based on  the reference, but we point out that  the results obtained with the lower redshift by \cite{Keeney2013} appear much more reasonable.

In the sample,  22 galaxies (32\%) were  classified as LIRGs (including five galaxies with known spectroscopic redshift). 
This means that  LIRGs and ULIRGs   in total compose 56\% of  the bright ADF--S sample at redshift $>$0.05.
 
The  control sample contains 30 normal galaxies (12 of them have spectroscopic redshift).

We note that in our sample of LIRGs and ULIRGs, only one LIRG ($\rm{ID_{ADFS}}$=46) and one ULIRG ($\rm{ID_{ADFS}}$=212) have only eight photometric measurements. 
The average number of photometric points for the final sample of 39 objects is equal to 12. 
It means that the spectra are well defined by the measurements, and the models are well fitted. 

\begin{figure}[ht]
        \centering
         \includegraphics[ height=0.25\textheight, clip]{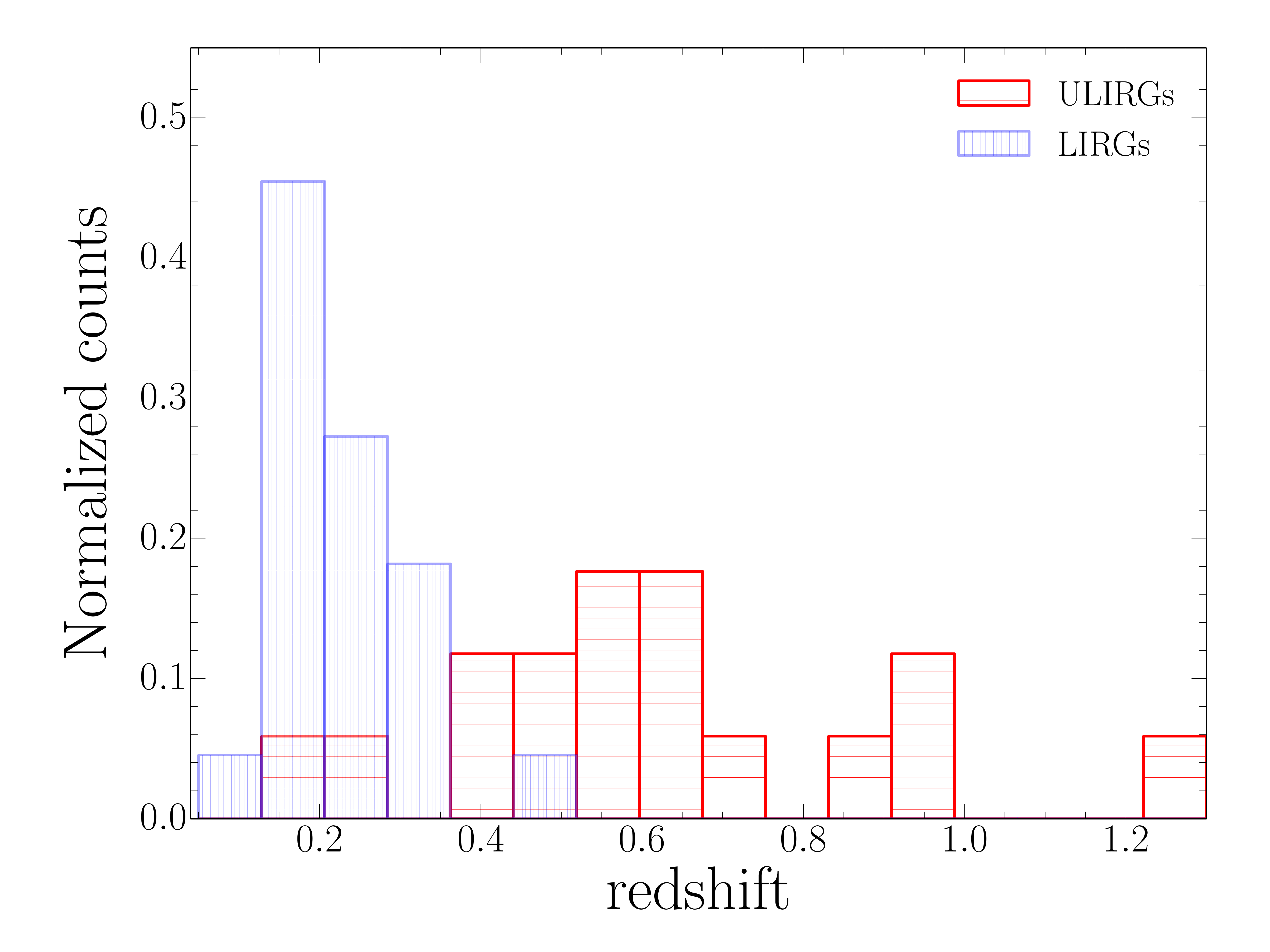}
         \caption{Normalized redshift distribution of {22 } ADF--S LIRGs (vertical striped blue  histogram) and 17 ULIRGs (horizontal striped red histogram).}
         \label{fig:new1}
\end{figure}

Redshift distribution of  ADF--S galaxies with  $\rm{L_{dust}>10^{11}\mbox{ }[L_{\odot}]}$ is shown in Fig.~\ref{fig:new1}. 
The mean values of redshift for ADF--S LIRGs and ULIRGs are equal to  0.23$\pm$0.08 and 0.61$\pm$0.26, respectively. 
This implies that ADF--S LIRGs are mainly objects from the local Universe, while ULIRGs are more commonly found at higher redshift. 
A similar distribution of LIRGs and ULIRGs was found by \cite{Lin16} for galaxies selected at 70~$\mu$m Spitzer MIPS band: the majority of ULIRGs were found on a higher redshift than LIRGs and star-forming galaxies. 

We have also checked the angular deviation between the found LIRGs and ULIRGs and their optical counterparts. 
Three LIRGs have angular distances between optical counterparts and AKARI measurements at 90~$\mu$m larger than 20'': 20.36'', 24.36'', and 30.06'' for $\rm{ID_{ADF--S}}$ equal to 6, 93, and 203, respectively. 
One ULIRG ($\rm{ID_{ADF--S}}$=59) has an angular distance equal to 21.24''. 
However, the mean angular distance for both LIRGs and ULIRGs is equal to 8.94'' with a standard deviation of 6.20 arcsec (median value 8.20'').  
We claim that only one source, a  LIRG with $\rm{ID_{ADF--S}}$=203, might be unreliable, but its SED is, overall, well fitted, and the obtained physical parameters are  reasonable (see Tab.~\ref{ULIRGS_wyniki}).

The final catalog of 39 LIRGs and ULIRGs used for our analysis is given in Table A.1 available at the CDS\footnote{Table A.1 is only available at the CDS via anonymous ftp to cdsarc.u-strasbg.fr (130.79.128.5) or via \url{http://cdsarc.u-strasbg.fr/viz-bin/qcat?J/A+A/xx/xx}}.
The catalog contains the following information:
Column 1 lists the ADF--S name of the source, Columns 2 and 3 give the coordinates, Column 4 provides the redshift, Column 5 lists the redshift references,  Column 6 gives 46  photometric flux densities and uncertainties for 20 bands spanning spectra from FUV (GALEX) to FIR (\textit{Herschel/SPIRE}), and Column 47 gives the name of the nearest optical counterpart. 
An example of two rows from the final catalog are shown in Tab.~\ref{ULIRGS_tabela}, at the end of this paper.

This significant percentage of galaxies, with $\rm{L_{dust}\geqslant10^{11} L_{\odot}}$ in the ADF--S sample, can be related to the fact that most of our galaxies (more than 55\%) were also detected at  24~$\mu$m, and the 24~$\mu$m band is a  well-known tracer of active, star-forming galaxies  \citep[e.g.,][]{calzetti07}.  

Table~\ref{tabela1} shows the mean values of physical parameters calculated for  LIRGs and ULIRGs    obtained from the UV to FIR SED fitting, while Fig.~\ref{fig:5} shows their distributions.
The discussion, and comparison of the parameters listed in  Tables~\ref{tab:mean_parameters} and~\ref{ULIRGS_wyniki} are presented in the following subsections with corresponding citations.

\subsection{\texttt{CMCIRSED}: NIR to FIR SED-fitting}

The Caitlin M. Casey Infra Red Spectral Energy Distribution model (\texttt{CMCIRSED}\footnote{Model available on-line at \url{http://herschel.uci.edu/cmcasey/sedfitting.html}}) published by \cite{casey12}, uses the single temperature gray-body + mid-IR power law, which was demonstrated to work very well for the FIR galaxy spectrum (for the  wavelength range $\rm{8~\mu m < \lambda < 1000~\mu m}$).  
We perform the SED fitting with the \texttt{CMCIRSED} model as a double-check for ADF--S  LIRGs and ULIRGs, as well as  to compute the dust mass and dust temperature for our sample, which is not given by the CIGALE models we used for our analysis. 

The SED-fitting procedure of the \texttt{CMCIRSED} model requires two free parameters: the slope coefficient ($\alpha$), and emissivity ($\beta$). 
We performed many tests for $\alpha$ and $\beta$ to obtain the  optimal SEDs  for the ADF--S sample. 
For  $\alpha$ (a power law slope coefficient) we adopted a range from 1.5 to 2.5, and for the spectral emissivity index  ($\beta$) we used values from 1.2 to 1.8. 
For a more detailed description of these parameters we refer the reader to \cite{casey12}.

\begin{figure}[]
        \begin{center}
                \includegraphics[width=0.5\textwidth, height=0.25\textheight, clip]{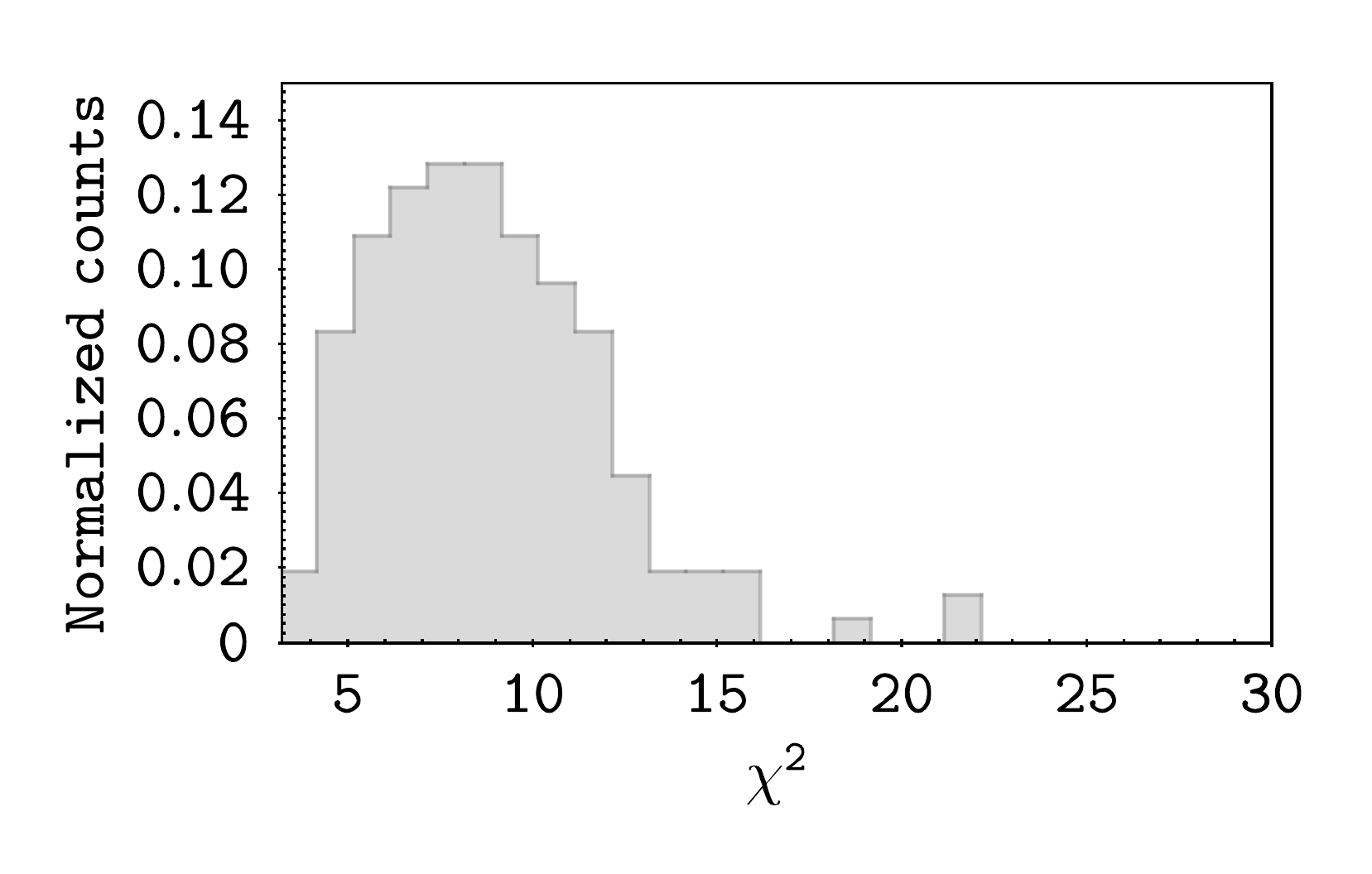}
        \end{center}
        \caption{The $\rm{\chi^2}$ distribution for the ADF-S sample obtained from \texttt{CMCIRSED} code. 
                One galaxy with $\rm{\chi^2>70}$ is not shown on this histogram.}
        \label{fig:15}
\end{figure}

\texttt{CMCIRSED} gives the residuals of the fits as the difference in the flux density at the rest-frame frequency between the data points and the best-fitting SED at 12, 25, 60, 100, and 850 $\mu$m wavelengths \citep{casey12}. 
To make both parts of the analysis (CIGALE and \texttt{CMCIRSED} SED fitting) homogeneous, we added an additional procedure to the \texttt{cmcirsed.pro} code that  computes the reduced $\chi^2$ value for the best-fit FIR spectrum. 
Fig.~\ref{fig:15} shows the distribution of the reduced $\chi^2$ values for the ADF--S sample.

\begin{figure*}[ht]
        \begin{center}
                \includegraphics[width=0.29\textwidth, height=0.22\textheight]{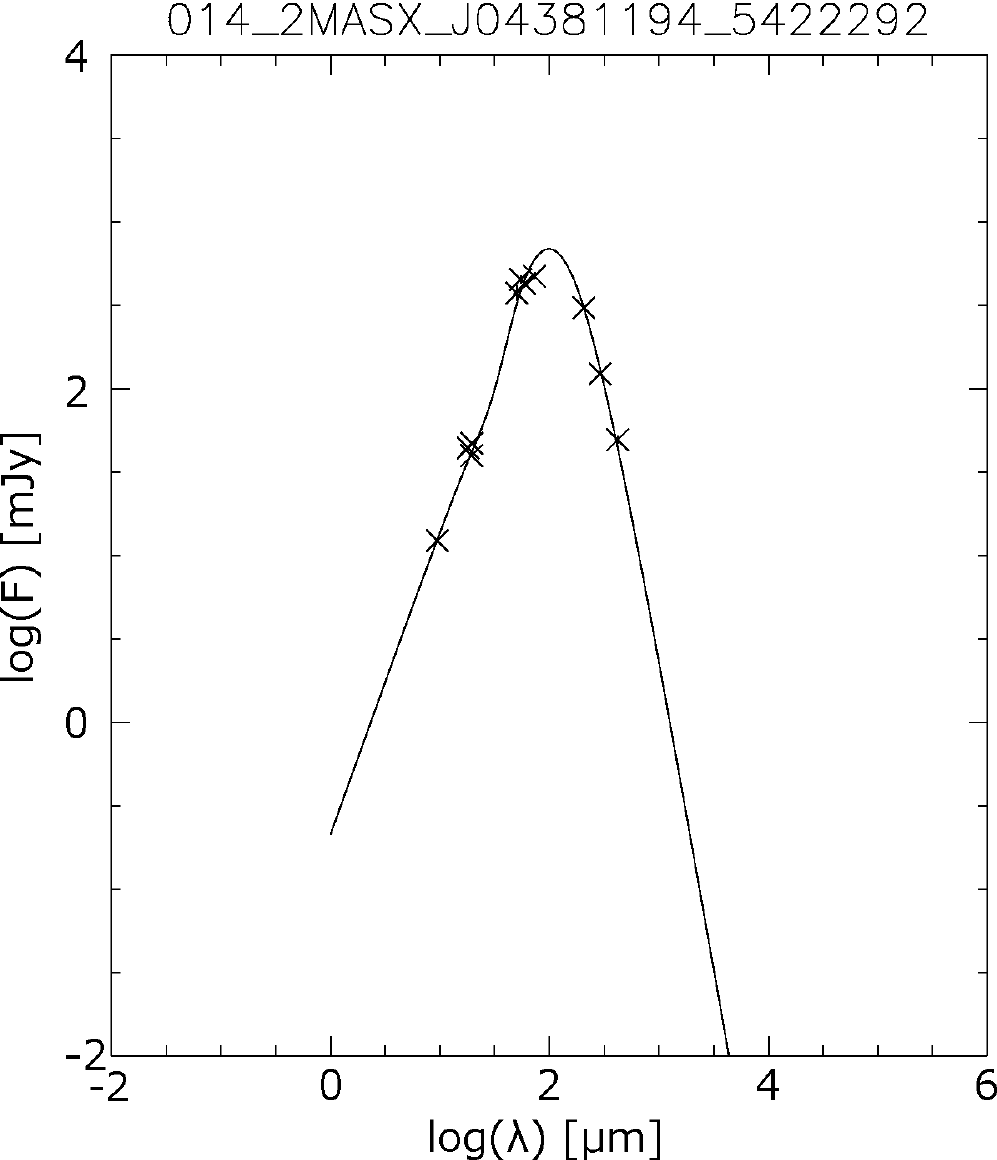}\includegraphics[width=0.29\textwidth, height=0.22\textheight]{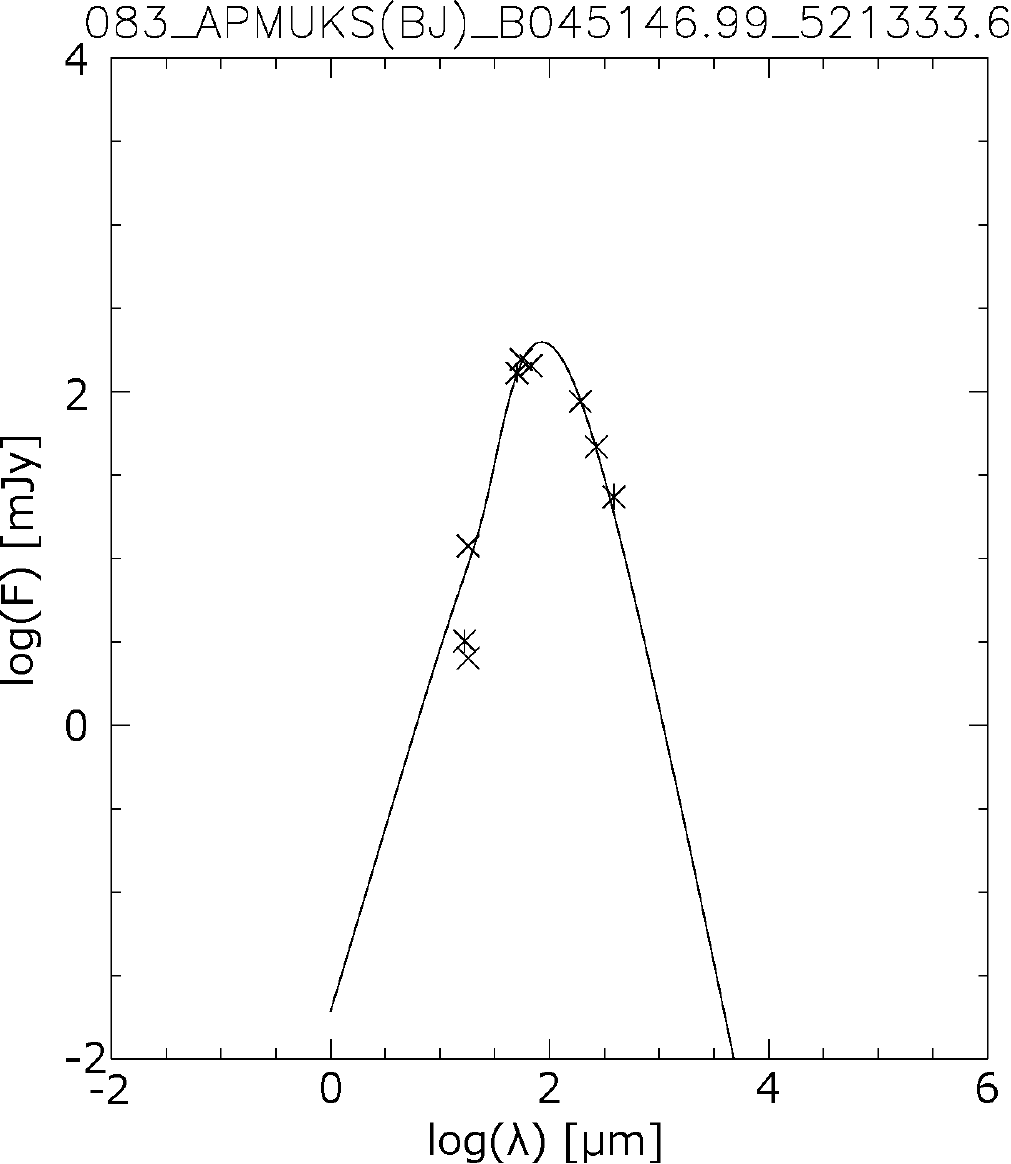}\includegraphics[width=0.29\textwidth, height=0.22\textheight]{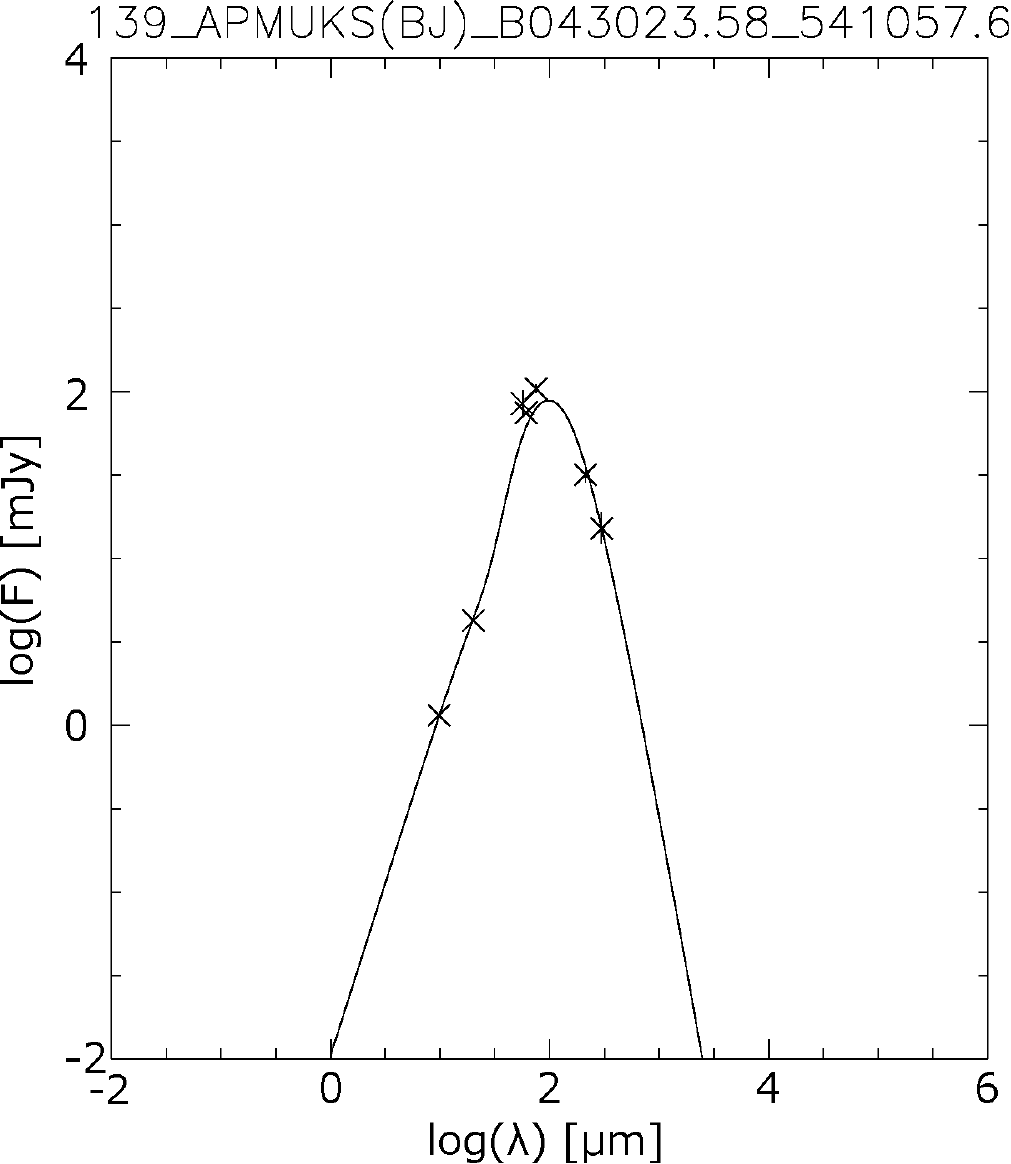}
        \end{center}
        \caption{Three examples of the best-fit models by \texttt{CMCIRSED} shown from left to right: ULIRG, LIRG, and a normal galaxy.}
        \label{fig:16}
\end{figure*}

Based on the presented histogram (Fig.~\ref{fig:15}) and the visual inspection of the obtained infrared SEDs,  we decided to restrict our analysis to the galaxies for which the reduced $\chi^2$ value was lower than  13.1 (according to the distribution, $\chi^2$ equal to 13 should be a threshold for this code, but the visual inspection of each individual SED increased the number of selected sources by one; an additional LIRG with a marginal value of $\chi^2$=13.08.  
We used 124 galaxies as the final sample for the subsequent analysis  which fulfilled this condition. 
Three examples of SED fits for the ADF--S galaxies are shown in  Fig.~\ref{fig:16}.

Based on the \texttt{CMCIRSED} model and an additional threshold for the reduced $\chi^2$ value, we found 12  ULIRGs and 13 LIRGs (25 galaxies in total, which corresponds to 22\% of the ADF--S galaxies with reliable \texttt{CMCIRSED} SED fits). 
For nine LIRGs and five ULIRGs found by CIGALE (14 objects in total) it was not possible to perform a satisfactory fitting using a  \texttt{CMCIRSED} code. 
For four of them, the number of IR measurements was simply too low to perform a reliable fit (three to four  photometric points). 
For three galaxies, the reduced $\chi^2$ was too high to use them for the further analysis, and for the remaining  objects, the fitted dust temperature was either lower than 10K (and at the same time, the uncertainties were equal to 0) or the estimated errors were much higher than 50\% of the dust temperature value.  
We excluded them from the final analysis.  
We stress that none of the LIRGs and ULIRGs identified by CIGALE were identified as normal galaxies by \texttt{CMCIRSED}, but rather \texttt{CMCIRSED} was unable to perform a reliable fit to their SEDs.  
This discrepancy between the results given by both codes is well  explained as the lack of \textit{Herschel/SPIRE} data for some galaxies, since far-infrared data  are  crucial for a code based on the FIR data only. 
CIGALE has the advantage over \texttt{CMCIRSED} in the cases where the IR measurements between 8 and 1000~$\mu$m  are scarce. 
In such cases CIGALE can estimate properties of galaxies based on the composed models from the stellar and NIR--MIR part of the spectrum, while the fitter, based on the IR data, has only  limited capabilities to deduce a real character of the analyzed object.

Infrared luminosity, dust temperature, and dust mass of galaxies, along with $\rm{\alpha}$ and $\rm{\beta}$ parameters, were derived from \texttt{the CMCIRSED} model. 
The {reliable}  LIRGs and ULIRGs, identified by this code, together with their physical parameters, are listed in Tab.~\ref{ULIRGS_wyniki}. 
Mean values for ULIRGs, LIRGs, and normal galaxies derived from the \cite{casey12} model are presented in Table~\ref{tab:mean_parameters}.

\section{ LIRGs and ULIRGs   main physical properties obtained from the SED fitting processes}
\label{sec:SEDresults}

\begin{figure*}[]
\begin{center}
\includegraphics[width=0.92\textwidth,clip]{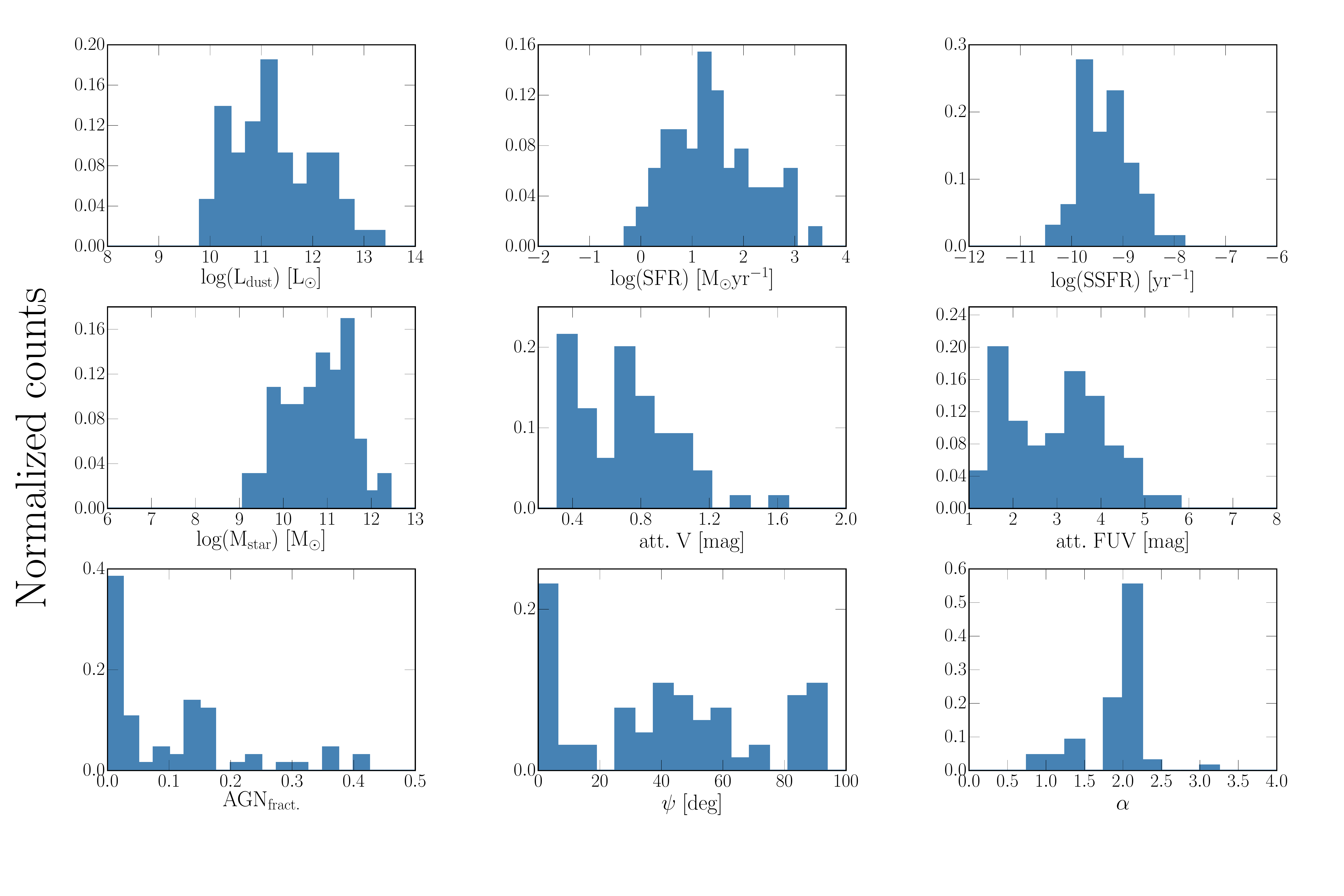}
\end{center}
\caption{Distribution of physical properties obtained by CIGALE for the ADF--S galaxies: dust luminosity ($\rm{L_{dust}}$), star formation rate (SFR), specific SFR, stellar mass ($\rm{M_{star}}$), dust attenuation in V and FUV bands,
 fractional contribution of  AGN to the MIR emission ($\rm{AGN_{fract}}$), AGN's torus angle with respect to the line of sight ($\psi$), IR spectral power-law slope $\alpha$. }
 \label{fig:5}
 \end{figure*}

\subsection{Dust luminosity}
\label{attenuation_V}

As  mentioned above, the dust luminosity of the ADF--S galaxies was calculated based on the \cite{dale14} model. 
The median dust luminosity is equal to $\rm{10^{10.30}\mbox{ }\pm {10^{1.07} [L_{\odot}]}}$. 
From the $\rm{L_{dust}}$  distribution  (Fig.~\ref{fig:5} top {left} panel) one can clearly recognize three peaks: 
(1) a very broad sample of normal star-forming  galaxies, with a maximum located at approximately $\rm{10^{10}\mbox{ }[L_{\odot}]}$, 
(2) a much narrower  distribution of galaxies with $\rm{log(L_{dust})}\mbox{ }\sim$ 11.3 $\rm{L_{\odot}}$, and 
(3) a group of ultra bright galaxies, with the peak of   $\rm{log(L_{dust})}$ distribution near 12.40 $\rm{[L_{\odot}]}$, and the tail shifted towards very bright sources (HLIRGs: Hyper-Luminous Infrared Galaxies). 

To examine how the extremely dusty component affects the stellar part of the spectra we have calculated their dust attenuation. 
With CIGALE we are able to calculate the amount of obscuration of stellar luminosity  for the old  stellar population using V and FUV filter measurements. 
As the correlation between ADF--S results and the mock catalog has a large scatter, we decided to check only the general trend for this relation.  
We have found that the dust attenuation for the old stellar population is not strong and has a similar value for normal galaxies, LIRGs, and ULIRGs, that is,  $\sim$ 0.7 [mag]. 
The same conclusion might be drawn from exemplary SEDs shown in Fig.~\ref{fig:2}, where it is difficult to distinguish between  unattenuated and attenuated stellar components for the old stellar population.  

Our results show that the shape of the spectrum  from the  old stellar population almost remains the same, and only in the  ultraviolet wavelengths, related to the emission from young stars, stellar spectra are obscured on the average level of 3 mag (Fig.~\ref{fig:9},  and Fig.~\ref{fig:2} -- example SEDs).   
Our findings are inconsistent with \cite{daCunha10} for 16 ULIRGs preselected from the IRAS~1~Jy survey, observed in the 5--38 $\mu$m range by Spitzer/IRS. 
To estimate the dust attenuation  \cite{daCunha10} used the strength of the 9.7 $\mu$m silicate feature from the Spitzer/IRS mid-infrared spectra, and  converted them  to the V-band optical depth by assuming dust optical properties and geometry. 
Da Cunha et al. (2010)  found that, for ULIRGs, the V-band optical depth is very large (mean $\tau^{Si}_{V}$ = 33.49 $\pm$ 7.14), and the shape of  the stellar spectrum is distorted by dust attenuation \citep[][ Fig.~3]{daCunha10}.

This alternative impact on the shape of the old stellar populations is due to the difference in the attenuation curve adopted in the MAGPHYS code used by \cite{daCunha10} and in the version of CIGALE used here. 
The MAGPHYS method leads to  a flatter attenuation curve than the Calzetti law adopted here \citep{2013MNRAS.432.2061C}  and implies a higher mean attenuation, which also affects the NIR spectrum and the contribution of the old stellar population (Lo Faro et al. in preparation).

\subsection{Stellar mass}
\label{subsec:stellarmass}
 
ADF--S galaxies are relatively massive with the median value of $\rm{M_{star}}$ equal to $\rm{10^{9.38}\mbox{ }\pm {10^{0.38} \mbox{ }[M_{\odot}]}}$. 
The mean $\rm{M_{star}}$ computed for ULIRGs, LIRGs, and galaxies with $\rm{log(L_{dust})<11\mbox{ }[L_{\odot}]}$ is equal to 11.51 $\pm$ 0.37, 10.35 $\pm$ 0.31, and {10.03 $\pm$ 0.50} $\rm{[M_{\odot}]}$, respectively. 
Stellar masses computed by CIGALE for the ADF--S sample are consistent with  stellar masses calculated for  LIRGs and ULIRGs   located at similar  redshifts, selected from infrared surveys and published by:
\begin{itemize}
 \item \cite{u12}; found 53 LIRGs and 11 ULIRGs from the  The Great Observatories All-sky LIRG Survey \cite[GOALS]{armus09} at redshift 0.012 $<$ z $<$ 0.083. They found mean stellar mass for LIRGs equal to $\rm{log(M_{LIRGs\mbox{ }star})}$=10.75$\pm$0.39 [$\rm{M_{\odot}}$], and for the ULIRGs: $\rm{log(M_{ULIRGs\mbox{ }star})}$=11.00$\pm$0.40 [$\rm{M_{\odot}}$],
 
 \item \cite{PereiraSantaella15}; described 29 local systems and individual galaxies with infrared luminosities between $10^{11}$ and $10^{11.8}$ $\rm{[L_{\odot}]}$, selected from IRAS Revised Bright Galaxy sample \citep{sanders03}. Among the  UV to FIR SED fitting results presented by \cite{sanders03} (Tab.~7),  we found 19 LIRGs  with a mean stellar mass equal to 10.82 $\pm$ 0.32 $\rm{[M_{\odot}]}$, and median value = 10.88 $\rm{[M_{\odot}]}$.
 
 \item Stellar masses estimated for ULIRGs are also consistent with \cite{Rothberg2013} dynamical masses calculated for  eight local (z$<$0.15) ULIRGs, randomly selected from a sample of 40 objects taken from IRAS 1 Jy Survey. The mean dynamical mass at $I$-band for ULIRGs obtained by \cite{Rothberg2013} is equal to 11.64$\pm$0.32.
 
 \end{itemize}
 
We have also compared our results with more distant galaxies, reported by: 
\begin{itemize} 
\item   \cite{giovannoli11}; 62 LIRGs from the Extended Chandra Deep Field South, selected at 24 $\rm{\mu}m$, on redshift $\sim$0.7 with the $\rm{log(M_{star})}$ between 10 and 12 [$\rm{M_{\odot}}$], and with a peak at 10.8 [$\rm{M_{\odot}}$].

\item \cite{melbourne08}; who calculated the stellar masses for 15 LIRGs from the GOODS-S HST treasury field; For this sample of LIRGs, with redshift$\sim$0.8, the  mean stellar mass equals log($\rm{M_{star}})\sim$10.51 [$\rm{M_{\odot}}$],

\item \cite{santos15}; for 12 XDCP J0044.0-2033 cluster members at redshift 1.58 observed by Herschel. For the sample of 12 ULIRGs, we have found the mean $\rm{log(M_{star})=11.06\mbox{ }\pm\mbox{ }0.32}\mbox{ }[M_{\odot}]$,

\item  we also compared our results with the stellar masses for 122 sub-millimeter galaxies (SMGs),  with  median photometric redshift equal to 2.83 reported by \cite{daCunha15}. 
We selected ULIRGs and LIRGs  from the SMG sample,  and found their mean log($\rm{M_{star}}$) for 88 ULIRGs equal to 10.88 $\pm$ 0.51 $\rm{[M_{\odot}]}$, while for 8 LIRGs: mean log($\rm{M_{star}})$ was equal to 9.77~$\pm$~0.52~$\rm{[M_{\odot}]}$.     
\end{itemize} 

In both cases (local, and more distant Universe), the stellar masses for the  LIRGs and ULIRGs   we found in  the ADF--S data are consistent with the literature. 
Both our results and results reported in the literature suggest that LIRGs and ULIRGs are more massive than normal galaxies, and that their stellar masses increase with infrared luminosity.

\subsection{IR power-law slope}
\begin{figure}[h]
        \includegraphics[width=0.5\textwidth, height=0.25\textheight, clip]{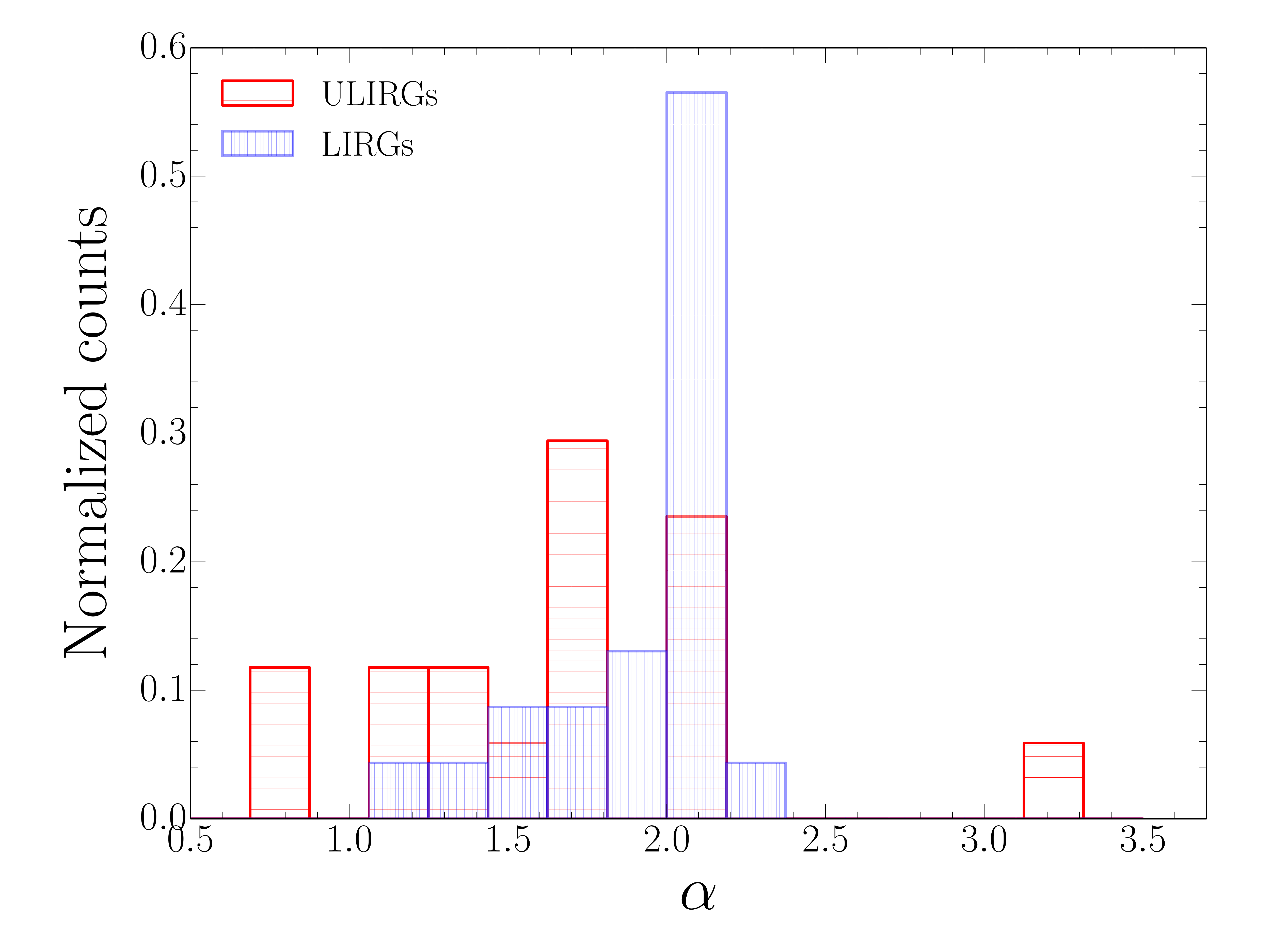}
        \caption{Histogram of the  $\alpha$ parameter from \cite{dale14} model. 
        The horizontal striped red histogram represents ULIRGs while the vertical striped blue histogram corresponds to LIRGs. }
        \label{fig:7}
\end{figure}

\begin{figure*}[t]
        \centering
        \begin{minipage}[t]{0.47\linewidth}
                \includegraphics[width=1\textwidth, height=0.3\textheight, clip]{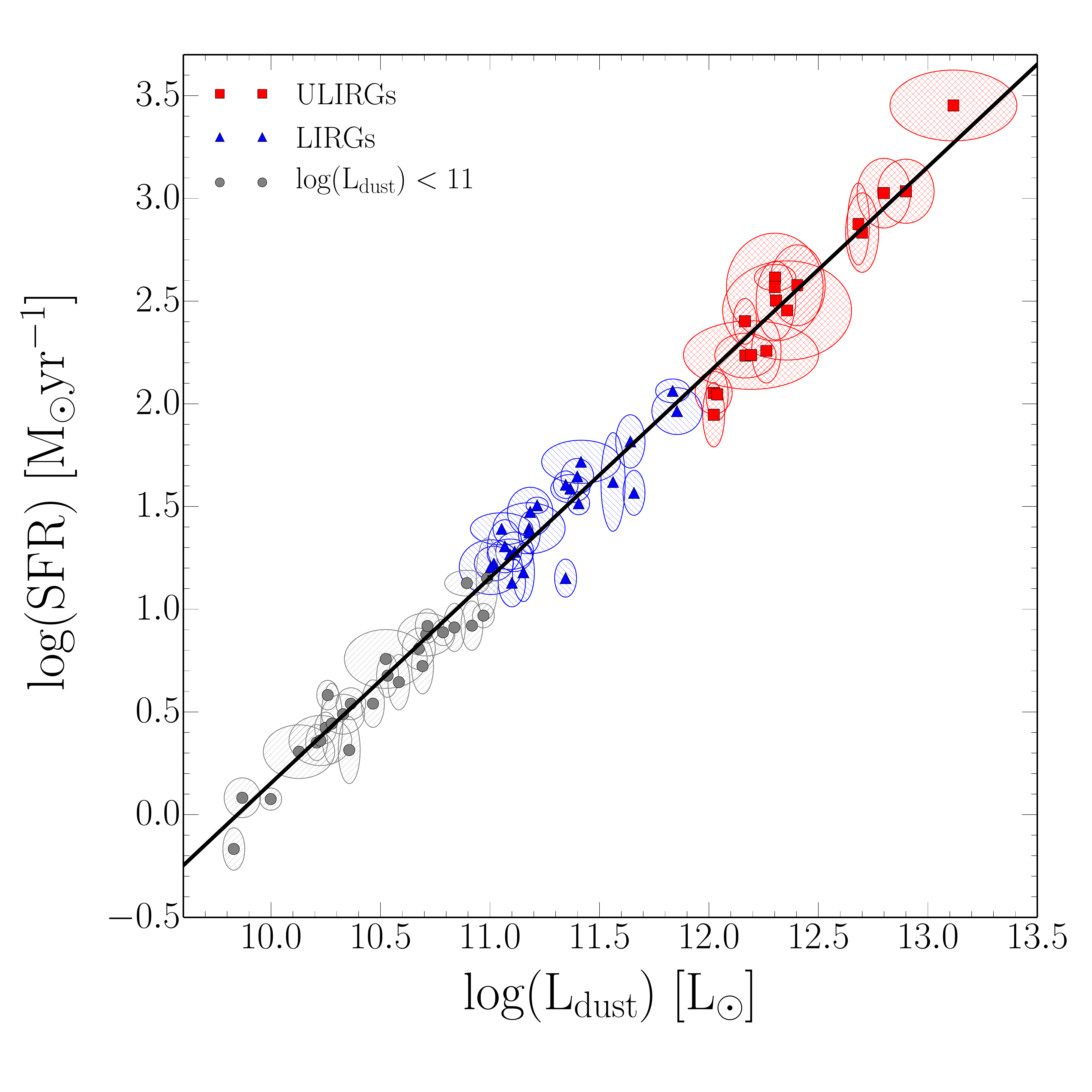}
                \caption{Dust luminosity versus  SFR for ADF--S sample.
                Filled red squares represent ULIRGs,  filled blue triangles correspond to LIRGs, and the filled gray circles  represent normal galaxies with $\rm{log(L_{dust})<11\mbox{ }[L_{\odot}}]$. 
                The radius of an ellipse represents error bars for dust luminosity, and star formation rate, respectively.
                The black line corresponds to the \cite{Kennicutt1998} star-formation rate, i.e.,  the dust luminosity relation.}
                \label{fig:8}
        \end{minipage}
        \quad
        \begin{minipage}[t]{0.47\linewidth}
                \includegraphics[width=1\textwidth, height=0.3\textheight, clip]{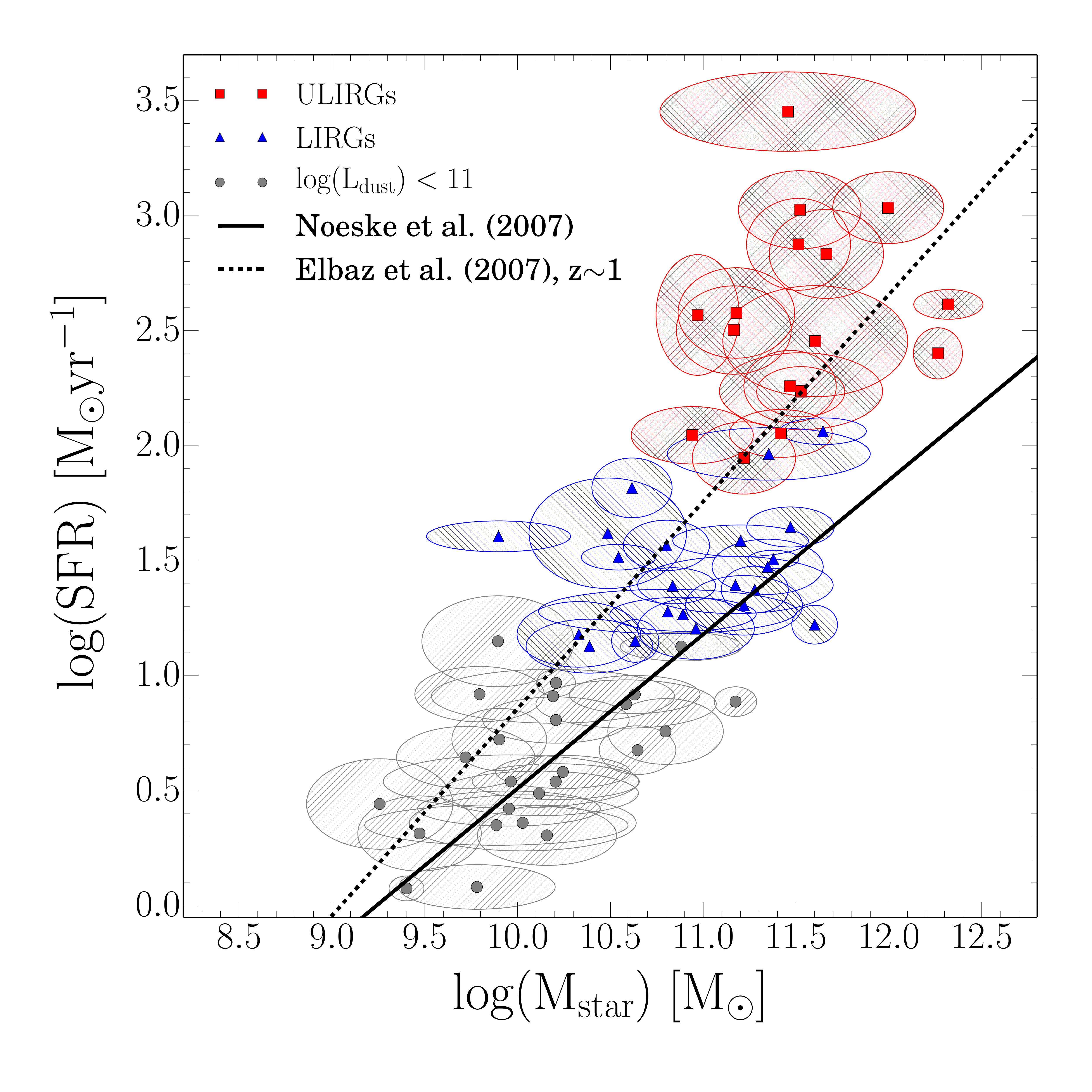}
                \caption{Stellar mass versus SFR for  
                ULIRGs (filled red ellipses), for LIRGs (filled blue ellipses), and for normal galaxies (filled gray ellipses).
                The black solid line represents $\rm{log(M_{star})}$ : log(SFR) relation  given by \cite{Noeske2007} for galaxies at redshift $\simeq0.6$, when dashed line corresponds to \cite{Elbaz07} relation for galaxies at z$\simeq$1.}
                \label{fig:9}
        \end{minipage}
\end{figure*}

Figure~\ref{fig:7} shows the distribution of $\alpha$ parameter \citep{dale02} for the ADF--S  LIRGs and ULIRGs  . 
It demonstrates that our sample consists mostly of  galaxies  active in star formation. 
Only five ULIRGs have  $\alpha$ $>$ 2. 
One of them is the most distant galaxy from our sample, a quasar, with known spectroscopic z=1.23, with $\alpha$ = 3.15 $\pm$ 0.47. 
Unfortunately, for this galaxy  we have no \textit{Herschel/SPIRE} data which implies that the temperature of dust may be not well evaluated.  

Nevertheless, more than 80\% of ULIRGs, and 95\% of LIRGs are characterized by a low value of $\alpha$. 
Taking into account a high reliability of the $\alpha$ parameter  (the Pearson product-moment correlation coefficient with mock catalog is equal to 0.95, Fig.~\ref{fig:3}), it is possible to  draw the conclusion that  LIRGs and ULIRGs   are very active in star formation. 
Furthermore, based on the  distribution of $\alpha$ parameter between different groups of sources, we can conclude that  LIRGs and ULIRGs   are characterized by a stronger starburst activity than normal galaxies.

The mean value of $\alpha$ is lower for ULIRGs than for LIRGs, but taking into account the uncertainties we cannot distinguish both groups based on this parameter.
Values of $\alpha$ computed for ULIRGs, LIRGs and normal galaxies are listed in Tab.~\ref{tab:mean_parameters}.
 
\subsection{Star-formation rate}

We checked the distribution and the median values of the star-formation rate (SFR) for our sample.
Values of SFR for  ADF--S   LIRGs and ULIRGs   vary from 13.45 to almost 2900  $\rm{[M_{\odot} yr^{-1} ]}$ (the latest for the extreme case of a quasar at z$_{\rm{spec}}$=1.23), with the median value equal to {65.58} $\rm{[M_{\odot} yr^{-1} ]}$. 

For ULIRGs, LIRGs, and the rest of the galaxies from our sample, the mean  $\rm{log(SFR)_{ULIRG}}$,  $\rm{log(SFR)_{LIRG}}$, and $\rm{log(SFR)_{normal}}$ are equal to 2.53 $\pm$ 0.39, 1.47 $\pm$ 0.20, and {0.47 $\pm$ 0.21} $\rm{[M_{\odot}yr^{-1}]}$, respectively (Tab.~\ref{tabela1}).  
Our results are consistent with the \cite{Kennicutt1998} relation between SFR and dust luminosity:
\begin{equation}
\rm{SFR\mbox{ }[M_{\odot yr^{-1}}]=4.5\times10^{-44} \times L_{dust} [erg\mbox{ }s^{-1}]},
\label{eq:Kennicutt98}
\end{equation}
where $\rm{L_{dust}}$ refers to the infrared luminosity integrated over the full mid and far-IR spectrum (8-1000 $\rm{\mu}$m). 
Figure~\ref{fig:8} shows the $\rm{SFR-L_{dust}}$ distribution of ADF--S galaxies with an additional black line which represents Eq.~\ref{eq:Kennicutt98}. 
We conclude that our results are in very good agreement with the  \cite{Kennicutt1998} relation, and the SFR increases with dust luminosity in the same way for normal galaxies, LIRGs, and ULIRGs.  
   
Our results for the LIRGs and ULIRGs are consistent with those presented in the literature, that is;  
\begin{itemize}
\item \cite{santos15}: $\rm{<log(SFR)_{ULIRGs}>}$ = 2.51 $\pm$ 0.22 $\rm{[M_{\odot}yr^{-1}]}$,
\item \cite{podigachoski15}: $\rm{<log(SFR)_{ULIRGs}>}$ = 2.55 $\pm$ 0.26 $\rm{[M_{\odot}yr^{-1}]}$,
\item  \cite{daCunha15}: $\rm{<log(SFR)_{ULIRGs}>}$ =2.55 $\pm$ 0.32, and $\rm{<log(SFR)_{LIRGs}>}$ = 1.69 $\pm$ 0.20 $\rm{[M_{\odot}yr^{-1}]}$, and 
\item \cite{howell10} (181 LIRGs and 21 ULIRGs on the median redshift 0.008 and a maximum redshift equal to 0.088):  $\rm{<log(SFR)_{ULIRGs}>}$ = 2.45 $\pm$ 0.14 $\rm{[M_{\odot}yr^{-1}]}$, and $\rm{<log(SFR)_{LIRGs}>}$ = 1.72 $\pm$ 0.69 $\rm{[M_{\odot}yr^{-1}]}$.    
\end{itemize}

Figure~\ref{fig:9} shows the correlation between SFR and stellar mass for the ADF--S sample. 
We have over-plotted the SFR--$\rm{M_{star}}$ relation found by \cite{Noeske2007} for All-Wavelength Extended Groth Strip International Survey galaxies (AEGIS) at redshift range from 0.2 to 0.7,  and \cite{Elbaz07} for Great Observatories Origins Deep Survey (GOODS), at $z\sim 1$. 
We can  compare these results to ours since both the sample selection and redshift range (0.2-0.7) are similar. 
We found that the majority of ADF--S ULIRGs are located above the observed linear correlation from \cite{Noeske2007} and \cite{Elbaz07}. 
LIRGs  are also shifted towards higher SFR, and the trend is consistent with \cite{Noeske2007}. 
The minimal value of log(SFR) for our LIRGs is  equal to 1.0 $\pm$ 0.2 $\rm{[M_{\odot} yr^{-1}]}$, and is consistent with their LIRG SFR limit obtained by \cite{Noeske2007}. 

\subsection{Specific star-formation rate}

We also calculated specific star-formation rate (SSFR), which is defined as the ratio between SFR  and the total stellar mass of a galaxy:
\begin{equation}
\rm{SSFR = SFR/M_{star}\mbox{ }[yr^{-1}].}
\end{equation}
The median values of SSFR for ADF--S ULIRGs, LIRGs, and normal galaxies are equal to -8.98 $\pm$ 0.47, -9.48 $\pm$ 0.45, and -9.56 $\pm$  0.38 $\rm{[yr^{-1}]}$, respectively (Tab.~\ref{tabela1}). 
These values indicate  that the SSFR increases with dust luminosity.  

We compared our results to the ones presented in the literature for LIRGs and ULIRGs selected in the infrared.
The median value of SSFR for ULIRGs in the GOALS field \citep{howell10} is equal to -9.41 [yr$^{-1}$], while \cite{u12} obtained a value of -9.17 [yr$^{-1}$]. Our results are consistent with both of them. 

On the other hand, \cite{daCunha15} found $\rm{SSFR_{ULIRGs}}=-8.32\mbox{ }\pm\mbox{ }0.49$ $\rm{[yr^{-1}]}$ which is also consistent with ours, but their SSFR for LIRGs is higher than the one obtained for the ADF--S ($\rm{SSFR_{LIRGs}}=-8.08\mbox{ }\pm\mbox{ }0.38$ $\rm{[yr^{-1}]}$). 
This discrepancy might be caused by a small number of objects (\citealp{daCunha15} sample consists of only eight LIRGs) and their  much-higher redshift (mean photometric redshift equal to 1.90). 
Another possibility is the use of a flatter attenuation curve than the Calzetti law (last paragraph of Section~\ref{attenuation_V}, and Lo~Faro et al., in preparation).

The mean SSFR calculated for 19 local LIRGs presented by \cite{PereiraSantaella15} is equal to -10.60 $\pm$ 0.45 $\rm{[yr^{-1}]}$, and this value is lower than the other values from the literature. 
It might be caused by the different sample selection or  statistical differences between samples. 
The results obtained for the ADF--S sample are statistically consistent with the results from \cite{howell10,u12}, and \cite{PereiraSantaella11}.

\begin{figure}[h]
        \begin{center}
                \includegraphics[width=0.5\textwidth, height=0.3\textheight, clip]{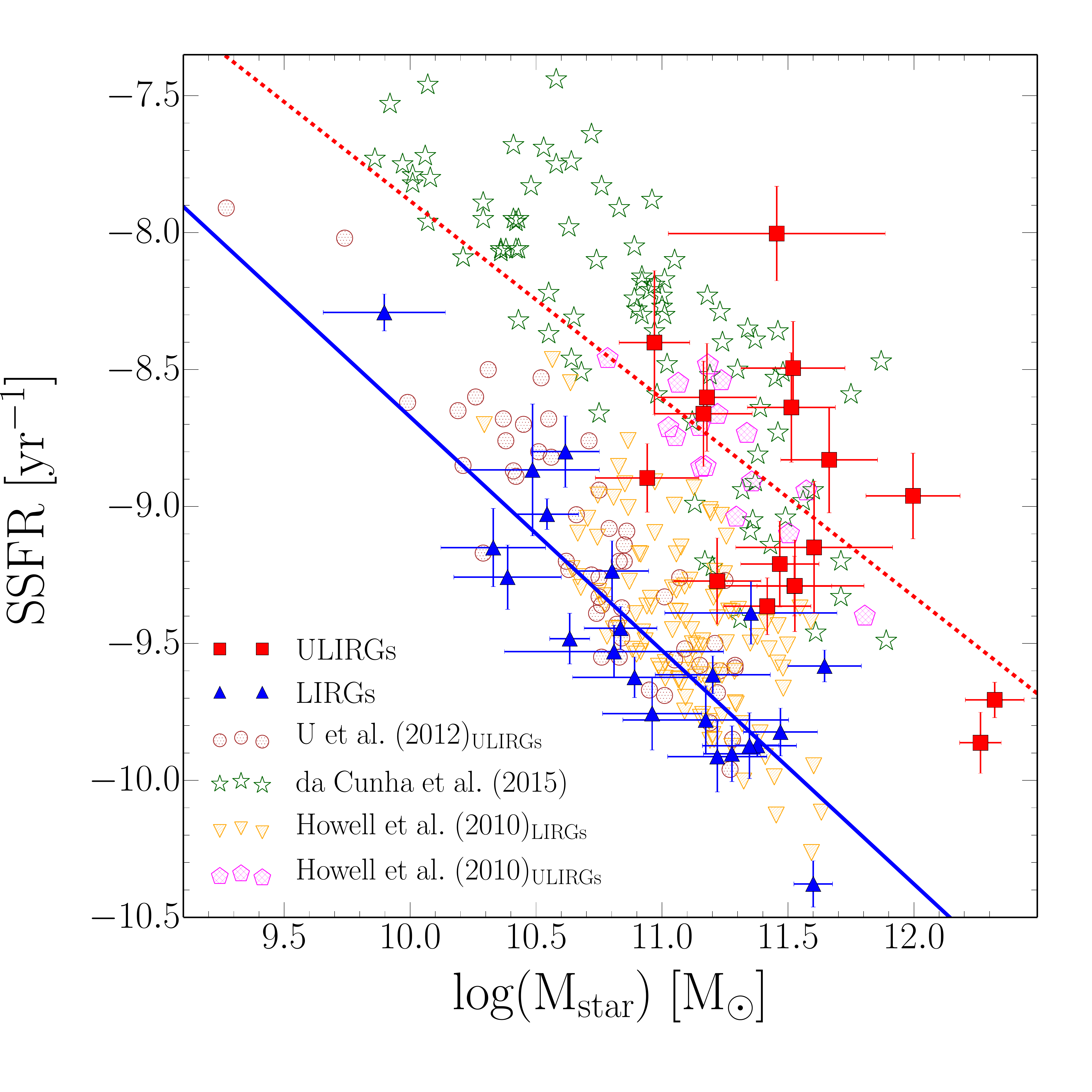}
        \end{center}
        \caption{Relationship between SSFR and stellar mass for ULIRGs and LIRGs derived from this work. 
        Lines represent the linear fit for ULIRGs (red dotted line) and LIRGs (blue solid line) from the ADF--S sample. Filled red squares and filled blue triangles correspond to the ULIRGs and LIRGs from the ADF--S sample, respectively. 
        Filled blue circles represent the \cite{u12} LIRG sample, open inverted triangles represent the \cite{howell10} LIRG sample, and open black squares represent the \cite{howell10} ULIRG sample; open magenta diamonds correspond to the \cite{daCunha15} sample of ULIRGs.}
        \label{fig:10}
\end{figure}

Figure~\ref{fig:10} shows the relationship between stellar mass and SSFR for ULIRGs and LIRGs. 
This figure presents ADF--S results in comparison with  IR-selected samples found in the literature. 
We found that the SSFR/$\rm{M_{star}}$ relationship is similar for different samples \citep{howell10, daCunha15, u12}, with a slope equal to -0.72 and -0.92, for ULIRGs, and LIRGs, respectively.  
We have also isolated an additional control sample of normal galaxies (with $\rm{10<log(L_{dust})<11} \mbox{ }[L_{\odot}]$, 25 objects in total), and for them the calculated slope between stellar mass and SSFR is equal to -0.75.  
We find that SSFR decreases with increasing stellar mass, in agreement with what was found  by  \cite{cowie96}, \cite{brinchmann04}, \cite{buat07}, \cite{iglesias07}, and \cite{malek14}, and that the slope of this relationship  is similar for normal galaxies, LIRGs, and ULIRGs.   

\subsection{Equivalent width of the dust part of the spectra vs physical properties of  LIRGs and ULIRGs   }

\begin{figure*}[ht]
        \begin{center}
                \includegraphics[width=1\textwidth,clip]{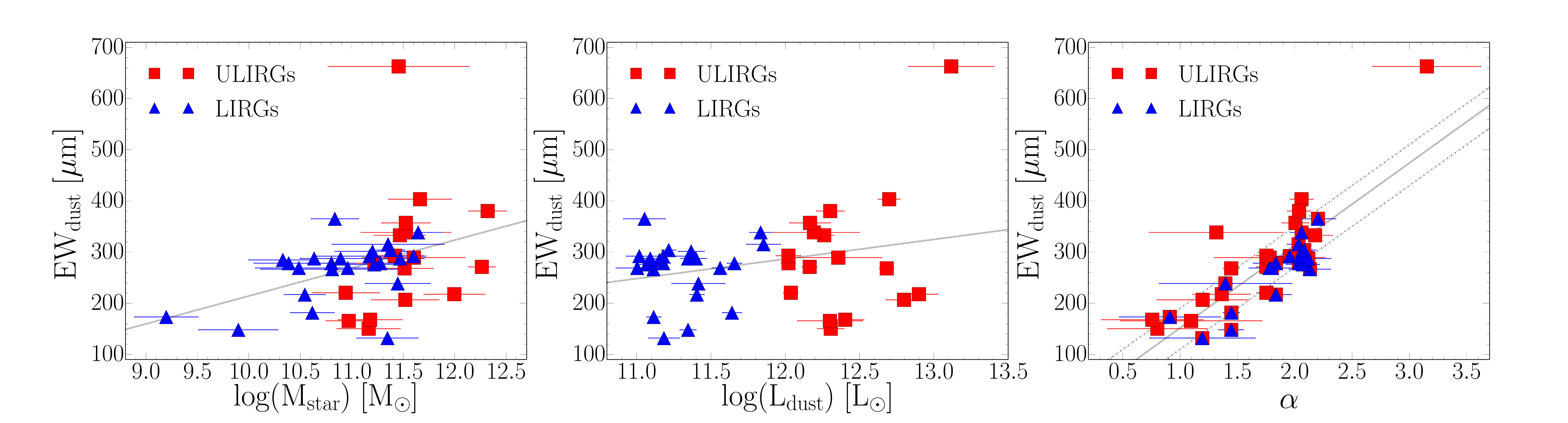}
        \end{center}
        \caption{Relationship between dust spectra  peak equivalent width and stellar mass (\textit{left panel}), dust luminosity (\textit{central panel}), and \cite{dale14} $\alpha$ parameter (\textit{right panel}). 
        Filled red squares represent ULIRGs, and filled blue triangles correspond to  LIRGs. 
        The most extreme value of $\rm{EW_{dust}}$ was found for the most distant object in the ADF--S sample ($z$=1.23): a quasar HE 0435-5304. 
        The dashed line represents the linear fit to the data.
        In the $\alpha$--$\rm{EW_{dust}}$ relation (\textit{right panel})  2$\sigma$ uncertainties for the fitted line (Eq.~\ref{eq:ew_alpha}) were added (solid gray lines). }
        \label{fig:13}
\end{figure*}

Based on the best fitted model found by CIGALE, we calculated artificial half equivalent widths of the dust peaks of the spectra ($\rm{EW_{dust}}$). 
We defined this value  as the difference between the wavelength for the maximal value of the dust peak ($\lambda_{max}$) and the 24 $\mu$m ($\lambda_{24\mu m}$)  for the rest-frame spectra multiplied by two:
\begin{equation}
\rm{EW_{dust}=2\times(\lambda_{max}-\lambda_{24\mu m})}.
\end{equation} 
We checked how the physical properties of LIRGs and ULIRGs  are related to the width of the dust emission peak. 
Our results show that stellar mass and dust luminosity increase with the increasing $\rm{EW_{dust}}$  (Fig.~\ref{fig:13}){, but the scatter  is too large to propose any strict relations.}  
The same plot includes  the $\alpha$ parameter from \cite{dale14} as a function of $\rm{EW_{dust}}$. 
$\rm{EW_{dust}}$ correlates very well with  \cite{dale14} $\alpha$ parameter.
Thus, the $\alpha$--$\rm{EW_{dust}}$ relation can be written as:
\begin{equation}
\rm{\alpha=0.6{\pm 0.3}+(0.0042\times EW_{dust})},
\label{eq:ew_alpha}
\end{equation}
where 0.3 uncertainty corresponds to the 2$\sigma$ level (more than 95\% of  LIRGs and ULIRGs   are within that distance from the line determined by the Eq.~\ref{eq:ew_alpha}).
This implies that  the wider  the dust emission peak of the spectrum,  the more quiescent a galaxy is. 
The notable exception is, again, the most distant galaxy in the sample (quasar HE 0435-5304), as the star-formation rate for this object is equal to 2834.36 $\pm$ 978.20 $\rm{[M_{\odot}yr^{-1}]}$, and $\alpha$=2.83 $\pm$ 0.59.

We also measured the distribution of the $\lambda_{max}$ parameter itself (Fig.~\ref{fig:z}). 
We found that the majority of LIRGs and ULIRGs are cold (i.e.,  characterized by the IR peak at wavelengths longer than 90~$\mu$m). 
This finding is in agreement with \cite{symeonidis11}.

\subsection{Dust temperatures \&  dust mass  for ADF--S  LIRGs and ULIRGs  }
\label{sec:gosia}

\begin{figure*}[htp!]
        \begin{center}
                \includegraphics[width=0.45\textwidth]{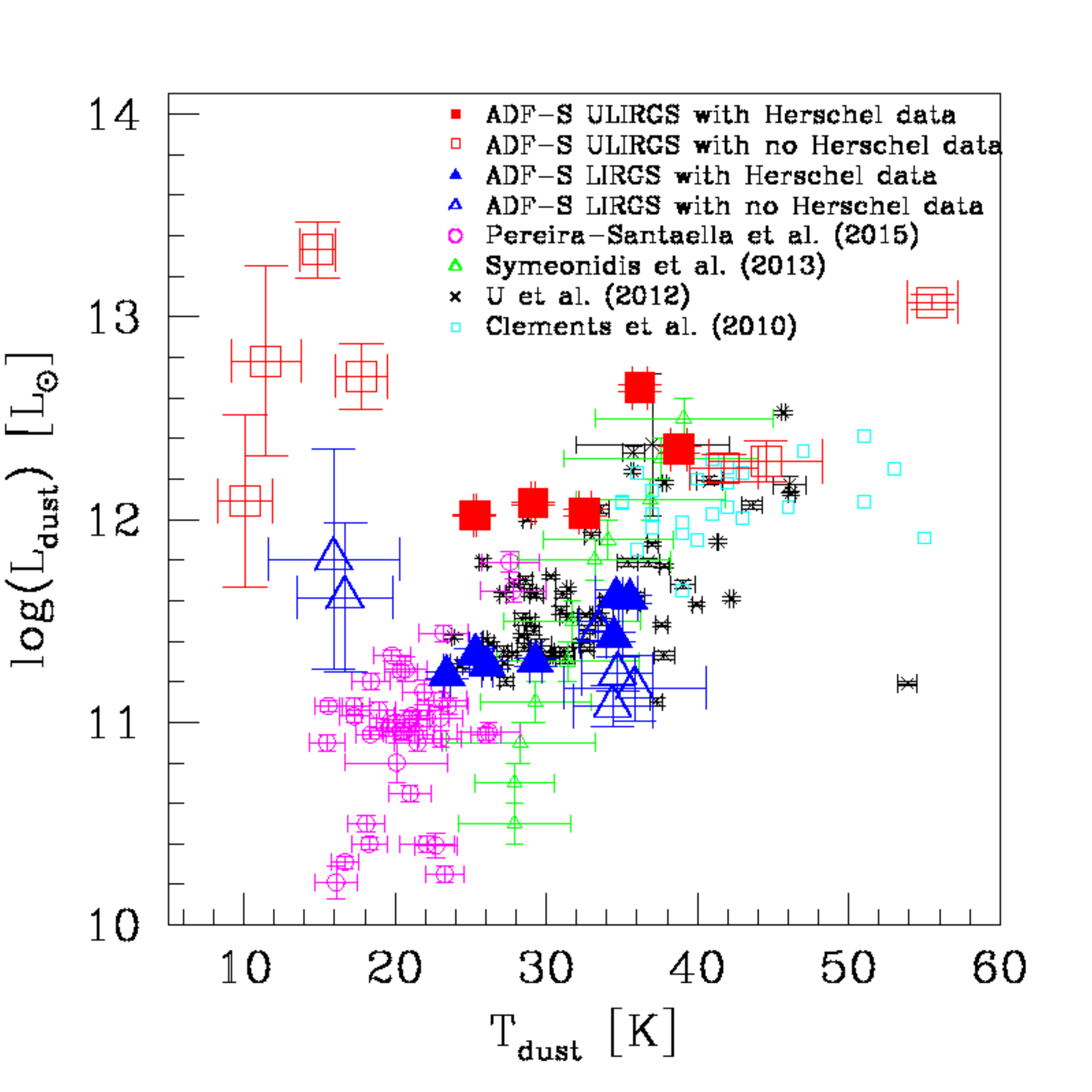}\includegraphics[width=0.45\textwidth]{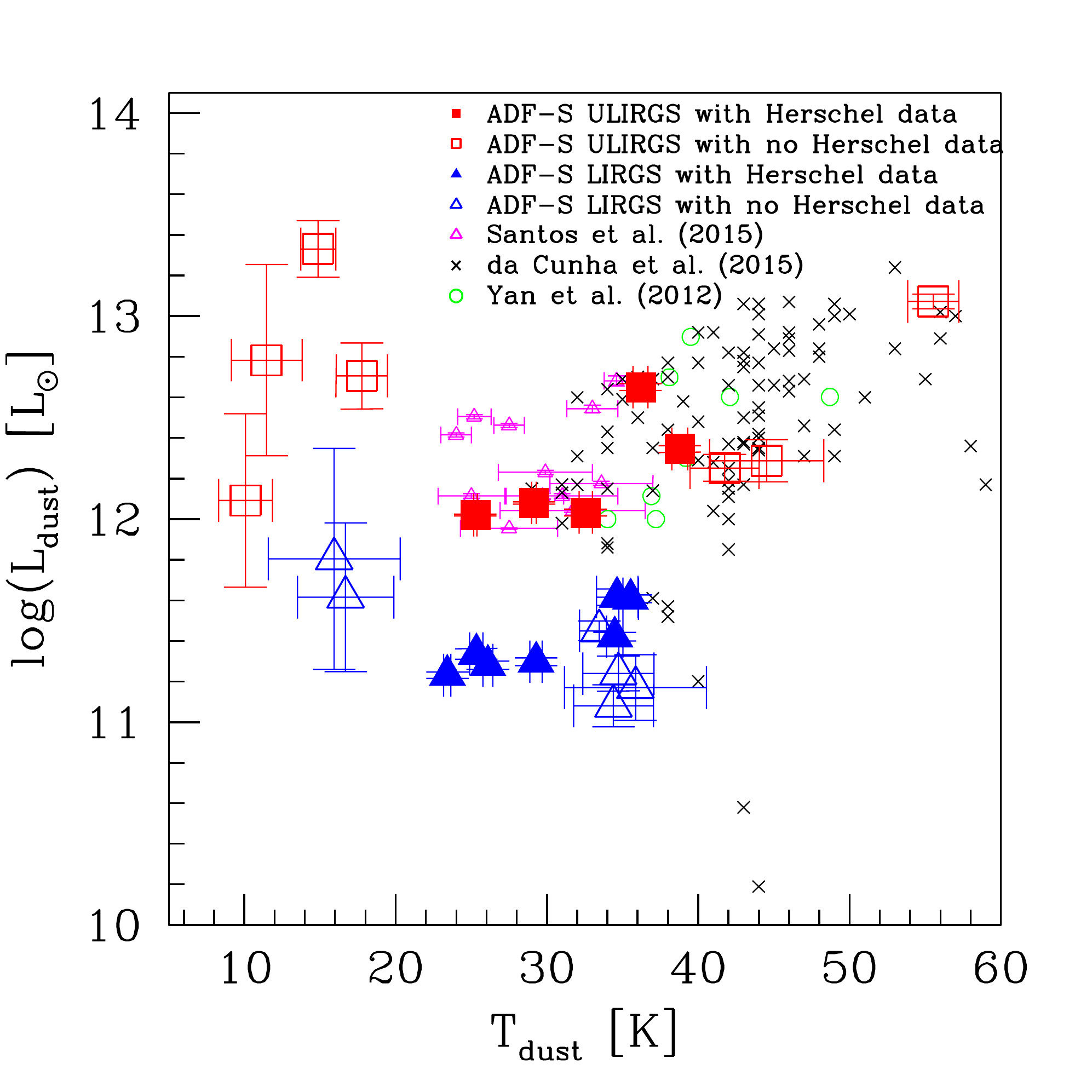}\\
        \end{center}
        \caption{
        Relation of dust temperature versus dust luminosity for local (\textit{left panel}), and  higher redshift  LIRGs and ULIRGs   (\textit{right panel}). 
        Red squares represent ULIRGs, and blue triangles - LIRGs from the ADF--S sample (empty symbols represent object without Herschel data).  
        \textit{Left panel:} 
        Open magenta circles denote  \cite{PereiraSantaella15} sample of LIRGs, open green triangles stand for \cite{symeonidis13},  black "x-s" --  \cite{u12} sample of LIRGs and ULIRGs, and open cyan squares -- \cite{clements10} sample of ULIRGs.  at redshift range $\rm{z<1}$. 
        \textit{Right panel}: 
        open magenta traingles represent \cite{santos15} sample of LIRGs and ULIRGs, black "x-s" -- LIRGs and ULIRGs from \cite{daCunha15}, and green open circles represent \cite{yan14} ULIRGs sample.
        The errorbars of \cite{daCunha15} data were not shown since they are very large and make the plot unreadable.}  
        \label{fig:18}
        
        \begin{center}
                \includegraphics[width=0.45\textwidth]{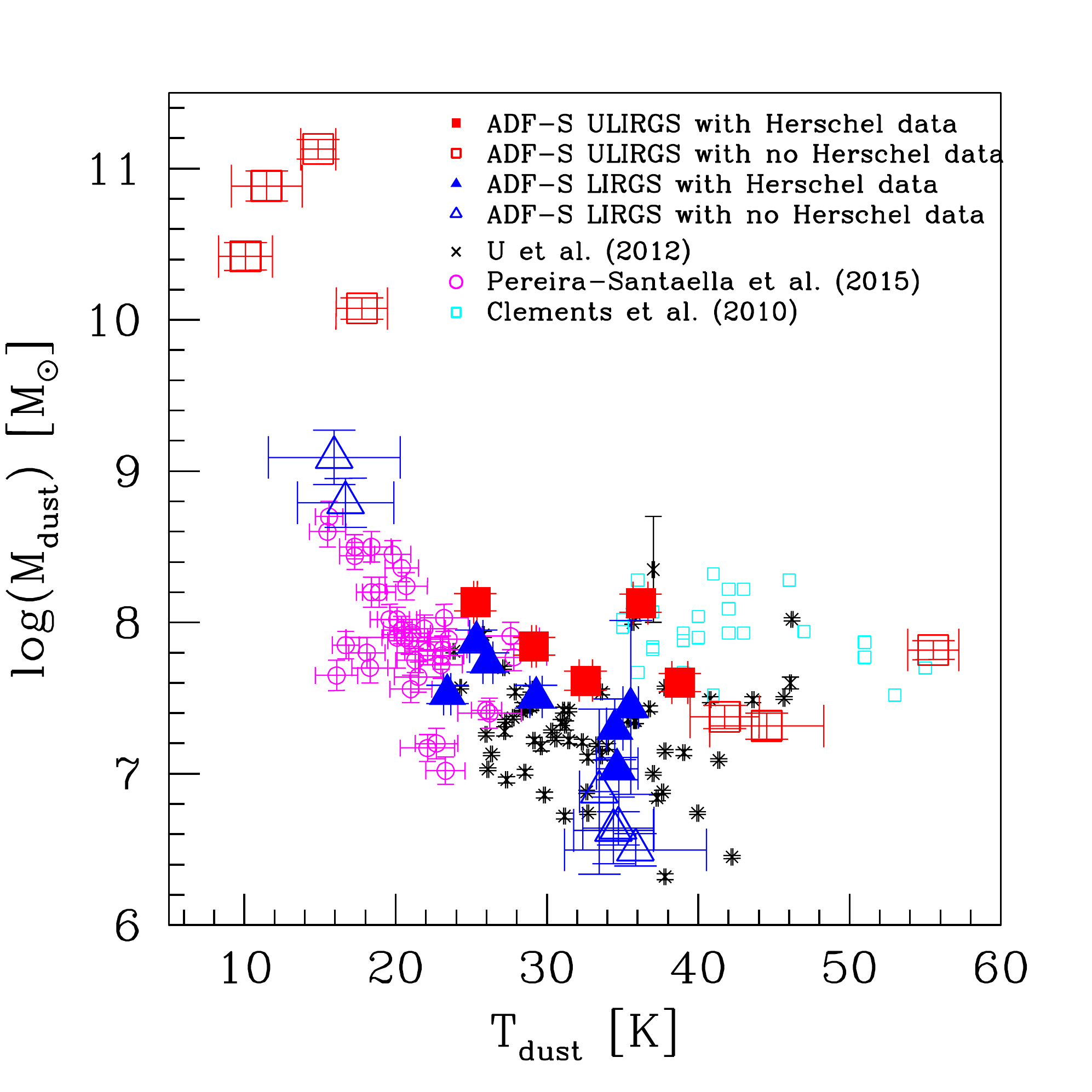}\includegraphics[width=0.45\textwidth]{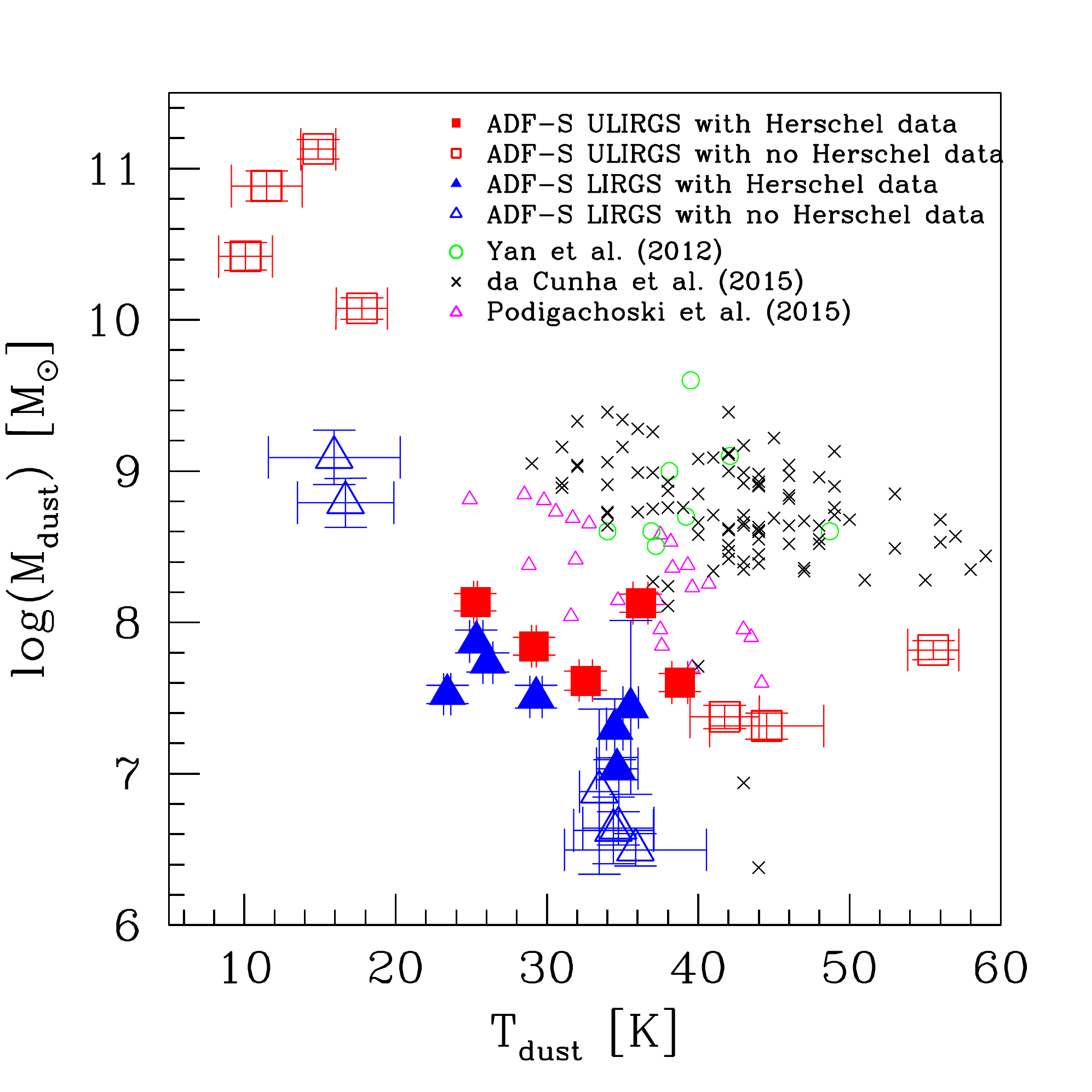}\\
        \end{center}
        \caption{
                Dust luminosity versus dust mass for local (\textit{left panel}) and hight redshift \textbf LIRGs and ULIRGs   (\textit{right panel}). Red squares represent LIRGs, while blue triangles - ULIRGs from the ADF--S sample (empty symbols present object without Herschel data). 
                \textit{Left panel:} 
                black crosses represent LIRGs and ULIRGs from \cite{u12}, and open pink circles correspond to the \cite{PereiraSantaella15} sample of LIRGs.
                \textit{Right panel:} 
                open green circles  represent \cite{yan14} ULIRGs, black "x-s" -- LIRGs and ULIRGs from \cite{daCunha15}, open pink triangles represent \cite{podigachoski15} sample of LIRGs and ULIRGs,  and open cyan squares -- \cite{clements10} sample of ULIRGs.
                The errorbars of \cite{daCunha15} data were not shown since they are very large and make the plot unreadable.}
                \label{fig:19}
        \end{figure*}

We used the \cite{casey12} model, and the \texttt{CMCIRSED}  pipeline, in order to obtain the independent identification of  LIRGs and ULIRGs   from the ADF--S sample, and, for objects characterized by  $\rm{log(L_{dust})>=11 \mbox{ }[L_{\odot}]}$ by both methods, to calculate $\rm{T_{dust}}$ and $\rm{M_{dust}}$.
Figs.~\ref{fig:18}, and~\ref{fig:19} show the dust luminosity/dust temperature, and the dust temperature/dust mass relationships 
for 25  objects identified as  LIRGs and ULIRGs    both by  CIGALE, and \texttt{CMCIRSED}, and the corresponding measurements from  the literature. 
All ADF--S  LIRGs and ULIRGs   are marked by red squares (ULIRGs) and blue traingles (LIRGs); empty symbols represent objects without \textit{Herschel} counterparts.

\begin{table*}[]
        \begin{center} 
                \caption[]{Main physical parameters for ULIRGs, LIRGs and galaxies with total infrared luminosity  $\mathrm{L_{TIR}<10^{11} L_{\odot}}$ from the ADF--S sample derived by CIGALE and \texttt{CMCIRSED}. 
                For CIGALE, we have listed numbers of sources, mean redshift, logarithm of stellar mass, dust luminosity, AGN fraction, AGN torus angle with respect to line of sight $\psi$, logarithm of star-formation rate, specific SFR, and  $\alpha$ parameter as defined by the \cite{dale14} model. 
                For \texttt{CMCIRSED} we have presented  the number of objects, dust luminosity, dust temperature, and the dust mass.}
                \label{tabela1}
                \begin{tabular}{c| r| r| r } \hline \hline
                        \label{tab:mean_parameters}
                        \multirow{2}{*}{paramater} & \multirow{2}{*}{ULIRGs} & \multirow{2}{*}{LIRGs} & \multirow{2}{*}{normal galaxies}\\ 
                        &        &       &  \\ \hline
                        \multicolumn{4}{c}{CIGALE} \\\hline 
                        \# \tablefootmark{a} & 17 & {22} & 30\\
                        $\rm{<z>}$ & 0.61 $\pm$ 0.26 & 0.23 $\pm$ 0.08 & 0.08 $\pm$ 0.02\\
                        $\rm{log(M_{star})}\mbox{ }[M_{\odot}]$ & 11.51 $\pm$ 0.37 & 10.95 $\pm$ 0.31 & 10.03 $\pm$ 0.50 \\
                        $\rm{log(L_{dust})}\mbox{ }[L_{\odot}]$ & 12.40 $\pm$ 0.32 & 11.31 $\pm$ 0.25 &  10.35 $\pm$ 0.32 \\
                        $\rm{AGN_{fraction}}$ [\%] & 19.12 $\pm$ 6.76 & 12.54 $\pm$ 3.59 & 12.82 $\pm$ 1.72 \\
                        $\psi$ & 19.28 $\pm$ 7.25 & 43.85 $\pm$ 9.14 & 55.29 $\pm$ 6.78 \\
                        $\psi_{AGN\_10}$\tablefootmark{b} & 19.73 $\pm$ 6.64& 46.41 $\pm$ 6.25 & 55.29  $\pm$ 18.40 \\
                        log(SFR) $\rm{[M_{\odot}yr^{-1}]}$   & 2.53 $\pm$ 0.39 & 1.47 $\pm$ 0.20 & 0.47 $\pm$ 0.21 \\
                        SSFR $\rm{[yr^{-1}]}$   & -8.98 $\pm$ 0.47 & -9.48 $\pm$ 0.45 & -9.56 $\pm$ 0.38 \\
                        $\alpha$ & 1.66 $\pm$ 0.27 &  1.92 $\pm$ 0.14 & 2.06 $\pm$ 0.28 \\ 
                        \hline
                        \multicolumn{4}{c}{\texttt{CIMCIRSED}} \\\hline 
                        \#\tablefootmark{a} & 12 & 13 & 17 \\
                        $ \rm{\log(L_{dust})\mbox{ }[L_{\odot}]} $ & $ 12.47\pm0.06 $ & $ 11.40\pm0.05 $ & $ 10.46\pm0.02 $ \\
                        $ \rm{T_{dust}\mbox{ }[K]} $ & $ 29.83\pm0.50 $ & $ 29.22\pm0.64 $ & $ 29.25\pm0.39 $ \\
                        $ \rm{\log(M_{dust})\mbox{ }[M_{\odot}]} $ & $ 8.69\pm0.02 $ & $ 7.46\pm0.07 $ & $ 6.43\pm0.05 $ \\ \hline
                \end{tabular} 
                \tablefoot{\tablefoottext{a}{number of sources;} \tablefoottext{b}{AGN torus angle with line of sight for sources with AGN contribution equalling more than 10\%;} }
        \end{center}
\end{table*}

Our first conclusion is  that the power-law + graybody fitting for objects without measurements at wavelengths longer than 160~$\mu$m (without \textit{Herschel/SPIRE} data) results in suspiciously  low temperature values (lower than 20 K), and thus should be treated as  unreliable. 
The lack of long-wavelength data forbids a proper estimation of the dust peak in the spectrum \citep[e.g.,][]{u12}. 
This might be caused by the fact that the SED-fitting in such a case is  performed only for radiation emitted by the large dust grains \citep{li05}, and not all the dust located in a galaxy.
As the \texttt{CMCIRSED} can be applied to the infrared data only, we decided to focus on the  LIRGs and ULIRGs   with both AKARI and \textit{Herschel/SPIRE} measurements (filled markers), to avoid possible underestimations of the dust temperature. 
Unfortunately, only five ULIRGs and seven LIRGs fulfill this condition, and thus our sample is reduced by $\sim$ 60\%. 
Nevertheless to estimate the $\rm{T_{dust}}$ -- $\rm{L_{dust}}$ relation we prefer to use the best quality data (which are not affected by possible bias caused by the incomplete SED).

We compared our results with the known samples of objects located in the similar redshift range ($\rm{z<1}$):
\begin{itemize}
        \item \cite{u12}, who used the \texttt{CMCIRSED} model  to calculate $\rm{T_{dust}}$, and $\rm{M_{dust}}$ values,
        \item \cite{symeonidis11,symeonidis13} who presented the Herschel sample with  $\rm{T_{dust}}$  estimated from the graybody emission. 
        The sample consists of 1159 sources:   LIRGs and ULIRGs  along with normal galaxies, at redshift range  $\rm{0.1<z<2.0}$,
        \item \cite{PereiraSantaella15}: here $\rm{T_{dust}}$ was calculated by fitting the graybody emission,  and $\rm{M_{dust}}$ was estimated using a power-law function,
        \item \cite{clements10} sample of local ULIRGs, detected at 850~$\mu$. The temperature used for our comparison was calculated based on the standard modified blackbody fit to SCUBA 850~$\mu$m fluxes, and IRAS 60,  100~$\mu$m and 450~$\mu$m (SCUBA) fluxes where available.   
\end{itemize}
To compare our galaxies with more distant onces, we have also used samples presented by:
\begin{itemize}
        \item \cite{podigachoski15}, who presented a sample of high-redshift ($z>1$) radio-loud AGNs from the Revised Third Cambridge Catalogue of radio sources; the values of  $\rm{T_{dust}}$ were estimated based on the gray-body emission,  and $\rm{M_{dust}}$ was calculated in the same way as in the \texttt{CMCIRSED} model. 
        \item \cite{yan14}; (redshifts between one and two) $\rm{T_{dust}}$ was based on the \cite{Siebenmorgen07} templates, and the stellar mass was computed in the same way as the  CMCIRSED model, 
        \item \cite{santos15}; z$\sim$1.58, $\rm{T_{dust}}$ was calculated by the graybody emission function fitting,
        \item \cite{daCunha15}; $\rm{z_{median}=2.83}$,  $\rm{T_{dust}}$ and $\rm{M_{dust}}$ were calculated by MAGPHYS model \citep{daCunha10,daCunha15}.
\end{itemize}

Both local and high redshift samples used for comparison were selected so that they were based on  a similar sample selection (IR bands) and an analogous method of estimation of the dust mass and temperature as for the ADF--S data.

\subsubsection{Luminosity/temperature relation}

We find that the dust temperature increases with the total infrared luminosity, and the relation is similar for ULIRGs and LIRGs (Fig.~\ref{fig:18}).
For the local Universe we compared our results with samples presented by \cite{u12}, \cite{symeonidis13},  \cite{PereiraSantaella11}, and \cite{clements10} (\textit{Fig.~\ref{fig:18}, left panel}). 
According to  \cite{symeonidis13}, the dust temperature of ULIRGs and LIRGs increases with dust luminosity. 
The relation found by \cite{symeonidis13}  is steeper than the $\rm{T_{dust}}$ versus $\rm{L_{dust}}$ relation for ADF--S,  \cite{u12}, and \cite{clements10} samples, however in all cases the increase of dust temperature with dust mass is very weak or even insignificant. 
As \cite{symeonidis13} suggested, the temperature-mass relation is shaped by the dust mass, and not the increase in  the dust heating. 
This theory can be supported by the distribution of an $\alpha$ parameter (Fig.~\ref{fig:7}) for the \cite{dale14} templates for LIRG and ULIRG samples. 
The distribution of the increasing global heating index $\alpha$ is very similar, and the difference between mean values of $\alpha$  between LIRGs and ULIRGs (Tab.~\ref{tabela1}) is  not significant. 
At the same time, the $\alpha$ distribution for normal galaxies  is significantly different.  

The dust luminosity/dust temperature relation for higher $z$ is shown in the right panel of Fig.~\ref{fig:18}.  Here we present a comparison of ADF--S $\rm{T_{dust}}$ versus $\rm{L_{dust}}$ relation with  LIRGs and ULIRGs   from \cite{yan14}, \cite{santos15}, and \cite{daCunha15}. 
It is clearly visible that the shape of the relation is preserved for higher redshifts.  
We conclude that  LIRGs and ULIRGs   in the local Universe are usually colder than similar objects at higher redshift.  
Our conclusion is in agreement with findings presented by  \cite{symeonidis13}, who claims that, in general, the temperature of  LIRGs and ULIRGs   is approximately constant, and the dust temperature is related to the dust mass. 
This finding  that  LIRGs and ULIRGs   with low dust mass are found only in the local Universe might be a selection effect. 

\subsubsection{Dust mass/temperature relation}

We tested whether or not our conclusions presented above are confirmed by the $\rm{T_{dust}}$ versus $\rm{M_{dust}}$ relation (Fig.~\ref{fig:19}), and show this relation for the ADF--S sample in comparison with other local samples of  LIRGs and ULIRGs   \citep[][as for these samples,  the dust mass {and the dust  temperature are}  available]{u12, PereiraSantaella15,clements10}. 
We found that more dusty galaxies have lower dust temperature. 
In other words, the dust temperature increases with decreasing dust mass. 
This relation holds most significantly for LIRGs.
For ULIRGs the $\rm{T_{dust}}$ versus $\rm{M_{dust}}$ relation is not that obvious, and both ADF--S and \cite{clements10} samples show no dependence on the dust mass. 
Comparing our data to  \cite{yan14}, \cite{podigachoski15}, and \cite{daCunha15} samples at redshift $>$1,  we find that this behavior is preserved for higher redshift. 
However,  the  $\rm{T_{dust}}$ versus $\rm{M_{dust}}$ relation for both local and $z>1$ samples  shows a more general relation: dust mass and dust temperature increasing with redshift, but we claim that this relation might be caused by a selection effect where for higher redshifts we can detect only IR bright galaxies.

\section{Fractional contribution of AGN mid-infrared emission and types of AGNs}
\label{sec:AGNs}

\begin{figure}[ht]
        \centering
                \begin{center}
        \includegraphics[width=0.5\textwidth, height=0.25\textheight, clip]{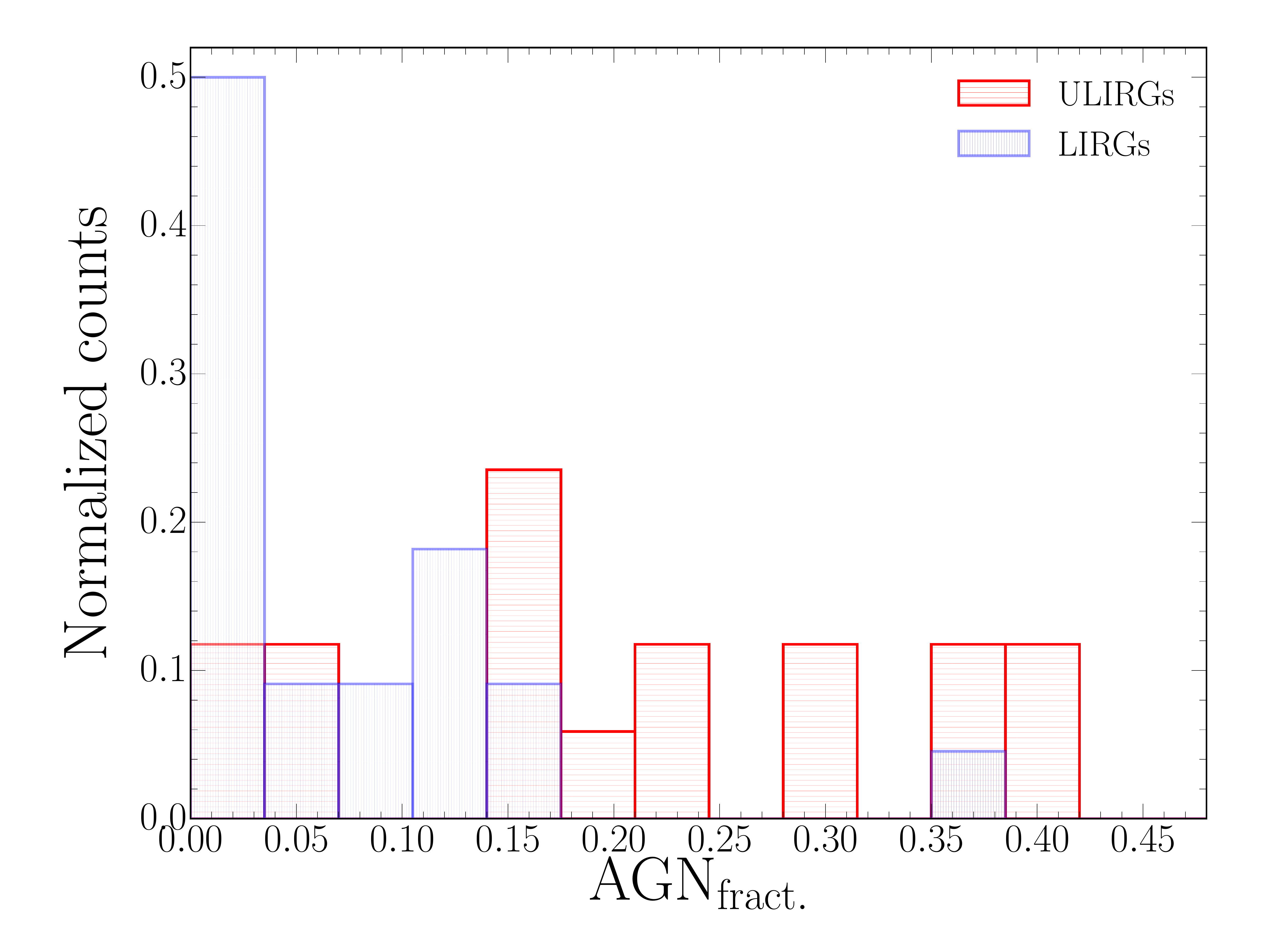}
                \end{center}    
                \caption{
                Fractional contribution of AGN to the MIR emission of  LIRGs and ULIRGs. 
                The horizontal striped red histogram represents ULIRGs while the vertical striped blue histogram corresponds to LIRGs.}
                \label{fig:11}
\end{figure}
\begin{figure}[ht]
        \begin{center}
        \includegraphics[width=0.5\textwidth, clip]{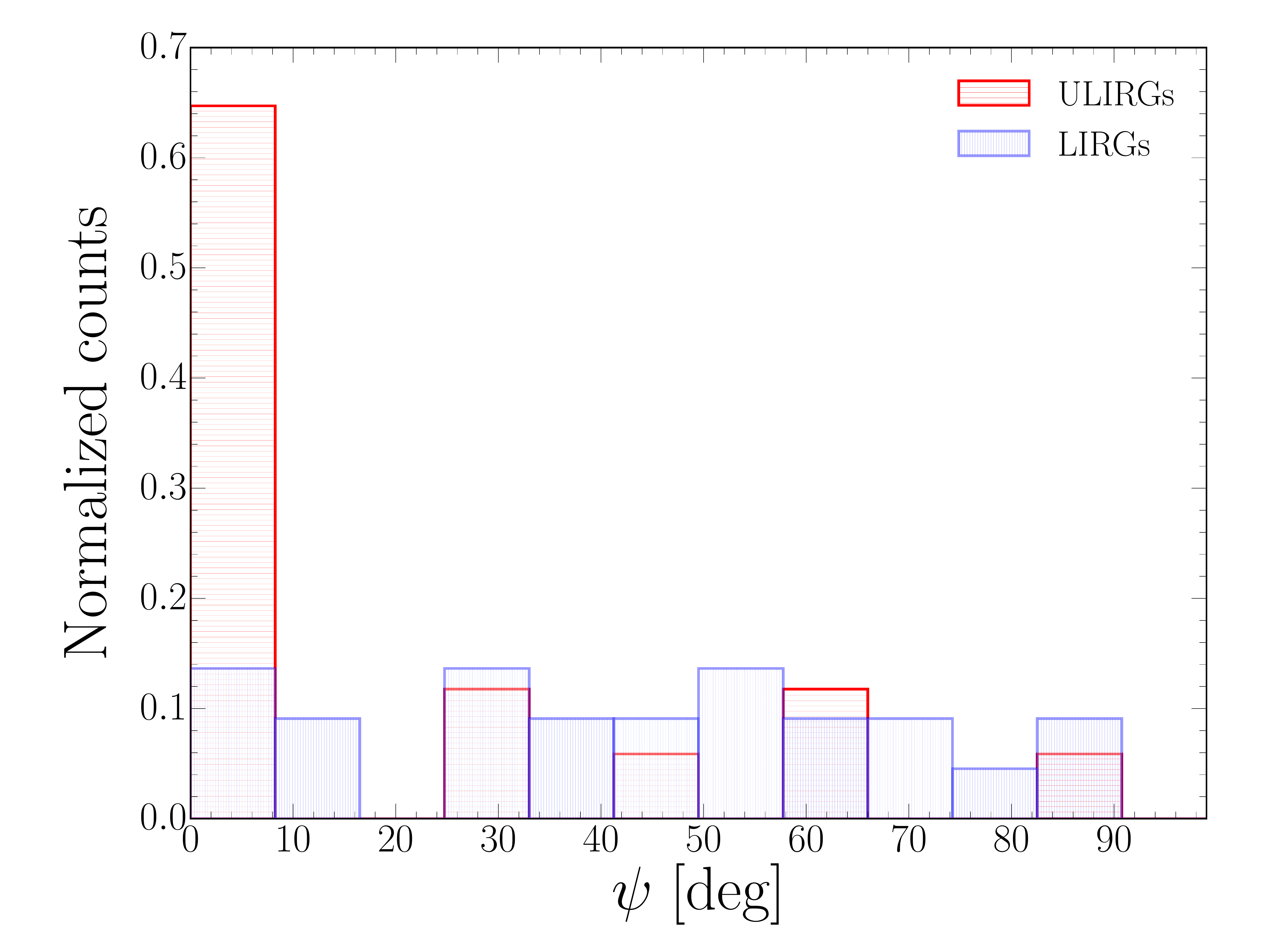}   
        \end{center}    
                \caption{
                Angle with respect to the line of sight ($\psi$) for the ADF--S  LIRGs and ULIRGs sample. 
                The striped red histogram represents ULIRGs while the vertical striped blue histogram represents LIRGs. }
                \label{fig:12}
\end{figure}

As the role of AGNs in  LIRGs and ULIRGs   is still not entirely clear, one of the main aims of  our analysis was to estimate the AGN contribution to the  LIRGs' and ULIRGs' infrared emission and which types of AGNs are related to  LIRGs' and ULIRGs'   activity. 
As we have shown, the ADF--S  LIRGs and ULIRGs   are comparable to  LIRGs and ULIRGs   selected by other authors in IR wavelengths, and our sample can be representative for the whole population. 

We have derived the fractional contribution of the AGN mid-infrared emission for \cite{fritz06} templates, which are built on  two components: (1) point-like isotropic emission of the central source, and (2) radiation from dust with a toroidal geometry in the vicinity of the central engine. 
The AGN emission is absorbed by the toroidal obscurer and re-emitted at 1--1000 $\rm{\mu}m$ wavelengths or scattered by the same obscurer. 
The \cite{fritz06} model describes different types of AGNs: Type 1, Type 2, and intermediate states.  
We perform SED fitting with a set of parameters of  \cite{fritz06} models as described by \cite{ciesla15}. 
We decided to use the same input list as \cite{ciesla15}, checking all possible biases, and the influence of input parameters on the final physical properties using the mock catalog.

Figure~\ref{fig:11} presents a fractional contribution of AGNs to  LIRGs and ULIRGs.  
We find that for 25  LIRGs and ULIRGs (more than 64\% of the sample) the AGN contribution is higher than 10\%.
In general, ULIRGs are  characterized by a higher fraction of AGNs than are LIRGs. 
Moreover, the $\rm{AGN_{frac.}}$ parameter increases with dust luminosity. 
{Our findings are consistent with results presented by \cite{Lin16} for LIRGs and ULIRGs selected at 70~$\mu$m.}
The majority of ADF--S ULIRGs show an AGN contribution to the MIR luminosity higher than 20\%. 
Based on \cite{ciesla15}, the fractional contribution of AGN emission  to the $\rm{L_{dust}}$  constrained by CIGALE is almost always overestimated  for a low fraction of Type~1 AGNs ($<$10\%), and underestimated for $\rm{AGN_{frac.}}>20\%$ for Type~1 and intermediate types of AGNs. 
As most LIRGs and ULIRGs    are characterized by  $\rm{AGN_{frac.}}>$10\% we conclude that the real contribution might be underestimated, and thus the values we have obtained may be treated as lower limits. 

\begin{table}[h]
\renewcommand{\arraystretch}{1.2}
\caption{Mean, median, and maximal values of the AGN-fraction for the ADF--S LIRGs and ULIRGs. All values are given in percentages. }
\begin{center}
\begin{tabular}{c| c| c} \hline\hline
\label{fractAGN}
 &   ULIRGs & LIRGs \\ \hline 
$\rm{mean(AGN_{frac.})}$ & 19.12 $\pm$ 6.76& 12.55 $\pm$ 3.59\\
$\rm{med.(AGN_{frac.})}$ & 19 $\pm$ 8 & 13 $\pm$ 3 \\
 $\rm{max.(AGN_{frac.})}$ & 41 $\pm$ 3 & 35 $\pm$ 4\\
\end{tabular}
\label{tab:3}
\end{center}
\end{table}

Previous studies of local  LIRGs and ULIRGs   also found a similar contribution of AGNs to ULIRGs \citep[$\sim$20 \%;][]{giovannoli11, fiorenza14, PereiraSantaella11,Petric11,Farrah07}.
The mean and median values of the AGN fraction for  LIRGs and ULIRGs   in our sample are presented in Table~\ref{tab:3}. 
The median fraction of AGNs in ULIRGs is 19\%  in accordance with previous studies. 
The AGN contribution to dust luminosity for LIRGs is slightly lower ($\sim$13\%). 
The mean and median values of $\rm{AGN_{frac.}}$ for  LIRGs and ULIRGs   are very similar, and their  uncertainties  are not high.
  
The maximum value of an AGN fraction among ULIRGs (41\%) was obtained for  galaxy  APMUKS(BJ) B043219.97-543942.2 ($\rm{ID_{ADFS}=59}$), with photometric redshift  equal to 0.50.
Among the LIRGs, we found a maximal AGN contribution (35 $\pm$ 4\%) for the Syfert galaxy RBS 0567 ($\rm{ID_{ADFS}=610}$) with a known spectroscopic redshift equal to 0.24.

 \begin{figure}[ht]
        \begin{center}
                \includegraphics[width=0.5\textwidth, height=0.3\textheight, clip]{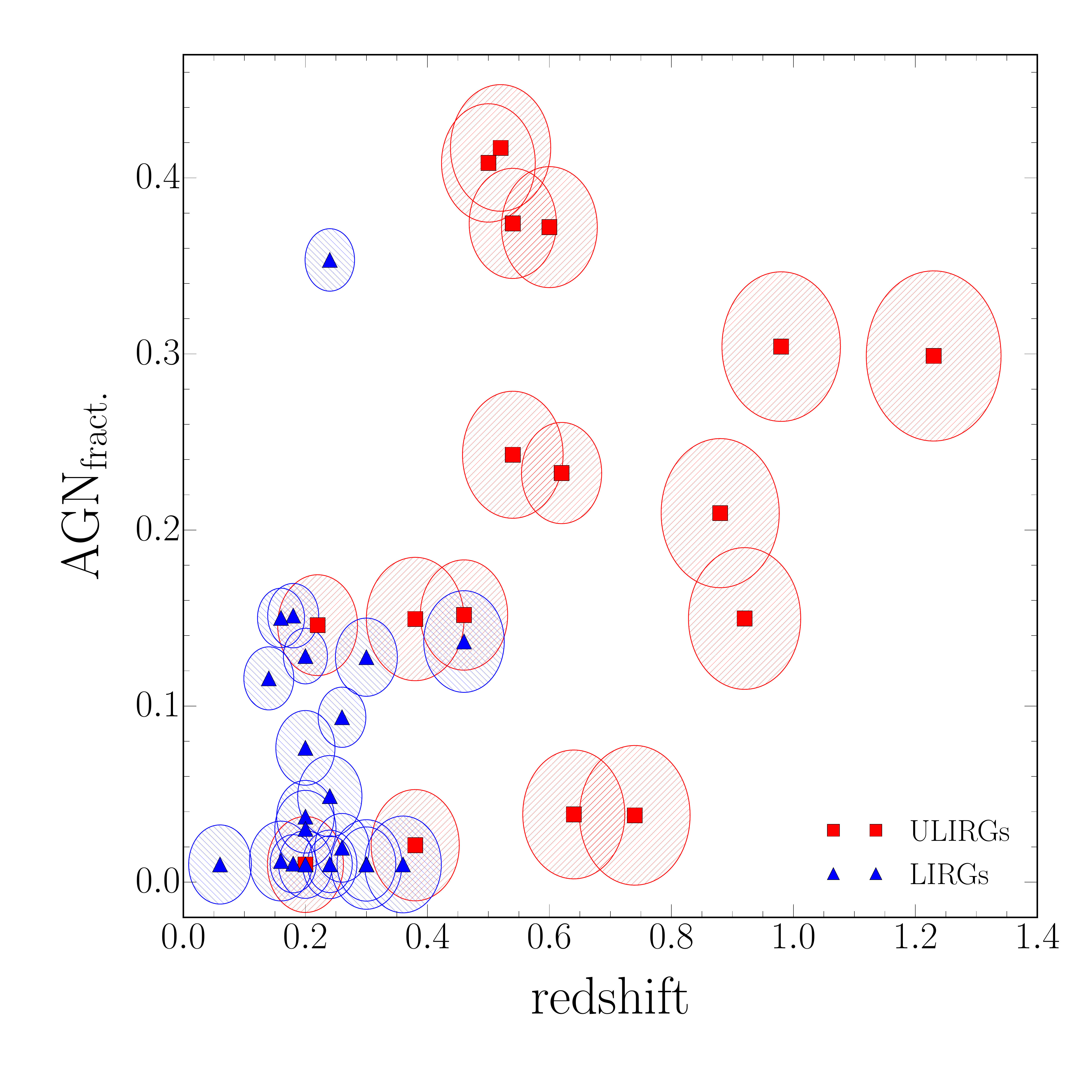}
        \end{center}
        \caption{
                Relationship between redshift and fractional contribution of AGN mid-infrared emission for ULIRGs and LIRGs from the ADF--S sample. 
                Filled red squares represent ULIRGs, while filled blue triangles represent LIRGs. The radii of circles correspond to the log(SFR) values for each galaxy (we used this as a scaling factor), that is, the larger the circle, the higher the measured SFR of the galaxy. }
        \label{fig:14}
 \end{figure}
 
 \begin{figure}[ht]
        \begin{center}
        \includegraphics[width=0.5\textwidth, height=0.3\textheight, clip]{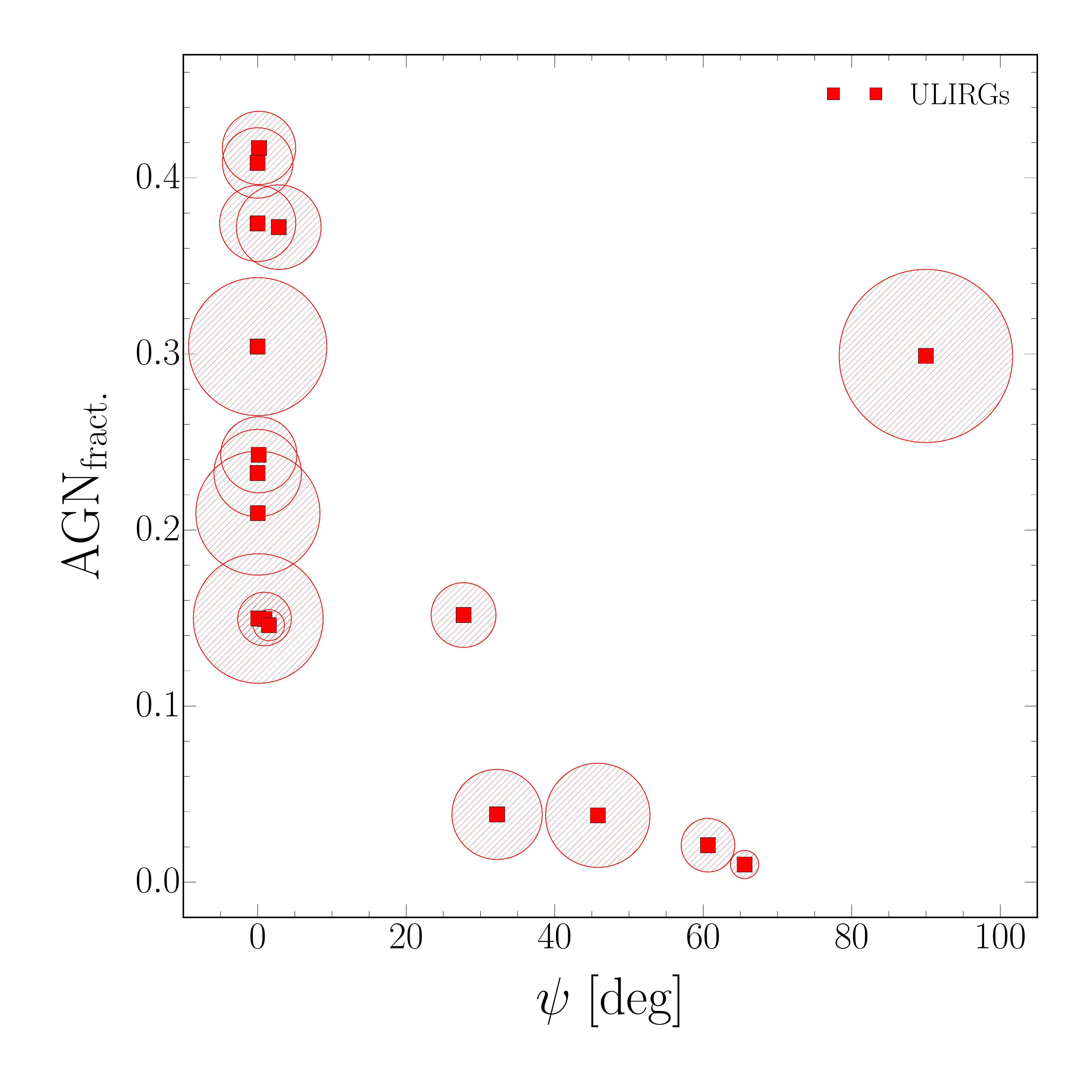}    
        \end{center}    
                \caption{{Relationship between the torus angle with respect to the  line of sight ($\psi$) and the fractional contribution of AGN mid-infrared emission for ULIRGs from the  ADF--S sample. 
                Radii of circles correspond to the redshift for each galaxy (used as a scaling factor; larger circles correspond to higher redshifts). 
                One outlier, the galaxy with torus angle equal to 89${^{o}}$}  , is the most distant galaxy in our sample with $\rm{z_{spec}=1.23}$ (a detailed discussion of this specific quasar can be found in Sec.~\ref{sec:Reliability_check}).}
                \label{fig:aa}
 \end{figure}
 
Based on results obtained from the AGN component, we conclude that we can distinguish different types of AGN components for LIRGs and ULIRGs. 
Fig.~\ref{fig:12} presents the distribution of $\psi$  for ADF--S  LIRGs and ULIRGs. 
It shows that LIRGs consist of both Type~1 and Type~2 AGNs, as well as intermediate types, whereas for ULIRGs, a majority of fractional contribution of AGN emission originates from Type~2 AGNs.  
\cite{Veilleux95} and \cite{Kim98b} showed that the percentage of type~1~AGNs increases with increasing infrared luminosity, and they found no contribution to the MIR luminosity from the Type~1 AGNs up to $\rm{log(L_{dust}/L_{\odot})<12.3}$.   
Their results are not in contradiction with our findings: Type~1~AGNs have, in general, a small influence on the MIR luminosity for the ADF-S LIRG sample (mean fractional contribution of type~1~AGN MIR emission for ADF--S LIRG sample is equal to 9~$\pm$~3\%). 
 
Our result shows that 70\% of ADF--S ULIRGs contain Type~2 AGNs, and for them,  the mean  fractional contribution of AGN to the mid-infrared emission is equal to 26$\pm$5\% and the obtained value is only slightly lower than the AGN contribution for $\sim$z<0.3 ULIRGs observed by Spitzer Infrared Spectrograph, and analyzed by \cite{Veilleux09} for which the total AGN contribution equals 35--40\%.  
This suggests that Type~1 AGNs make only a small contribution to the mid-infrared emission for galaxies with $\rm{L_{dust}>10^{12}\mbox{ }[L_{\odot}]}$, and the high luminosity of the dust part of the spectra is related to their star-forming activity. 
{Fig.~\ref{fig:aa} shows the relationship between the torus angle with respect to the line of sight and the fractional contribution of AGNs to the mid-infrared emission of ULIRGs. 
The general trend suggests that ULIRGs containing Type~2 AGNs are characterized by a higher AGN contribution. 
The one outstanding object with high $\rm{AGN_{frac}}$  and $\psi\sim$90 is the HLIRG and a known quasar with $\rm{log(L_{dust})=13.12\pm0.29\mbox{ }[L_{\odot}]}$. } 
 
Our measurements indicate that Type~2 and intermediate types of AGNs occur more often  in the ADF–S ULIRGs than Type~1 AGNs.
It has already been shown by other authors \citep[e.g.,][]{Lonsdale2006book} that warm ULIRGs tend to have optical spectral features indicating the presence of Type~1 AGNs while cool ULIRGs have characteristics of starburst or Type~2 AGNs. 
Cool MIR colors (the galaxy displaying cool MIR color has a ratio between fluxes measured at 25 and 60~$\mu$m lower than 0.2; as we did not use IRAS bands for our analysis, we have measured the ratio between the modeled fluxes) and positions of dust peaks of the spectra located at wavelengths longer than 90~$\mu$m (Fig.~\ref{fig:z}) indicate that almost all the ADF-S ULIRGs are cool, and associating them with Type~2 AGNs is indeed consistent with previous observations.
  
The question emerges as to why it might be that cool ULIRGs have a higher chance of containing an edge--on AGN than a face--on type AGN. 
If we make an assumption that ULIRGs' IR emission is isotropic; such a finding could undermine the unification theory of AGNs according to which   Seyfert~1 and~2 galaxies are the same kind of objects, surrounded by an optically thick dust torus but viewed from different angles  \citep{Antonucci1993}.
Indeed, such an explanation of the observed preference of different kinds of ULIRGs to host different types of AGNs was suggested by   \citealp{veilleux97} and \citealp{Lee11}, for example.
  
A much more likely explanation may be the geometrical contribution from the host galaxy, as the host galaxy orientation may add some bias.
In particular, a non-isotropic dust distribution resulting in the (cold) dust column density along the line of sight being higher in the case of the egde--on objects might increase their infrared luminosity resulting in the detection of an object as an ULIRG. 
From a broader perspective, it might lead to a {unification scenario} for LIRGs and ULIRGs, in which seeing a galaxy as a cold/warm ULIRG or LIRG could be related to the geometry of the object rather than, or together with, its physical properties. However, confirming or rejecting such a hypothesis requires further observations of both ULIRGs' host galaxies and their central AGNs.
 
\section{Conclusions}
\label{sec:Results} 

We analyzed the physical properties of a sample of 178  infrared-selected galaxies detected at 90 $\mu$m. 
For our analysis we used one of the deep AKARI surveys, ADF--S, and  the catalogs published by \cite{malek10} and \cite{malek13},  completed by  spectroscopic redshifts and \textit{Herschel/SPIRE} data. 

The CIGALE v.0.5  \citep[][Burgarella et al. in preparation]{noll09}   program for fitting spectral energy distribution was used  to estimate basic physical properties of the ADF-S sources: stellar mass, dust luminosity, star-formation rate, dust heating intensity, AGN fraction of mid-infrared emission, and the AGN's viewing angle with respect to  the line of sight. 
SEDs were fitted for all {78 }galaxies, and for {69} sources (20  with spectroscopic and 49 with photometric redshifts) the quality of the fits was satisfactory and the obtained physical parameters were deemed reliable. 

Additionally, we used a graybody + power law \texttt{CMCIRSED} SED modeling code for the final sample of {69} ADF--S galaxies, to obtain independent measurements of dust luminosity, and to calculate dust mass and dust temperature for  LIRGs and ULIRGs. 
With the use of  \texttt{CMCIRSED,}  we find 13 LIRGs and 12  ULIRGs (this discrepancy between the results given by both codes might be expounded as the lack of \textit{Herschel/SPIRE} data for some galaxies, while very far infrared data  are  crucial for a code based on the FIR data only). 
Thus, galaxies with $\rm{log(L_{dust})\geq11\mbox{ }[L_{\odot}]}$ compose  25\% of our sample.

Based on the UV-FIR and FIR only SED fitting, we find that:

\begin{enumerate}
        
\item The ADF--S sample mainly consists of galaxies active in star formation.

\item Using CIGALE, among {69} ADF--S galaxies,  we find  17 ULIRGs and {22} LIRGs, spanning over the redshift range 0.06$<$z$<$1.2. 
The majority of them have the IR peak placed at wavelengths longer than 90~$\mu$m (so called cold  LIRGs and ULIRGs, \cite{symeonidis11}),

\item  Galaxies bright in the FIR are relatively massive (mean $\rm{log(M_{stars})}$ computed for ULIRGs and LIRGs, and galaxies with $\rm{log(L_{dust})<11\mbox{ }[L_{\odot}]}$ is equal to 11.51 $\pm$ 0.37, 10.90 $\pm$ 0.56, and {10.03 $\pm$ 0.50} $\rm{[M_{\odot}]}$, respectively). 
Our results are consistent with stellar masses calculated for LIRGs and ULIRGs located at similar redshifts, selected from IR surveys \citep{u12,PereiraSantaella11,giovannoli11,melbourne08,santos15,daCunha15,Rothberg2013}.

\item The  mean  log(SFR)  for ULIRGs, LIRGs, and the rest of ADF--S galaxies is equal to 2.53 $\pm$ 0.39, 1.48 $\pm$ 0.24, and {0.47 $\pm$ 0.21} $\rm{[M_{\odot}yr^{-1}]}$, respectively, and these results are consistent with the \cite{Kennicutt1998} relation. 
Our results are consistent with those presented in the literature, that is, \cite{santos15}, \cite{podigachoski15}, \cite{daCunha15}, and \cite{howell10}. 
The SFR increases with $\rm{L_{dust}}$ in the same way as  normal galaxies, LIRGs, and ULIRGs. 

\item ADF--S "normal" star-forming galaxies follow the  general SFR--$\rm{M_{star}}$ correlation as defined by \cite{Noeske2007}.
However,  the majority of ADF--S  LIRGs and ULIRGs   are located above the correlation line and they are shifted toward higher SFR. 

\item We calculated the specific SFR and found that SSFR  decreases with increasing stellar mass, with a similar slope  as the SSFR--$\rm{M_{star}}$ relation for the whole ADF--S sample  (-0.72, -0.92, and {-0.75}, for ULIRGs, LIRGs, and normal galaxies, respectively). 

\item We calculated the artificial equivalent width of the dust component, defined as $\rm{EW_{dust}=2\times(\lambda_{max}-\lambda_{24\mu m})}$. 
We find that the stellar mass, the dust luminosity, and the $\alpha$ parameter from \cite{dale14} for  LIRGs and ULIRGs   increase with increasing width of dust component, and find a steep correlation between $\alpha$ and $\rm{EW_{dust}}$ with a slope  equal to 0.6$\pm$0.3 + (0.0042$\times \rm{EW_{dust}}$) .
\end{enumerate}

A dust temperature/mass analysis was performed for  LIRGs and ULIRGs   selected by both fitting methods (CIGALE and \texttt{CMCIRSED}). 
As the visual inspection of \texttt{CMCIRSED} SEDs shows a  difference in quality of the fitted spectra between galaxies with and without \textit{Herschel/SPIRE} data (lack of measurements after the peak of the dust emission), we based this part of the analysis  on the 12  LIRGs and ULIRGs   with AKARI--\textit{Herschel} data and found that  the $\rm{T_{dust}/L_{dust}}$ relationship for ADF--S  LIRGs and ULIRGs  agrees with the literature. 
The relation is rather flat, and flatter than the same relation estimated by \cite{symeonidis13}. 
As was suggested by \cite{symeonidis13}, this relation is shaped by the dust mass, and not the increase in dust heating.

Based on the UV--FIR SED fitting method, we conclude that the median fractional contribution of AGN mid-infrared emission for ADF--S ULIRGs  is 19\%. 
The AGN contribution in the dust luminosity of LIRGs is slightly lower, and equal to 13\%. 
We find that the ADF--S  LIRG sample consists both of Type~1, and Type~2 AGNs, as well as intermediate types. 
At the same time, for ULIRGs selected at 90~$\mu$m AKARI band,  the majority of fractional contribution of AGN emission originates from Type~2 AGNs. 


\begin{acknowledgements}
We thank the anonymous referee for useful comments and suggestions.
This work is based on observations with AKARI, a JAXA project with the participation of ESA. 
Research was conducted in the scope of the HECOLS International Associated Laboratory.
This research has made use of the SIMBAD database, operated at CDS, Strasbourg, France, the NASA/IPAC Extragalactic Database (NED), which is operated by the Jet Propulsion Laboratory, and the California Institute of Technology, under contract with the National Aeronautics and Space Administration.
This research has made use of the NASA/ IPAC Infrared Science Archive, which is operated by the Jet Propulsion Laboratory, California Institute of Technology, 
under contract with the National Aeronautics and Space Administration.
This research has made use of data from HerMES project (http://hermes.sussex.ac.uk/). 
HerMES is a Herschel Key Programme utilising Guaranteed Time from the SPIRE instrument team, ESAC scientists and a mission scientist.
The HerMES data was accessed through the Herschel Database in Marseille (HeDaM - http://hedam.lam.fr) operated by CeSAM and hosted by the Laboratoire d'Astrophysique de Marseille.
This work have been supported by the Polish National Science Centre (grants UMO-2012/07/B/ST9/04425 and UMO-2013/09/D/ST9/04030), and the European Associated Laboratory Astrophysics Poland-France HECOLS. 
MM acknowledges support from NASA grants NNX08AU59G and NNX09AM45G for analysis of GALEX data in the AKARI Deep Fields. 
TTT  has been supported by the Grant-in- Aid for the Scientific Research Fund (24111707), and by the Strategic Young Researches Overseas Visits Program for Accelerating Brain Circulation  commissioned by the Ministry of Education, Culture, Sports, Science and Technology (MEXT) of Japan.
\end{acknowledgements}

\twocolumn
\bibliographystyle{aa}
\bibliography{KMbib}

\appendix
\section{The influence of the use of photometric redshifts on the final results}
\label{app:A}
To investigate how the photometric redshift uncertainty can affect our results, we performed an additional test: for the final sample of our galaxies we ran the same analysis using redshift values as a redshift $\pm$ redshift accuracy, which equals 0.05 (hereafter $\rm{sample_{z-0.05}}$ and $\rm{sample_{z+0.05}}$). 
We calculated the values of dust luminosities,  star-formation rates, stellar masses, specific star-formation rates, AGN fractions, torus angles with respect to the lines of sight ($\psi$), and the slopes of the infrared parts of the spectra ($\alpha$). 

\begin{figure}[h]
        \begin{center}
        \includegraphics[width=0.5\textwidth, clip]{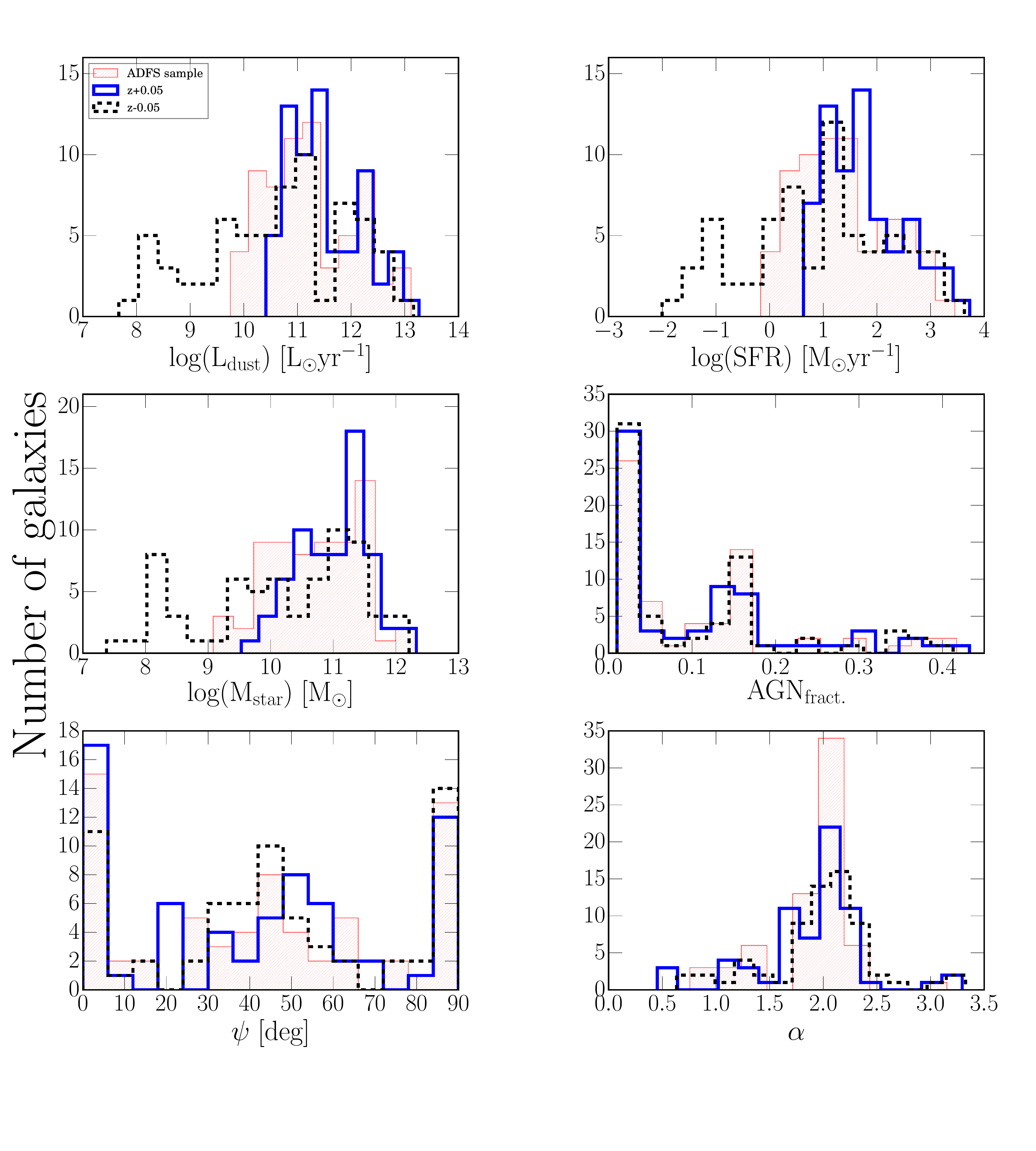}
        \end{center}    
                \caption{Distribution of dust luminosity, star-formation rate, stellar mass, fractional contribution of AGN to the MIR emission, AGN's torus angle with respect to the line of sight ($\psi$), and IR spectral power-law slope $\alpha$ for an original ADF--S sample (sloping red histogram ), and $\rm{sample_{z+0.05}}$ and $\rm{sample_{z-0.05}}$ samples (open blue histogram, and dashed black histogram, respectively).}
                \label{fig:A1}
 \end{figure}

From these two new samples we obtained the same number of HLIRGs (one from $\rm{sample_{z-0.05}}$ and $\rm{sample_{z+0.05}}$) as for the original ADF-S sample. 
$\rm{Sample_{z-0.05}}$ gives us 12 ULIRGs, 15 LIRGs, and 38 normal galaxies, while $\rm{sample_{z-0.05}}$ gives 18 ULIRGs, 29 LIRGs, and 18 normal galaxies. 
It is not surprising that with lower redshifts the CIGALE fitter found less ULIRGs and LIRGs, while for higher redshifts we obtained one more ULIRG and seven LIRGs. 
In general, we observe the trend that bright infrared galaxies at higher redshifts are more often LIRGs than for the lower redshifts, but most of them ($\sim$70\%) remain in the same luminosity range as that of the original sample. 
The average physical properties obtained from $\rm{sample_{z-0.05}}$ and $\rm{sample_{z+0.05}}$ are consistent with those obtained from the original ADF--S sample and all trends we found remain unchanged. 
Fig.~\ref{fig:A1} shows the distribution of physical properties obtained by CIGALE for the ADF--S original sample as well as for  $\rm{sample_{z-0.05}}$ and $\rm{sample_{z+0.05}}$. 
For galaxies with redshifts calculated as ADF--S redshifts minus redshift accuracy, the general tendency to lower stellar masses and fainter dust luminosity is visible. 
All galaxies with $\rm{log(L_{dust})<9.5\mbox{ }[L_{\odot}]}$,  $\rm{log(M_{star})<8.8\mbox{ }[M_{\odot}]}$, and $\rm{log(SFR)<-0.1}$ have new redshifts$\rm{_{z-0.05}}$ $\sim$ 0.

\begin{figure}[h]
        \begin{center}
        \includegraphics[width=0.25\textwidth, clip]{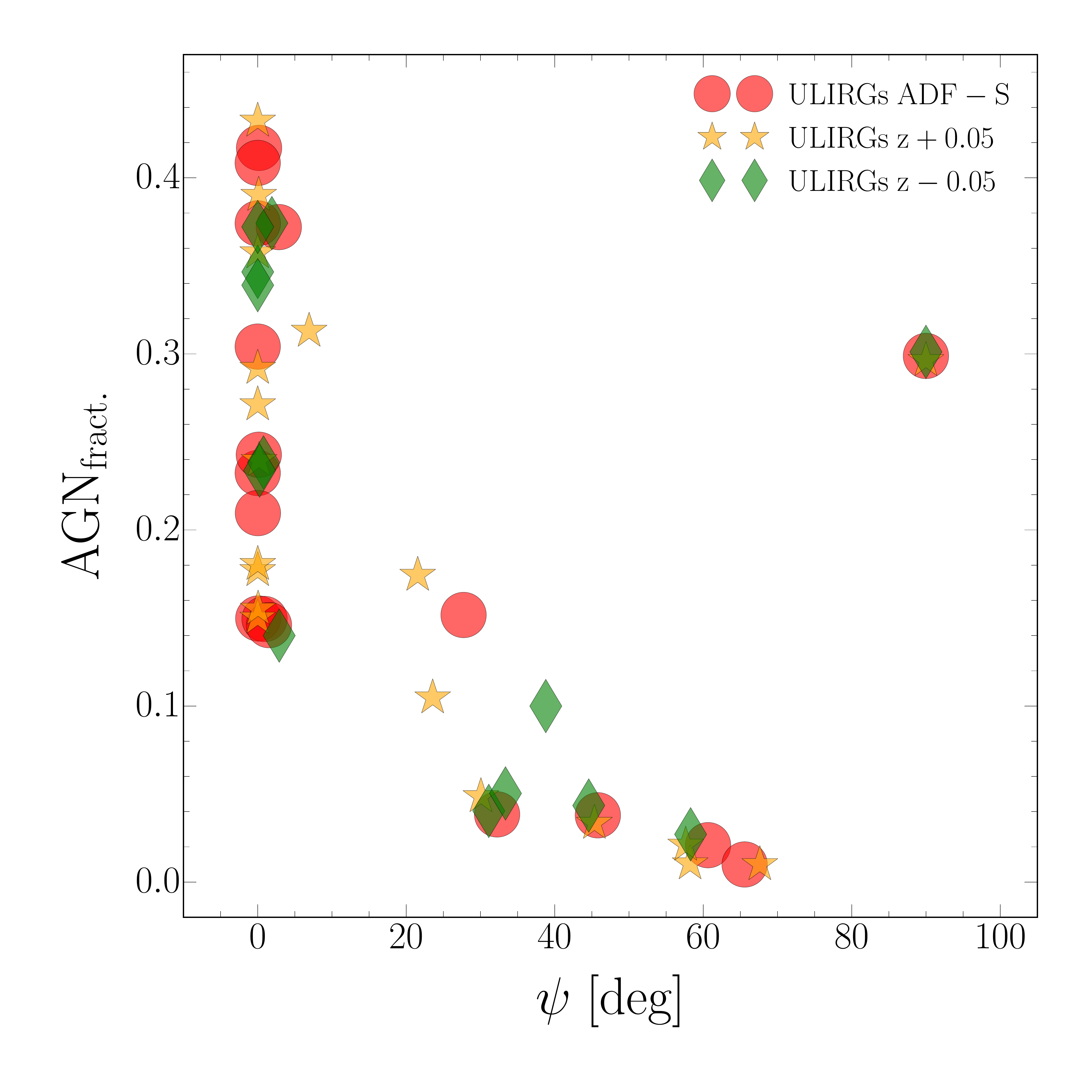}\includegraphics[width=0.25\textwidth, clip]{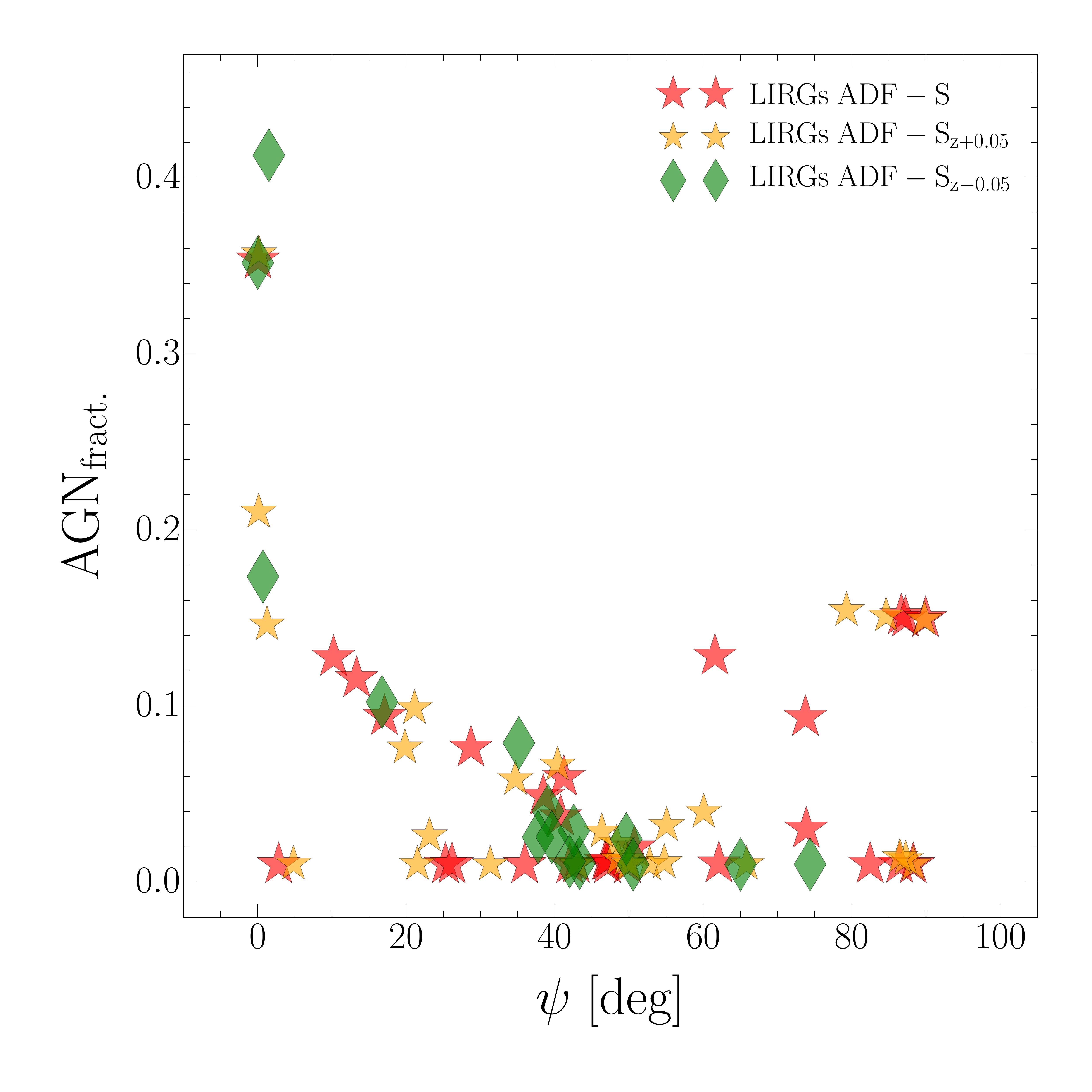}
        \end{center}    
                \caption{Relationship between the torus angle with respect to the line of sight ($\psi$) and the fractional contribution of AGN mid-infrared emission for ULIRGs (\textit{left panel}) and LIRGs (\textit{right panel}) for the sample of original ADF-S galaxies (red circles), the ADF-S sample with redshift increased by 0.05 (orange stars) and the ADF-S sample with redshift  reduced  by value 0.05.}
                \label{fig:A23}
 \end{figure}

The physical properties obtained for both samples are consistent with those presented in the paper (except for galaxies with new redshifts at approximately 0). 
Most importantly, a general trend between the fractional contribution of AGN mid-infrared emission and the types of AGNs, which was found for ULIRGs and LIRGs from the original ADF-S sample, is preserved, which is shown in Fig.~\ref{fig:A23}. 
We conclude that the photometric redshift inaccuracy has a marginal influence on the general trends found in our analysis. 
 
\begin{table*}
        \begin{center}          
                \caption{An example of two rows from the final catalog of 39 LIRGs and ULIRGs selected from the ADF--S sample$^a$. The upper limits are presented as errors without given values for a flux. The full catalog is available from the CDS server. }\label{ULIRGS_tabela}
                \centering
                \begin{tabular}{r|r|r|l}
                        \hline \hline
                        \multicolumn{1}{c|}{Column} & \multicolumn{1}{|c|}{Example 1} & \multicolumn{1}{|c|}{Example 2} &  \multicolumn{1}{|c}{Content}\\ \hline
                ADFS   & 2 &  13 &  ADF-S identification number \\
                RAdeg  & 67.90437  &    68.59812 &   ADF-S source right ascension (J2000.0) \\
                DEdeg  & -53.28733 & -54.69248  &  ADF-S source declination (J2000.0) \\
                z      & 0.058811 &  0.2 &       Redshift of the counterpart \\
                r\_z  & a &  b&       References for redshifts See Note (1) \\
                FUV    & -10000.0  & -10000.0 & Flux density at GALEX FUV  (0.15 $\mu$m)  band\\
                e\_{FUV} & -10000.0 & -10000.0 & rms uncertainty on GALEX FUV band \\
                NUV       & -10000.0 & -10000.0 &  Flux density at GALEX NUV   (0.23 $\mu$m)  band \\
                e\_{NUV} & -10000.0 & -10000.0 &  rms uncertainty on GALEX NUV band \\
                D\_I     & 1.19     & -10000.0 & Flux density at DENIS I  (0.79 $\mu$m)  band \\
                e\_{D\_I}&  0.0656  & -10000.0 &  rms uncertainty on DENIS I band \\
                2\_J      & 2.11     & 1.91  & Flux density at 2MASS J (1.25 $\mu$m)  band \\
                e\_{2\_J}& 0.191    &  0.137 & rms uncertainty on 2MASS J band \\
                D\_J      &   2.81   & -10000.0 &  Flux density at DENIS J   (1.24 $\mu$m)  band \\
                e\_{D\_J}& 0.287    & -10000.0 & rms uncertainty on DENIS J band\\
                2\_H      & 2.9     & 2.84     & Flux density at 2MASS H  (1.65 $\mu$m)  band \\
                e\_{2\_H}& 0.286    &   0.251  & rms uncertainty on 2MASS H band\\
                2\_Ks     &  3.46 & 3.17 &  Flux density at 2MASS Ks   (2.17 $\mu$m)  band\\
                e\_{2\_Ks}& 0.236 & 0.213 & rms uncertainty on 2MASS Ks band\\
                D\_Ks      & 4.1 &  -10000.0 & Flux density at DENIS Ks  (2.16 $\mu$m)  band \\
                e\_{D\_I} & 0.644 & -10000.0 & rms uncertainty on DENIS Ks band\\
                W1         &   4.43 &  3.14 & Flux density at WISE W1  (3.38 $\mu$m)  band\\
                e\_{W1}   & 0.0939 & 0.0666 &  rms uncertainty on WISE W1 band\\
                W2         & 3.4  &  2.16 & Flux density at WISE W2  (4.62 $\mu$m)  band \\
                e\_{W2}   & 0.0658 & 0.0418 &  rms uncertainty on WISE W2 band \\
                W3         & 25.1 & 12.0 &  Flux density at WISE W3   (12.81 $\mu$m)  band \\
                e\_{W3}   & 0.346 & 0.166  & rms uncertainty on WISE W3 band \\
                MIPS1      &  -10000.0 & -10000.0 & Flux density at Spitzer MIPS1   (23.84 $\mu$m)  band\\
                e\_{MIPS1} & -10000.0 & -10000.0 &  rms uncertainty on Spitzer MIPS1 band\\
                N60         &  1150.0  & 397.0 &  Flux density at AKARI N60  (66.69 $\mu$m) band\\
                e\_{N60}   &   41.8 & 20.4 & rms uncertainty on AKARI N60 band \\
                MIPS2       & -1000.0 & 420.0 & Flux density at Spitzer MIPS2  (72.55 $\mu$m)  band \\
                e\_{MIPS2} & -1000.0& 3.57 &  rms uncertainty on Spitzer MIPS2 band\\
                WIDEs       &  1450.0 & 479.0  & Flux density at AKARI WIDE-S (89.14 $\mu$m)  band \\
                e\_{WIDEs} & 47.9 & 16.6 &  rms uncertainty on AKARI WIDE-S band \\
                WIDEl       & -10000.0 & 917.0&  Flux density at AKARI WIDE-L (150 $\mu$m) band \\
                e\_{WIDEl} & -10000.0 & 65.9 &  rms uncertainty on AKARI WIDE-L band \\
                N160        & 1260.0&  -10000.0 &  Flux density at AKARI N160 (163.10 $\mu$m) band\\
                e\_{N160}   &   212.0 & -10000.0  &  rms uncertainty on AKARI N160 band\\
                S250        & -10000.0& 493.0 &  Flux density at Herschel SPIRE 250  (251.50 $\mu$m)  band\\
                e\_{S250} & -10000.0&  2.4 & rms uncertainty on  Herschel SPIRE 250 band\\
                S350 & -10000.0& -10000.0 & Flux density at Herschel SPIRE 350   (352.80 $\mu$m)  band\\
                e\_{S350} &-10000.0 & 36.1  &  rms uncertainty on  Herschel SPIRE 350 band\\
                S500       & -10000.0 & 70.0 & Flux density at Herschel SPIRE 500   (510.60 $\mu$m)  band\\
                e\_{S500} & -10000.0& 2.8 &  rms uncertainty on  Herschel SPIRE 500 band \\
                Name       & ESO 157-IGA040 &  2MASX      J04342317-5441331 &      Name of the nearest optical counterpart\\
                \hline
                \end{tabular}
                \newline
                \begin{samepage}
                        \begin{center}
                         $^a$ All the fluxes are expressed in mJy and the coordinates in degrees. \\
                         {Note: 1}       a: \cite{Fisher95}, b: \cite{malek14}, c: \cite{jones04}, d: \cite{Thomas98},  e: \cite{Wisotzki00}.
                        \end{center}
                \end{samepage}
        \end{center}
\end{table*}

\begin{sidewaystable*}
        \begin{center}  
                \caption{Main physical parameters  of the {LIRG and ULIRG} sample derived based on the CIGALE SED-fitting code (columns 3--8) and CMCIRSED code (columns 9--12).}\label{ULIRGS_wyniki}
                \centering
                \begin{tabular}{r|r|r|r|r|r|r|r|r|r|r|r|r}
                        \hline \hline
                        &  & \multicolumn{7}{c|}{CIGALE} & \multicolumn{4}{|c}{CMCIRSED}\\ \cline{3-13}
                        \multicolumn{1}{c|}{$\rm{ID
                        }$} &
                        \multicolumn{1}{|c|}{$\rm{z}$} &
                        \multicolumn{1}{c|}{$\chi^2_{red}$} &
                        \multicolumn{1}{c|}{$\rm{log(L_{dust})}$ $\rm{[L_\odot]}$} &
                        \multicolumn{1}{c|}{SFR $\rm{[M_{\odot}yr^{-1}]}$} &
                        \multicolumn{1}{c|}{$\rm{log(M_{star})}$ $\rm{[M_{\odot}]}$} &
                        \multicolumn{1}{c|}{$\rm{AGN_{frac}}$} &
                        \multicolumn{1}{c|}{$\psi \mbox{ }\rm{[deg
                        ]}$} &
                        \multicolumn{1}{c|}{$\alpha$} &
                        \multicolumn{1}{c|}{$\chi^2$} &
                        \multicolumn{1}{c|}{$\rm{log(L_{dust})}$ $\rm{[L_\odot]}$} &
                        \multicolumn{1}{c|}{$\rm{T_{dust}}$ $\rm{[K]}$} &
                        \multicolumn{1}{c}{$\rm{log(M_{dust})}$ $\rm{[M_{\odot}]}$}\\

                         \cline{2-13
                         }
                        & \multicolumn{1}{c|}{(1)} & \multicolumn{1}{|c|}{(2)} &
                         \multicolumn{1}{|c|}{(3)} & \multicolumn{1}{|c|}{(4)} & \multicolumn{1}{|c|}{(5)} & \multicolumn{1}{|c|}{(6)} &
                         \multicolumn{1}{|c|}{(7)} & \multicolumn{1}{|c|}{(8)} & \multicolumn{1}{|c|}{(9)} &
                        \multicolumn{1}{|c|}{(10)} & \multicolumn{1}{|c|}{(11)} &
                        \multicolumn{1}{|c}{(12)} \\\hline

                        \hline
                        2 & 0.06$^a$ &  2.85 & $\rm{11.35\pm0.06}$ & $\rm{40.37\pm5.42}$ & $\rm{9.90\pm0.39}$ &    $\rm{0.05\pm0.00}$ &26.16 & $\rm{1.57\pm0.16}$ &  $\rm{}$ & $\rm{}$ & $\rm{}$& $\rm{}$
                        \\
                        13 & 0.20$^b$ & 4.67 & $\rm{12.02\pm0.05}$ & $\rm{88.47\pm27.79}$ & $\rm{11.22\pm0.28}$ & $\rm{0.05\pm0.00}$ &35.96  & $\rm{1.75\pm0.09}$ & 9.38 & $\rm{12.02\pm0.00}$ & $\rm{25.27\pm0.12}$ & $\rm{8.13\pm0.06}$\\
                        14 & 0.22$^b$ & 1.44 & $\rm{12.04\pm0.05}$ & $\rm{111.03\pm27.73}$ & $\rm{10.94\pm0.33}$ & $\rm{0.13\pm0.02}$ & 65.57 & $\rm{1.75\pm0.09}$ & 11.50 & $\rm{12.08\pm0.01}$ & $\rm{29.15\pm0.15}$ & $\rm{7.84\pm0.06}$\\
                        24 & 0.16$^b$ & 4.29 & $\rm{11.56\pm0.06}$ & $\rm{41.64\pm20.00}$ & $\rm{10.49\pm0.42}$ & $\rm{0.05\pm0.01}$ & 47.01 & $\rm{1.87\pm0.15}$ & 10.14 & $\rm{11.45\pm0.05}$ & $\rm{33.47\pm1.30}$ & $\rm{6.88\pm0.55}$\\
                        26 &  0.38$^b$ & 0.32 & $\rm{12.31\pm0.09}$ & $\rm{318.36\pm122.24}$ & $\rm{11.16\pm0.31}$ & $\rm{0.17\pm0.08}$ & 0.92 & $\rm{1.26\pm0.32}$ &  & $\rm{}$ & $\rm{}$ & $\rm{}$\\
                        46 & 0.20$^b$ & 2.29 & $\rm{11.66\pm0.05}$ & $\rm{36.86\pm8.09}$ & $\rm{10.80\pm0.23}$ & $\rm{0.05\pm0.00}$ & 73.87 & $\rm{2.06\pm0.10}$ & 5.03 & $\rm{12.09\pm0.43}$ & $\rm{10.08\pm1.79}$ & $\rm{10.42\pm0.09}$\\
                        47 & 0.54$^b$ & 0.63 & $\rm{12.40\pm0.13}$ & $\rm{377.66\pm148.54}$ & $\rm{11.18\pm0.31}$ & $\rm{0.25\pm0.08}$ & 0.15  & $\rm{0.96\pm0.49}$ & 5.07 & $\rm{12.25\pm0.07}$ & $\rm{41.74\pm2.28}$ & $\rm{7.38\pm0.08}$\\
                        59 & 0.50$^b$  & 1.54 & $\rm{12.16\pm0.05}$ & $\rm{252.03\pm56.06}$ & $\rm{12.26\pm0.13}$ & $\rm{0.41\pm0.03}$ & 0.01 & $\rm{1.76\pm0.09}$ &  & $\rm{}$ & $\rm{}$ & $\rm{}$\\
                        60 & 0.14$^b$  & 1.34 & $\rm{11.11\pm0.09}$ & $\rm{19.03\pm3.66}$ & $\rm{10.81\pm0.70}$ & $\rm{0.14\pm0.05}$ & 13.32  & $\rm{1.93\pm0.38}$ &  & $\rm{}$ & $\rm{}$ & $\rm{}$\\
                        61 & 0.98$^b$ & 0.81 & $\rm{12.90\pm0.13}$ & $\rm{1083.40\pm338.29}$ & $\rm{12.00\pm0.30}$ & $\rm{0.29\pm0.07}$ & 0.01 & $\rm{1.09\pm0.43}$ & 6.60 & $\rm{13.07\pm0.04}$ & $\rm{55.54\pm1.68}$ & $\rm{7.82\pm0.06}$\\
                        63 & 0.38$^b$ & 1.56 & $\rm{12.26\pm0.07}$ & $\rm{181.03\pm56.66}$ & $\rm{11.47\pm0.25}$ & $\rm{0.05\pm0.03}$ & 60.64 & $\rm{2.07\pm0.10}$ & 12.67 & $\rm{12.78\pm0.47}$ & $\rm{11.47\pm2.34}$ & $\rm{10.88\pm0.10}$\\
                        74 & 0.46$^b$  & 0.35 & $\rm{12.19\pm0.31}$ & $\rm{172.68\pm57.66}$ & $\rm{11.53\pm0.44}$ & $\rm{0.19\pm0.17}$ & 27.72  & $\rm{1.68\pm0.28}$ &  & $\rm{}$ & $\rm{}$ & $\rm{}$\\
                        93 & 0.02$^b$ & 2.12 & $\rm{11.41\pm0.05}$ & $\rm{32.78\pm3.66}$ & $\rm{10.54\pm0.20}$ & $\rm{0.13\pm0.01}$ & 28.71 & $\rm{1.85\pm0.14}$ &  & $\rm{}$ & $\rm{}$ & $\rm{}$\\
                        96 & 0.88$^b$ & 1.63 & $\rm{12.80\pm0.12}$ & $\rm{1060.77\pm360.17}$ & $\rm{11.52\pm0.33}$ & $\rm{0.23\pm0.08}$ & 0.03 & $\rm{0.66\pm0.40}$ & 6.09 & $\rm{13.33\pm0.14}$ & $\rm{14.88\pm1.16}$ & $\rm{11.13\pm0.06}$\\
                        102 & 0.52$^b$ & 1.54 & $\rm{12.30\pm0.22}$ & $\rm{370.22\pm194.25}$ & $\rm{10.97\pm0.22}$ & $\rm{0.20\pm0.12}$ & 0.18 & $\rm{0.68\pm0.42}$ & 5.38 & $\rm{12.71\pm0.16}$ & $\rm{17.77\pm1.70}$ & $\rm{10.07\pm0.07}$\\
                        146 & 0.10$^c$ & 1.41 & $\rm{11.18\pm0.05}$ & $\rm{23.69\pm4.82}$ & $\rm{11.28\pm0.18}$ & $\rm{0.05\pm0.00}$ & 54.05 & $\rm{2.06\pm0.10}$ & 7.48 & $\rm{11.24\pm0.09}$ & $\rm{34.72\pm2.36}$ & $\rm{6.64\pm0.11}$\\
                        151 & 0.54$^b$ &  1.38 & $\rm{12.17\pm0.14}$ & $\rm{171.70\pm37.43}$ & $\rm{11.53\pm0.24}$ & $\rm{0.33\pm0.08}$ & 0.01 & $\rm{2.06\pm0.10}$ & 11.97 & $\rm{12.35\pm0.02}$ & $\rm{38.78\pm0.52}$ & $\rm{7.60\pm0.06}$\\
                        154 & 0.24$^b$ & 1.51 & $\rm{11.40\pm0.07}$ & $\rm{44.33\pm7.73}$ & $\rm{11.47\pm0.24}$ & $\rm{0.16\pm0.07}$ & 38.45 & $\rm{2.06\pm0.10}$ & 10.74 & $\rm{11.42\pm0.02}$ & $\rm{34.48\pm0.54}$ & $\rm{7.29\pm0.20}$\\
                        165 & 0.36$^b$ & 3.73 & $\rm{11.85\pm0.11}$ & $\rm{92.13\pm20.98}$ & $\rm{11.35\pm0.55}$ & $\rm{0.06\pm0.04}$ & 51.86  & $\rm{2.05\pm0.10}$ &  & $\rm{}$ & $\rm{}$ & $\rm{}$\\
                        175 & 0.20$^b$ & 1.47 & $\rm{11.18\pm0.10}$ & $\rm{29.76\pm7.15}$ & $\rm{11.35\pm0.30}$ & $\rm{0.22\pm0.10}$ & 40.78 & $\rm{0.82\pm0.50}$ &  & $\rm{}$ & $\rm{}$ & $\rm{}$\\
                        198 & 0.60$^b$ &  0.77 & $\rm{12.36\pm0.29}$ & $\rm{284.73\pm137.47}$ & $\rm{11.60\pm0.50}$ & $\rm{0.14\pm0.16}$ & 2.84 & $\rm{1.74\pm0.12}$ & 5.02 & $\rm{12.29\pm0.10}$ & $\rm{44.52\pm3.78}$ & $\rm{7.31\pm0.09}$\\
                        203 & 0.13$^c$ & 1.96 & $\rm{11.64\pm0.07}$ & $\rm{65.58\pm17.02}$ & $\rm{10.62\pm0.22}$ & $\rm{0.06\pm0.03}$ & 2.82 & $\rm{1.45\pm0.07}$ & 10.73 & $\rm{11.62\pm0.04}$ & $\rm{34.64\pm1.38}$ & $\rm{7.03\pm0.07}$\\
                        206 & 0.92$^b$ & 1.88 & $\rm{12.68\pm0.05}$ & $\rm{750.48\pm299.76}$ & $\rm{11.51\pm0.28}$ & $\rm{0.13\pm0.01}$ & 0.07  & $\rm{1.45\pm0.07}$ & 9.73 & $\rm{12.65\pm0.02}$ & $\rm{36.19\pm0.50}$ & $\rm{8.13\pm0.06}$\\
                        212 &  0.74$^b$ & 0.26 & $\rm{12.70\pm0.08}$ & $\rm{681.61\pm263.71}$ & $\rm{11.66\pm0.31}$ & $\rm{0.05\pm0.02}$ & 45.80 & $\rm{2.03\pm0.19}$ &  & $\rm{}$ & $\rm{}$ & $\rm{}$\\
                        215 & 0.20$^b$ & 2.33 & $\rm{11.10\pm0.06}$ & $\rm{13.46\pm3.16}$ & $\rm{10.39\pm0.34}$ & $\rm{0.13\pm0.01}$ & 61.57 & $\rm{1.87\pm0.30}$ &  & $\rm{}$ & $\rm{}$ & $\rm{}$\\
                        222 & 0.62$^b$ & 3.96 & $\rm{12.02\pm0.08}$ & $\rm{113.09\pm23.50}$ & $\rm{11.42\pm0.28}$ & $\rm{0.19\pm0.08}$ & 0.02 & $\rm{1.76\pm0.09}$ &  & $\rm{}$ & $\rm{}$ & $\rm{}$\\
                        234 &  0.24$^b$ & 1.80 & $\rm{11.35\pm0.05}$ & $\rm{14.17\pm2.62}$ & $\rm{10.63\pm0.13}$ & $\rm{0.05\pm0.01}$ & 82.47 & $\rm{2.06\pm0.10}$ & 9.06 & $\rm{11.30\pm0.02}$ & $\rm{29.28\pm0.41}$ & $\rm{7.51\pm0.08}$\\
                        251 & 0.18$^d$ & 2.78 & $\rm{11.15\pm0.05}$ & $\rm{15.13\pm4.32}$ & $\rm{10.33\pm0.33}$ & $\rm{0.05\pm0.01}$ & 62.10 & $\rm{2.06\pm0.10}$ & 9.24 & $\rm{11.23\pm0.01}$ & $\rm{23.42\pm0.24}$ & $\rm{7.53\pm0.06}$\\
                        252 & 0.16$^b$ & 1.03 & $\rm{11.00\pm0.14}$ & $\rm{16.03\pm4.27}$ & $\rm{10.96\pm0.31}$ & $\rm{0.20\pm0.08}$ & 87.23 & $\rm{1.76\pm0.09}$ & 8.15 & $\rm{11.08\pm0.10}$ & $\rm{34.40\pm2.64}$ & $\rm{6.63\pm0.22}$\\
                        260 & 0.46$^b$ & 2.36 & $\rm{11.83\pm0.08}$ & $\rm{115.59\pm13.40}$ & $\rm{11.65\pm0.23}$ & $\rm{0.18\pm0.07}$ & 4.73 & $\rm{2.08\pm0.10}$ & 9.05 & $\rm{12.03\pm0.02}$ & $\rm{32.55\pm0.44}$ & $\rm{7.62\pm0.06}$\\
                        265 &  0.18$^b$ &  0.73 & $\rm{11.07\pm0.08}$ & $\rm{20.25\pm5.24}$ & $\rm{11.22\pm0.31}$ & $\rm{0.13\pm0.03}$ & 86.67 & $\rm{2.21\pm0.27}$ &  & $\rm{}$ & $\rm{}$ & $\rm{}$\\
                        338 & 0.24$^b$ &  4.67 & $\rm{11.05\pm0.1}$
                        & $\rm{24.61\pm3.85}$ & $\rm{10.84\pm0.23}$ & $\rm{0.11\pm0.03}$ & 25.28 & $\rm{2.45\pm0.31}$ & 13.08 & $\rm{11.28\pm0.02}$ & $\rm{26.10\pm0.34}$ & $\rm{7.73\pm0.06}$\\
                        360 &  0.26$^b$ & 2.04 & $\rm{11.18\pm0.16}$ & $\rm{24.79\pm6.15}$ & $\rm{11.17\pm0.53}$ & $\rm{0.13\pm0.09}$ & 50.71 & $\rm{1.98\pm0.16}$ & 6.63 & $\rm{11.62\pm0.37}$ & $\rm{16.70\pm3.18}$ & $\rm{8.79\pm0.16}$\\
                        402 & 0.64$^b$ & 3.30 & $\rm{12.30\pm0.09}$ & $\rm{411.46\pm53.08}$ & $\rm{12.32\pm0.19}$ & $\rm{0.06\pm0.04}$ & 32.24 & $\rm{2.10\pm0.15}$ &  & $\rm{}$ & $\rm{}$ & $\rm{}$\\
                        443 & 0.30$^b$ & 2.58 & $\rm{11.37\pm0.09}$ & $\rm{38.62\pm5.31}$ & $\rm{11.20\pm0.37}$ & $\rm{0.27\pm0.06}$ & 10.31 & $\rm{2.09\pm0.10}$ & 9.91 & $\rm{11.61\pm0.02}$ & $\rm{35.53\pm0.51}$ & $\rm{7.44\pm0.58}$\\
                        466 & 0.26$^b$ & 1.52 & $\rm{11.02\pm0.09}$ & $\rm{16.68\pm2.82}$ & $\rm{11.60\pm0.12}$ & $\rm{0.14\pm0.05}$ & 73.76 & $\rm{1.89\pm0.16}$ & 5.71 & $\rm{11.17\pm0.16}$ & $\rm{35.86\pm4.70}$ & $\rm{6.50\pm0.11}$\\
                        610 & 0.24$^e$ & 0.29 & $\rm{11.09\pm0.10}$ & $\rm{18.51\pm2.74}$ & $\rm{10.89\pm0.39}$ & $\rm{0.35\pm0.04}$ & 0.03 & $\rm{2.11\pm0.17}$ & 5.16 & $\rm{11.80\pm0.54}$ & $\rm{15.94\pm4.36}$ & $\rm{9.09\pm0.18}$\\
                        618 & 0.30$^b$ &  2.93 & $\rm{11.22\pm0.05}$ & $\rm{32.06\pm2.54}$ & $\rm{11.38\pm0.14}$ & $\rm{0.05\pm0.00}$ & 42.90 & $\rm{2.21\pm0.26}$ & 6.75 & $\rm{11.34\pm0.03}$ & $\rm{25.33\pm0.45}$ & $\rm{7.87\pm0.07}$\\
                        837 & 1.23$^f$ & 1.03 & $\rm{13.12\pm0.29}$ & $\rm{2834.36\pm978.20}$ & $\rm{11.46\pm0.69}$ & $\rm{0.38\pm0.08}$ & 89.90
                         & $\rm{3.15\pm0.47}$ &  & $\rm{}$ & $\rm{}$ & $\rm{}$\\
                        \hline\end{tabular}
                \newline
                \begin{samepage}
                        {Note:} Columns 2--8: CIGALE, Columns 9--12: CMCIRSED code; (1) redshift: a) \cite{Fisher95}, b) \cite{malek14}, c) \cite{jones04}, d)  \cite{sedgwick11}, e) \cite{Thomas98}, f) \cite{Wisotzki00}, ; (2) reduced $\chi^2$;  (3) dust luminosity;  (4) star formation rate; (5) stellar mass; (6) fractional contribution of AGN mid-infrared emission; (7) $\psi$ - the torus angle with respect to the line of sight \citep{fritz06}; (8) $\alpha$ parameter from the \cite{dale14} model; (9)  reduced $\chi^2$; (10) dust luminosity; (11) dust temperature; (12) dust mass.
                \end{samepage}
        \end{center}
\end{sidewaystable*}

\end{document}